\documentclass{article}

\usepackage[utf8]{inputenc}
\usepackage{amsmath}
\usepackage{amssymb}
\usepackage{graphicx}
\usepackage{dsfont}  
\usepackage{authblk}
\usepackage{hyperref}
\usepackage{siunitx}
\usepackage{xcolor}
\usepackage{enumitem}

\usepackage[font=small,skip=2pt]{caption} 
\usepackage{subfig} 

\usepackage{algorithm}
\usepackage{algpseudocode}
\usepackage{bbm}
\usepackage{siunitx}

\usepackage{float}
\usepackage{booktabs}
\usepackage{adjustbox}
\usepackage{tikz}
\usepackage{bm}

\usepackage{booktabs,tabularx,makecell,siunitx,ragged2e,array}

\usepackage{titlesec}
\titlespacing*{\subsubsection}
{0pt}{0.5em}{0.1em}

\usepackage[biblabel,nosort]{cite}

\DeclareMathOperator*{\argmin}{arg\,min}

\usetikzlibrary{shapes, arrows, positioning,decorations.markings}
\usetikzlibrary{positioning}
\usetikzlibrary{decorations.pathreplacing}
\usetikzlibrary{arrows.meta}
\usetikzlibrary{shapes.geometric}
\usetikzlibrary{calc}
\tikzstyle{vecArrow} = [thick, decoration={markings,mark=at position
	1 with {\arrow[semithick]{open triangle 60}}},
double distance=1.4pt, shorten >= 5.5pt,
preaction = {decorate},
postaction = {draw,line width=1.4pt, white,shorten >= 4.5pt}]
\tikzstyle{innerWhite} = [semithick, white,line width=1.4pt, shorten >= 4.5pt]
\hypersetup{hidelinks}
\usepackage[tmargin=1in,bmargin=1in,lmargin=0.75in,rmargin=0.75in]{geometry}

\newcommand{\R}{\ensuremath{\mathbb{R}}}
\newcommand{\Z}{\ensuremath{\mathbb{Z}}}
\newcommand{\C}{\ensuremath{\mathbb{C}}}

\providecommand{\keywords}[1]
{
	{
		\textbf{\textit{Keywords---}} #1
	}
}

\newcommand\blfootnote[1]{%
	\begingroup
	\renewcommand\thefootnote{}\footnote{#1}%
	\addtocounter{footnote}{-1}%
	\endgroup
}

\newcommand{\epssymbol}{\varepsilon}
\newcommand{\epsPlane}{\epssymbol_\mathrm{r}^\mathrm{2D}}
\newcommand{\epsMax}{\epssymbol_\mathrm{max}}
\newcommand{\epsVol}{\epssymbol_\mathrm{r}^\mathrm{3D}}

\newcommand{\ScenarioMask}{\textit{mask-only}}
\newcommand{\ScenarioDisk}{\textit{disks-only}}
\newcommand{\ScenarioSquare}{\textit{squares-only}}
\newcommand{\ScenarioDiskAndSquare}{\textit{disks-squares}}
\newcommand{\ScenarioAll}{\textit{mask-disks-squares}}

\newcommand{\windowPlane}{W^\mathrm{2D}}
\newcommand{\windowPlaneDiscrete}{W_\mathrm{d}^\mathrm{2D}}

\newcommand{\windowVolume}{W^\mathrm{3D}}
\newcommand{\windowVolumeDiscrete}{W_\mathrm{d}^\mathrm{3D}}

\newcommand{\EField}{\mathbf{E}}
\newcommand{\HField}{\mathbf{H}}

\newcommand{\JField}{\mathbf{J}_\mathrm{source}}

\newcommand{\FNOL}{FNO-L2{}}
\newcommand{\FNOConstr}{FNO-SD{}}
\newcommand{\WaveYConstr}{3D-WaveY-Net-SD{}}

\newcommand{\rvec}{\mathbf{r}}
\newcommand{\sourceJ}{\mathbf{J}_0}
\newcommand{\propDir}{\mathbf{k}}
\newcommand{\xvec}{\mathbf{x}}
\newcommand{\yvec}{\mathbf{y}}
\newcommand{\centroid}{\mathbf{c}}
\newcommand{\Centroid}{\mathbf{C}}
\newcommand{\modeVector}{\mathbf{k}}

\newcommand{\EFieldDisrete}{\mathbf{E}_\mathrm{d}}
\newcommand{\HFieldDisrete}{\mathbf{H}_\mathrm{d}}

\newcommand{\EFieldPhasorDisrete}{\widehat{\mathbf{E}}_\mathrm{d}}
\newcommand{\HFieldPhasorDisrete}{\widehat{\mathbf{H}}_\mathrm{d}}

\newcommand{\EFieldPhasorDisreteReal}{\widehat{\mathbf{E}}_\mathrm{d,split}}
\newcommand{\HFieldPhasorDisreteReal}{\widehat{\mathbf{H}}_\mathrm{d,split}}

\newcommand{\EFieldPhasorDisreteRealAlt}{{\mathbf{E}}_\mathrm{d}}
\newcommand{\HFieldPhasorDisreteRealAlt}{{\mathbf{H}}_\mathrm{d}}

\newcommand{\NA}{\textemdash}

\newcommand{\LS}[3]{\text{LS}\!\bigl(#1\allowbreak,\,#2\allowbreak,\,#3\bigr)}
\newcommand{\nm}{\,\mathrm{nm}} 

\newcolumntype{L}{>{\RaggedRight\arraybackslash}p{0.050\linewidth}}
\newcolumntype{P}{>{\Centering\arraybackslash}p{0.16\linewidth}}

\newcommand{\tikzPlot}[5]{
	\begin{tikzpicture}
		\node(im) {\includegraphics[width=#4\textwidth]{#1}};
		
		\node[inner sep=2pt, above=0.25cm of im.south,anchor=north] (xlabel) {{#2}};
		
		\node[inner sep=2pt, rotate=90,left=-0.2cm of im.west,anchor=south] (ylabel) {{#3}};
		
		\node[left=0cm of im.north west,anchor=south east,yshift=-0.6cm,xshift=0.3cm]{\Large \textbf{#5} };
	\end{tikzpicture}
}

\newcommand{\tikzPlotAdvanced}[7]{
	
	\node(#6)[#7] {\includegraphics[width=#4\textwidth]{#1}};
	
	\node[inner sep=2pt, above=0.1cm of #6.south,anchor=north] (xlabel) {{#2}};
	
	\node[inner sep=2pt, rotate=90,left=-0.1cm of #6.west,anchor=south] (ylabel) {{#3}};
	
	\node[left=0.0cm of #6.north west,anchor=south east,yshift=-0.5cm,xshift=0.1cm](#6caption){\Large \textbf{#5} };
	
}

\usepackage{import}

\usetikzlibrary{positioning}
\usetikzlibrary{3d} 
\usetikzlibrary{shapes, arrows, positioning,decorations.markings}
\usepackage{xstring}
\usetikzlibrary{fit}
\usetikzlibrary{backgrounds}
\usetikzlibrary{calc}
\usetikzlibrary{decorations.pathreplacing}

\definecolor{inputcolor}{RGB}{0, 174, 79}
\definecolor{convcolor}{RGB}{189,215,238}
\definecolor{relucolor}{RGB}{255,192,0}
\definecolor{sumcolor}{RGB}{168,208,137}
\definecolor{residualBlockColor}{RGB}{31,78,120}
\definecolor{shufflecolor}{RGB}{237,125,49}
\definecolor{sigmoidcolor}{RGB}{1,176,241}

\definecolor{densecolor}{RGB}{254,255,2}

\newdimen\mydim

\newdimen\mydims

\newcommand\defaultWidth{0.4cm}
\newcommand\defaultHeight{2.2cm}

\newcommand\defaultWidthSmall{0.0003cm}

\newcommand{\defaultDistance}{0.3cm}

\newcommand{\defaultDistanceClose}{0.2cm}

\newcommand{\block}[9]{
	\node (#4) at #1 [inner sep=1pt,draw,thin,text width=#2,text height=#3,anchor=#5,fill=#7] {
		\centering
		\rotatebox{90}{
			\textcolor{#8}{
				\centering #6
			}	
	}};
	\node(#4o) [above=0cm of #4.north,anchor=south,inner sep=0pt]{\tiny 	#9
	};
	\node(#4left)[inner sep=0pt,left=0cm of #4.west,anchor=east]{};
	\node(#4right)[inner sep=0pt,right=0cm of #4.east,anchor=west]{};	
}

\newcommand{\inputLayer}{
	\block{(0,0)}{\defaultWidth}{\defaultHeight}{inputLayer}{center}{Input $\epsVol$}{inputcolor}{black}{}
}

\newcommand{\convLayer}[5]{
	\block{([xshift=#3]#1right.east)}{\defaultWidth}{\defaultHeight}{#2}{west}{Conv}{convcolor}{black}{#4}
	
	\IfEqCase{#5}{%
		{1}{\connect{#1}{#2}}%
		{0}{}%
	}	
	
}

\newcommand{\convLayerSmall}[5]{
	\block{([xshift=#3]#1.east)}{\defaultWidthSmall}{\defaultHeight}{#2}{west}{}{convcolor}{black}{#4}
	\IfEqCase{#5}{%
		{1}{\connect{#1}{#2}}%
		{0}{}%
	}	
}

\newcommand{\Dense}[5]{
	\block{([xshift=#3]#1right)}{\defaultWidth}{\defaultHeight}{#2}{west}{Dense(#4)}{densecolor}{black}{}
	\IfEqCase{#5}{%
		{1}{\connect{#1}{#2}}%
		{0}{}%
	}		
}

\newcommand{\ReLU}[3]{
	\block{([xshift=#3]#1.east)}{\defaultWidth}{\defaultHeight}{#2}{west}{ReLU}{relucolor}{black}{}
}

\newcommand{\GeLU}[3]{
	\block{([xshift=#3]#1.east)}{\defaultWidth}{\defaultHeight}{#2}{west}{GeLU}{relucolor}{black}{}
}

\newcommand{\Sigmoid}[4]{
	\block{([xshift=#3]#1.east)}{\defaultWidth}{\defaultHeight}{#2}{west}{Sigmoid}{sigmoidcolor}{black}{}
	\IfEqCase{#4}{%
		{1}{\connect{#1}{#2}}%
		{0}{}%
	}	
}

\newcommand{\ScaledTanh}[4]{
	\block{([xshift=#3]#1.east)}{\defaultWidth}{\defaultHeight}{#2}{west}{Scaled tanh}{sigmoidcolor}{black}{}
	\IfEqCase{#4}{%
		{1}{\connect{#1}{#2}}%
		{0}{}%
	}	
}

\newcommand{\LeakyReLU}[4]{
	\block{([xshift=#3]#1.east)}{\defaultWidth}{\defaultHeight}{#2}{west}{\small Leaky ReLU}{relucolor}{black}{}
	\IfEqCase{#4}{%
		{1}{\connect{#1}{#2}}%
		{0}{}%
	}	
}

\newcommand{\ReLUSmall}[3]{
	\block{([xshift=#3]#1.east)}{\defaultWidthSmall}{\defaultHeight}{#2}{west}{}{relucolor}{black}{}
}

\newcommand{\LeakyReLUSmall}[3]{
	\block{([xshift=#3]#1.east)}{\defaultWidthSmall}{\defaultHeight}{#2}{west}{}{relucolor}{black}{}
}

\newcommand{\SUM}[4]{
	\block{([xshift=#3]#1.east)}{\defaultWidth}{\defaultHeight}{#2}{west}{\tiny Elementwise sum}{sumcolor}{black}{}
	\IfEqCase{#4}{%
		{1}{\connect{#1}{#2}}%
		{0}{}%
	}
}

\newcommand{\SUMSmall}[4]{
	\block{([xshift=#3]#1.east)}{\defaultWidthSmall}{\defaultHeight}{#2}{west}{}{sumcolor}{black}{}
	\IfEqCase{#4}{%
		{1}{\connect{#1}{#2}}%
		{0}{}%
	}		
	
}

\newcommand{\PixelShuffle}[4]{
	\block{([xshift=#3]#1.east)}{\defaultWidth}{\defaultHeight}{#2}{west}{\tiny PixelShuffle}{shufflecolor}{black}{}
	
	\IfEqCase{#4}{%
		{1}{\connect{#1}{#2}}%
		{0}{}%
	}
}

\newcommand{\Upsampling}[4]{
	\block{([xshift=#3]#1.east)}{\defaultWidth}{\defaultHeight}{#2}{west}{\tiny Upsampling 1.25}{shufflecolor}{black}{}
	
	\IfEqCase{#4}{%
		{1}{\connect{#1}{#2}}%
		{0}{}%
	}
}

\newcommand{\Fourier}[5]{
	\block{([xshift=#3]#1.east)}{\defaultWidth}{\defaultHeight}{#2}{west}{Fourier layer}{sumcolor}{black}{#4}
	\IfEqCase{#5}{%
		{1}{\connect{#1}{#2}}%
		{0}{}%
	}
}

\newcommand{\residualBlock}[4]{

	\node(#2ref)[right=#3 of #1.east]{};
	\node(#2refright) [right=0cm of #2ref] {};
	\convLayer{#2ref}{#2conv1}{-0.0cm}{k3n64s1}{0}
	\ReLU{#2conv1}{#2conv1_relub}{0cm}
	\convLayer{#2conv1_relub}{#2conv2b}{\defaultDistanceClose}{k3n64s1}{1}
	\ReLU{#2conv2b}{#2conv2_relub}{0cm}	
	\SUM{#2conv2_relub}{#2sum}{\defaultDistanceClose}{1}	
	
	\begin{pgfonlayer}{background}
		\node(#2)[draw,  fill=residualBlockColor, 
		inner sep=0pt, outer sep=0pt,
		fit={([xshift=-0.3cm,yshift=0.2cm]#2conv1.north west) ([xshift=0.1cm,yshift=-0.05cm]#2sum.south east)}
		]{
		};
	\end{pgfonlayer}

	\path let \p{#1left}=(#1left),\p{#2.west}=(#2.west) in node[inner sep=0pt](#2left) at (\x{#2.west},\y{#1left}) {};
	
	\path let \p{#2left}=(#2left),\p{#2.east}=(#2.east) in node[inner sep=0pt](#2right) at (\x{#2.east},\y{#2left}) {};

	\IfEqCase{#4}{%
		{1}{\connect{#1}{#2}}%
		{0}{}%
	}
	\draw [->] ([xshift=-0.1cm]#2left.west) |-([shift={(-0.1cm,-1.5cm)}]#2.west)-| ([]#2sum.south);
	
}

\newcommand{\residualBlockSmall}[4]{
	
	\node(#2ref)[right=#3 of #1right.east,inner sep=0pt,anchor=west]{};
	\convLayerSmall{#2ref}{#2conv1b}{-0cm}{}{0}
	\ReLUSmall{#2conv1b}{#2conv1_relub}{0cm}	
	\convLayerSmall{#2conv1_relub}{#2conv2b}{\defaultDistanceClose}{}{1}
	\ReLUSmall{#2conv2b}{#2conv2_relub}{0cm}	
	\SUMSmall{#2conv2_relub}{#2sumb}{\defaultDistanceClose}{1}		
	
	\begin{pgfonlayer}{background}
		\node(#2)[draw,  fill=residualBlockColor, 
		inner sep=0pt, outer sep=0pt,
		fit={([xshift=-0.05cm,yshift=0.2cm]#2conv1b.north west) ([xshift=0.05cm,yshift=-0.05cm]#2sumb.south east)}
		]{
		};
	\end{pgfonlayer}
	
	\path let \p{#1left}=(#1left),\p{#2.west}=(#2.west) in node[inner sep=0pt](#2left) at (\x{#2.west},\y{#1left}) {};
	
	\path let \p{#2left}=(#2left),\p{#2.east}=(#2.east) in node[inner sep=0pt](#2right) at (\x{#2.east},\y{#2left}) {};
	
	\IfEqCase{#4}{%
		{1}{\connect{#1}{#2}}%
		{0}{}%
	}
	\draw [->] ([xshift=-0.1cm]#2left.west) |-([shift={(-0.1cm,-1.5cm)}]#2.west)-| ([]#2sumb.south);

}

\newcommand{\connect}[2]{	
	\draw[->] ([xshift=0.01cm]#1right) --	([xshift=-0.01cm]#2left);
}

\newcommand{\discriminatorBlock}[5]{

	\node(#2ref)[right=#3 of #1right]{};
	\node(#2refright) [right=0cm of #2ref] {};
	\convLayer{#2ref}{#2conv1}{-0.0cm}{#5}{0}
	\LeakyReLU{#2conv1}{#2conv1_relub}{0cm}{0}
	
	\node(#2left)[left=0cm of #2conv1.west,inner sep=0pt]{};
	\node(#2right)[right=0cm of #2conv1_relub.east,inner sep=0pt]{};
	
		\IfEqCase{#4}{%
			{1}{\connect{#1}{#2}}%
			{0}{}%
		}	
		
	}
	
	\newcommand{\discriminatorBlockSmall}[5]{
		
		\node(#2ref)[right=#3 of #1right]{};
		\node(#2refright) [right=0cm of #2ref] {};
		\convLayerSmall{#2ref}{#2conv1}{-0.0cm}{#5}{0}
		\LeakyReLUSmall{#2conv1}{#2conv1_relub}{0cm}
		
		\begin{pgfonlayer}{background}
			\node(#2)[draw,  
			inner sep=0pt, outer sep=0pt,
			fit={([xshift=-0.4cm,yshift=0.2cm]#2conv1.north west) ([xshift=0.4cm,yshift=-0.05cm]#2conv1_relub.south east)}
			]{
			};
		\end{pgfonlayer}

		\path let \p{#1left}=(#1left),\p{#2.west}=(#2.west) in node[inner sep=0pt](#2left) at (\x{#2.west},\y{#1left}) {};
		
		\path let \p{#2left}=(#2left),\p{#2.east}=(#2.east) in node[inner sep=0pt](#2right) at (\x{#2.east},\y{#2left}) {};
		\IfEqCase{#4}{%
			{1}{\connect{#1}{#2}}%
			{0}{}%
		}	
		
	}

\title{Physics-informed Neural Operators for Predicting 3D Electromagnetic Fields Transformed by Metasurfaces}
\author{Orkun Furat$^{\ast}$,Vinay Chakravarthi Gogineni, Henrik Bindslev, Esmaeil S. Nadimi

}
\date{}

\begin{document}

\maketitle
\vspace{-4em}
\begin{center}
	\it
	Applied AI and Data Science Unit, Maersk Mc-Kinney Moller Institute, Faculty of Engineering, \\ University of Southern Denmark, Odense, Denmark
\end{center}
\blfootnote{
	$^\ast$Corresponding author \\ \textit{Email address: ofu@mmmi.sdu.dk}
}

\begin{abstract}
	\noindent
	Metasurfaces, typically realized as arrays of nanopillars, transform electromagnetic (EM) fields depending on their geometry and spatial arrangement. For solving the inverse problem of designing new metasurfaces that transform EM fields in a desirable manner, it is often necessary to explore large design spaces through full-wave simulations that can be computationally demanding. In this work, we demonstrate that neural operators, which are artificial neural network architectures designed to learn operators between function spaces, can effectively approximate the differential operators underlying Maxwell's equations, enabling their use as fast and accurate 3D surrogate models that can predict 3D EM fields transformed by metasurfaces. To calibrate neural operators, we generate synthetic training data consisting of 3D metasurface geometries together with their associated 3D EM fields obtained by numerically solving Maxwell's equations. Using the generated synthetic data, we train physics-informed neural operators to minimize physical inconsistencies of predicted EM fields by incorporating residuals that capture deviations from Maxwell's equations. We observe that a training dataset consisting of fewer than 5000 examples already suffices to achieve reasonable results. In particular, our experiments show that the resulting 3D surrogate model achieves high predictive performance across a wide range of metasurface geometries, including types of structures not encountered during training. Notably, it predicts diffraction efficiencies with relative errors of \SI{3.9}{\percent} and provides a 67-fold speedup compared to conventional 3D simulations; moreover, relative errors of \SI{6.1}{\percent} and \SI{10.2}{\percent} are observed for the magnetic and electric field values, respectively. Overall, once trained, our 3D surrogate model can rapidly predict EM fields for previously unseen metasurface geometries, which can facilitate efficient gradient-based design of nanostructured materials for EM wave control. 
\end{abstract}
\keywords{Metasurface, Surrogate model, Physics-informed neural network, Neural Operator.
}

\section{Introduction}
Metasurfaces are increasingly deployed in various optical applications, including focusing, beam steering and holography, by enabling the manipulation of phase, amplitude and polarization \cite{Jeong:24,leng2024meta,hu2024metasurface,ghaffari2023integrated,ren2020complex}.
However, current designs often exhibit limitations in diffraction efficiency and functional bandwidth, highlighting the need for more effective designs and, consequently, for methodologies that enable the derivation of such improved designs \cite{li2024heterogeneousfreeformmetasurfacesplatform,muhammad2024radiationless}.
To address these challenges, it is essential to understand how the underlying geometry of metasurfaces transforms electromagnetic (EM) fields, which in turn allows for tailoring improved metasurfaces.
More precisely, the functionality of such  metasurfaces arises from the spatial arrangement of subwavelength scatterers, typically nanopillars with high refractive indices and tunable shapes, sizes, and orientations \cite{UenoLinYangAnMartin}.
Through these geometric degrees of freedom, the nanopillars can enable control over phase, amplitude and polarization of transformed EM fields \cite{PhysRevLett.118.113901}. 
Tailored designs of metasurface structures enable optical devices to achieve highly specific and application-driven optical properties, e.g., for holographic image formation \cite{ren2020complex}.
For the purpose of designing metasurfaces with desired properties, quantitative structure-property relationships that link the geometry of nanopillars and the associated EM field are of particular interest.

However, establishing such relationships  through physical experiments can be costly both in time and resources, especially when large numbers of metasurface designs must be investigated \cite{9806373}. For this reason, numerical simulations based on solving Maxwell's equations are often used  to investigate quantitative structure-property relationships \cite{ma2021deep}. 
For example, full-wave simulations can be deployed which are able to accurately resolve subwavelength features of each nanopillar as well as near-field interactions between adjacent nanopillars \cite{10018880}. To capture these different effects, simulations typically require
large computational domains at fine spatial discretization, which can substantially increase computational costs \cite{elsawy2019global}. Consequently, even single forward simulations  may require significant computational resources which can make exploring a large design space for an optimized metasurface infeasible with conventional solvers.

These limitations have motivated the deployment of so-called surrogate models that approximate the mapping from a metasurface geometry to the corresponding transformed EM field. A wide range of machine-learning approaches have been explored for deriving such surrogate models. In \cite{Li:19} fully connected neural networks have been deployed to predict resonance spectra in plasmonic nanostructures. Similar neural network architectures have been considered in \cite{an2019deep} for training a surrogate model that predicts EM fields transformed by differently sized nanopillars, followed by deploying the surrogate model in an inverse-design loop. Convolutional neural networks (CNNs), in particular U-Net-like architectures \cite{ronneberger2015}, have been deployed in \cite{wiecha2019deep} as surrogate models. Several studies have investigated surrogate models that incorporate physical knowledge into their training procedure. For example, neural-network-based surrogate models that take the geometries of adjacent nanopillars as input have been proposed to capture effects of EM coupling \cite{an2022mutual,zhelyeznyakov2021}. 
These approaches mitigate some of the limitations of surrogate models for single nanopillars, which can lead to an improved prediction performance, in particular, in the presence of non-identical neighboring nanopillars.

In addition to advances that can take the coupling of fields from multiple nanopillars into account, recently physics-informed neural networks (PINNs) \cite{RAISSI2019686} and related computational models have been explored as surrogate models for EM field simulations \cite{10203003}. Typically, these approaches incorporate residuals quantifying deviations from Maxwell’s equations (also called \emph{Maxwell residuals}) directly into the loss function to train the PINNs. Consequently, trained surrogate models are encouraged to satisfy underlying physical laws, even when training data has been limited or noisy. For example, in \cite{Chen2022Stanford} CNNs were trained with a loss function that quantified errors in the training data (consisting of computer-generated metasurfaces and simulated EM fields) and Maxwell-residual terms. This approach was demonstrated for two-dimensional structures (i.e., 3D metasurfaces with no variability in one in-plane direction) and was further combined with generative artificial intelligence (AI) techniques to enable freeform design of nanopillars.
In addition, recently the applicability of data-free PINNs as surrogate models for metasurfaces has been explored, i.e., neural networks that have been solely trained with a loss that quantifies residuals of differential operators. For example, in \cite{zhelyeznyakov2023large} a PINN-based framework is demonstrated that numerically solves Maxwell's equations via residual minimization.
This is conceptually similar to conventional FDTD solvers which also try to minimize such residuals; however, in the PINN-based framework, the values of EM fields on a discrete grid are parameterized via the weights of a neural network. Such data-free PINNs have, in particular, been employed for the inverse design of large metasurfaces.

Despite recent advances, existing surrogate models still have some limitations. Supervised neural networks typically require large datasets for training purposes, which necessitates large numbers of numerical simulations of EM fields \cite{ma2021deep}. Physics-informed approaches can reduce or even eliminate the need for training data. However, to the best of our knowledge, existing PINN-based surrogate models for predicting EM fields transformed by metasurfaces are restricted to simplified geometries. For example, in \cite{Chen2022Stanford,zhelyeznyakov2023large}, the metasurfaces consist of nanoridges that are invariant along one in-plane direction. This geometric symmetry permits Maxwell's equations to be solved on a 2D domain, enabling the deployment of computationally feasible 2D surrogate models. In contrast, metasurfaces comprising more general nanopillar geometries lack such symmetries and therefore require surrogate models that solve Maxwell's equations in 3D domains.

The transition from surrogate models that predict 2D EM fields to those capable of handling 3D metasurfaces to predict full 3D EM fields substantially increases memory and computational demands \cite{Chen2022Stanford}. 
Solutions to Maxwell's equations exhibit long-range spatial correlations, so accurate EM field  predictions should take into account information from distant parts of the simulation domain. However, accounting for such nonlocal interactions can further increase the  complexity of 3D surrogate models. For example,
conventional CNN-based surrogate models typically have a bounded, finite receptive field (sometimes also referred to as field of view) \cite{Rangel2024}, restricting their ability to exhibit long-range correlations. Expanding the receptive field of 3D CNNs by increasing their depth or kernel size leads to rapidly growing memory requirements, making such approaches computationally prohibitive. 

Recently in scientific computing, so-called neural operators have emerged as powerful surrogate models capable of learning mappings between function spaces, particularly, for approximating differential operators \cite{azizzadenesheli2024neural,JMLR:v24:21-1524}. Their ability to efficiently capture non-local correlations, to operate independently of discretization and to generalize across resolutions has made them attractive for training surrogate models for a wide range of physical phenomena.

In the present paper, we demonstrate how neural operators can be used as surrogate models to predict 3D EM fields transformed by 3D metasurfaces. The main contributions are as follows:
\begin{enumerate}[label=(\roman*)]
	\item A diverse set of 3D metasurfaces is generated using methods from stochastic geometry \cite{chiu2013stochastic}, and the corresponding 3D EM fields are computed using a finite-difference time-domain (FDTD) simulations \cite{mahlau2024flexibleframeworklargescalefdtd}.
	\item A neural-operator-based surrogate model is deployed for predicting full 3D EM fields of metasurfaces, with Maxwell residuals incorporated into the loss function to reward physical consistency.
	\item The proposed neural-operator approach is benchmarked against a 3D adaptation of an established CNN-based surrogate model \cite{Chen2022Stanford}.
	\item A quantitative analysis of prediction performance is carried out, including an investigation of how geometric descriptors of metasurfaces influence the prediction performance.
	\item Surrogate models are additionally trained on datasets which only exhibit single types of  geometries in order to assess their capability to generalize to previously unseen geometries.
	\item A runtime analysis is performed to evaluate computational efficiency of 3D surrogate models relative to FDTD simulations.
	\item The super-resolution capability of neural operators is demonstrated, showing that  3D EM fields can be predicted at higher spatial resolutions than those present in the training data.
\end{enumerate}
The results reported in our study paper indicate that the combination of fast inference, resolution flexibility and robust generalization across diverse geometries makes neural operators suitable surrogate models that could facilitate iterative inverse-design schemes for new, improved metasurfaces in 3D.

The remainder of this paper is structured as follows. In Section~\ref{sec:Methods}, we first describe the generation of metasurfaces (Section~\ref{sec:generation:metasurfaces}) and the numerical simulation of corresponding EM fields (Section~\ref{sec:simulation}). Then, in Section~\ref{Sec:Surrogate} we describe the considered neural-operator-based surrogate model, the physics-based loss function and the training procedure. In
Section~\ref{sec:results} the results of our method are presented and discussed. In particular, in Section~\ref{sec:results:training} an analysis of training dynamics of the surrogate models considered is presented, followed by a quantitative evaluation of their performance in Section~\ref{sec:quantitative}. Furthermore, in Section~\ref{sec:results:influence} we investigate how the geometry of metasurfaces  influences model performance to provide guidance on structural scenarios for which the considered models struggle to provide accurate predictions of EM fields. A runtime analysis that compares surrogate models with conventional numerical solvers is presented in Section~\ref{sec:results:runtime}.  Finally, in Section~\ref{sec:results:sr} we demonstrate that neural operators enable resolution-independent inference. The paper concludes with a summary in Section~\ref{sec:conclusions}.

\section{Methods}
\label{sec:Methods}
In this section, we describe the methods used in this study to develop neural-operator-based surrogate models that take 3D metasurfaces as input and predict the associated transformed 3D EM fields.
In Section~\ref{sec:generation:metasurfaces} differently structured metasurfaces are generated, using methods from stochastic geometry \cite{chiu2013stochastic}. 
In particular, so-called excursion sets of random fields are deployed as  a flexible, interpretable probabilistic model that can generate freeform nanopillars.
Moreover, packing algorithms are employed to arrange nanopillars with disk-shaped and square-shaped cross-sections without overlap \cite{guo2022two}. In total, five different structural scenarios are considered, and for each scenario approximately 1000 metasurfaces with distinct structural statistics are generated.
Section~\ref{sec:simulation} outlines the numerical procedure for computing the corresponding EM fields for each generated metasurface. Then, Section~\ref{Sec:Surrogate} introduces the surrogate model that is trained with pairs of metasurfaces and their simulated EM fields.

\subsection{Generation of metasurfaces}\label{sec:generation:metasurfaces}
\begin{figure}[h]
\subfloat[]{
	\begin{tikzpicture}
		\tikzPlotAdvanced{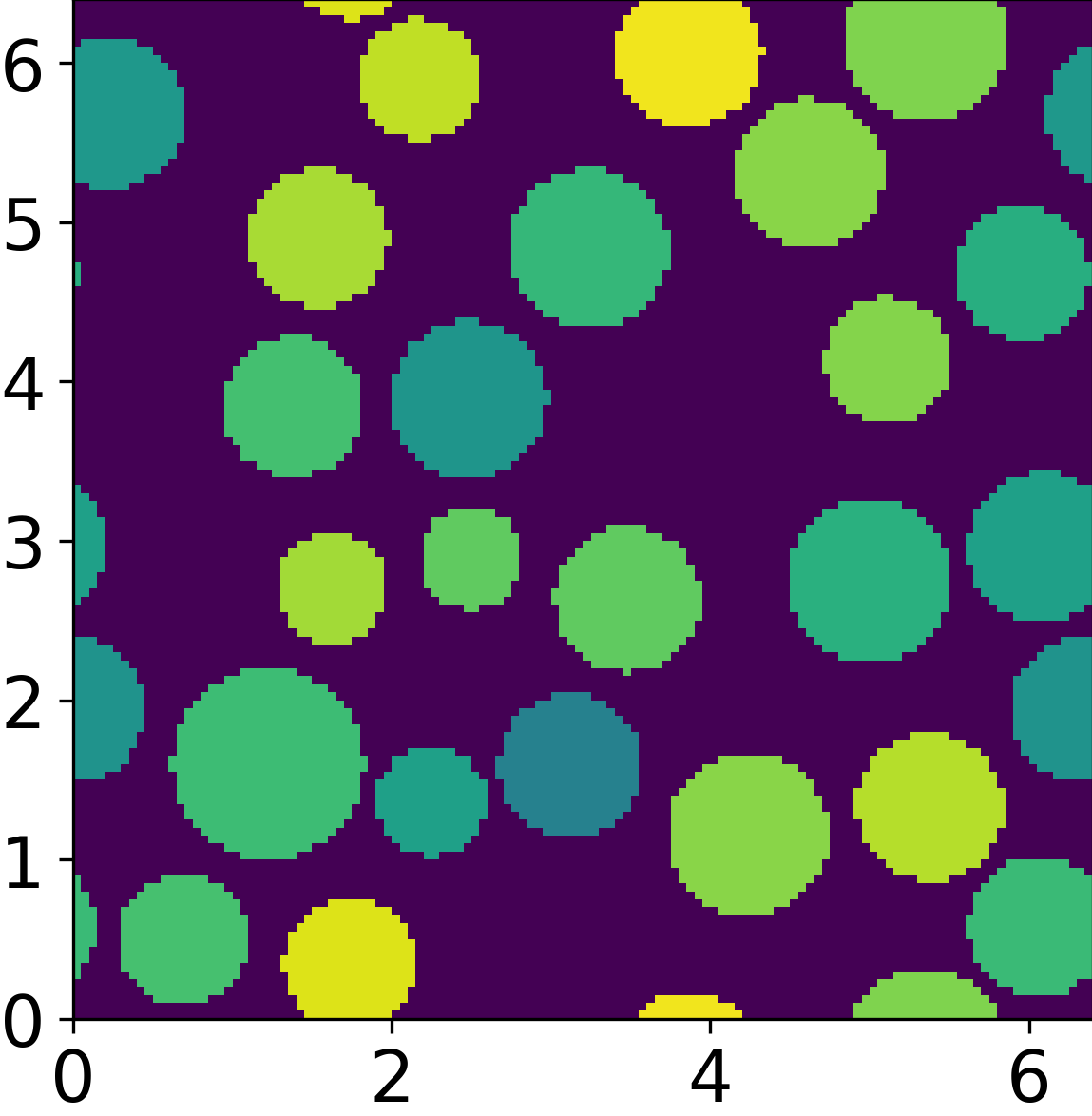}{$x$  [\si{\micro\meter}]}{$y$  [\si{\micro\meter}]}{0.29}{}{reforig}{inner sep=0pt}
	\end{tikzpicture}
}
\subfloat[]{
	\begin{tikzpicture}
		\tikzPlotAdvanced{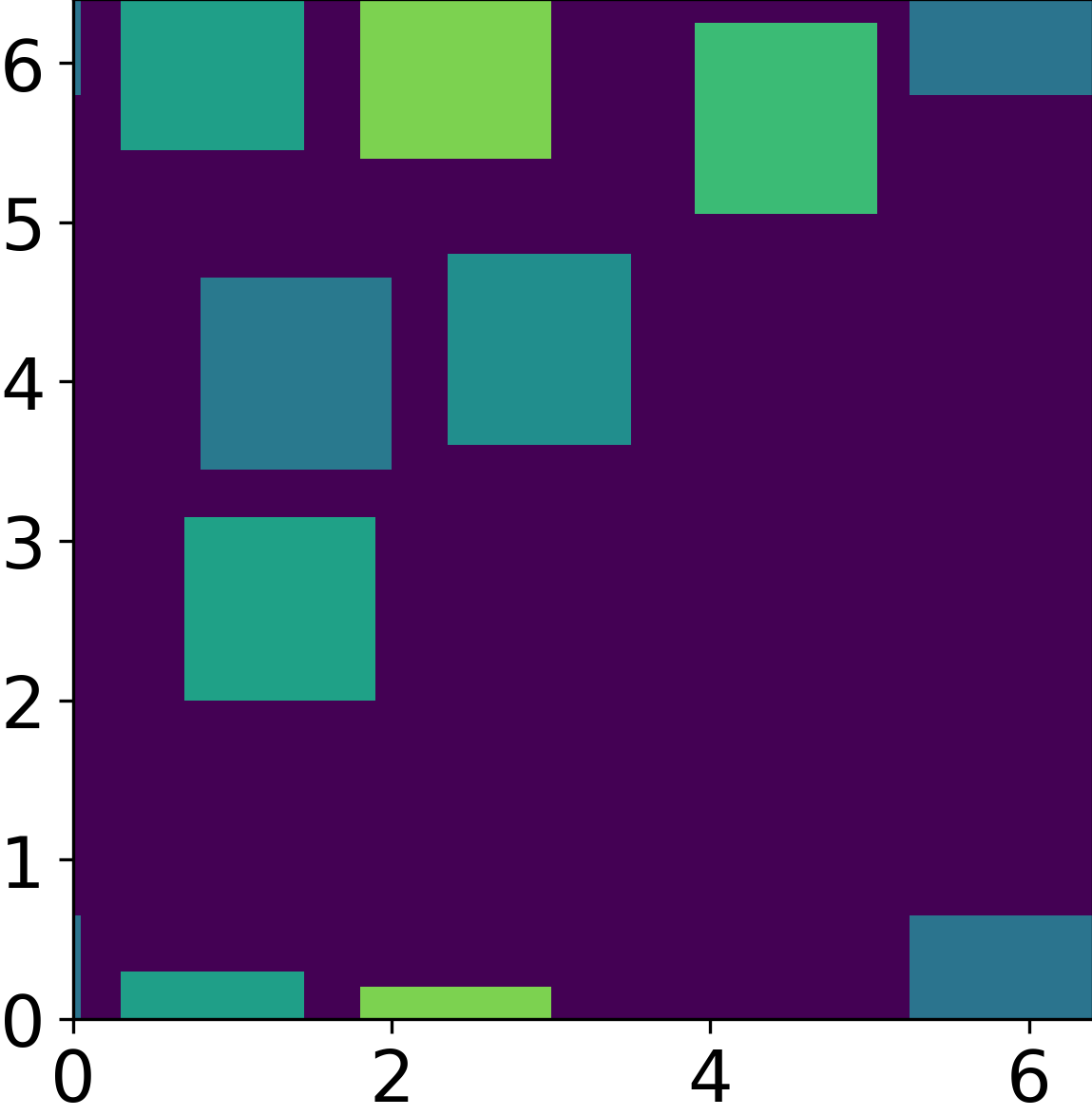}{$x$  [\si{\micro\meter}]}{$y$  [\si{\micro\meter}]}{0.29}{}{reforig}{inner sep=0pt}
	\end{tikzpicture}
}
\subfloat[]{
	\begin{tikzpicture}
		\tikzPlotAdvanced{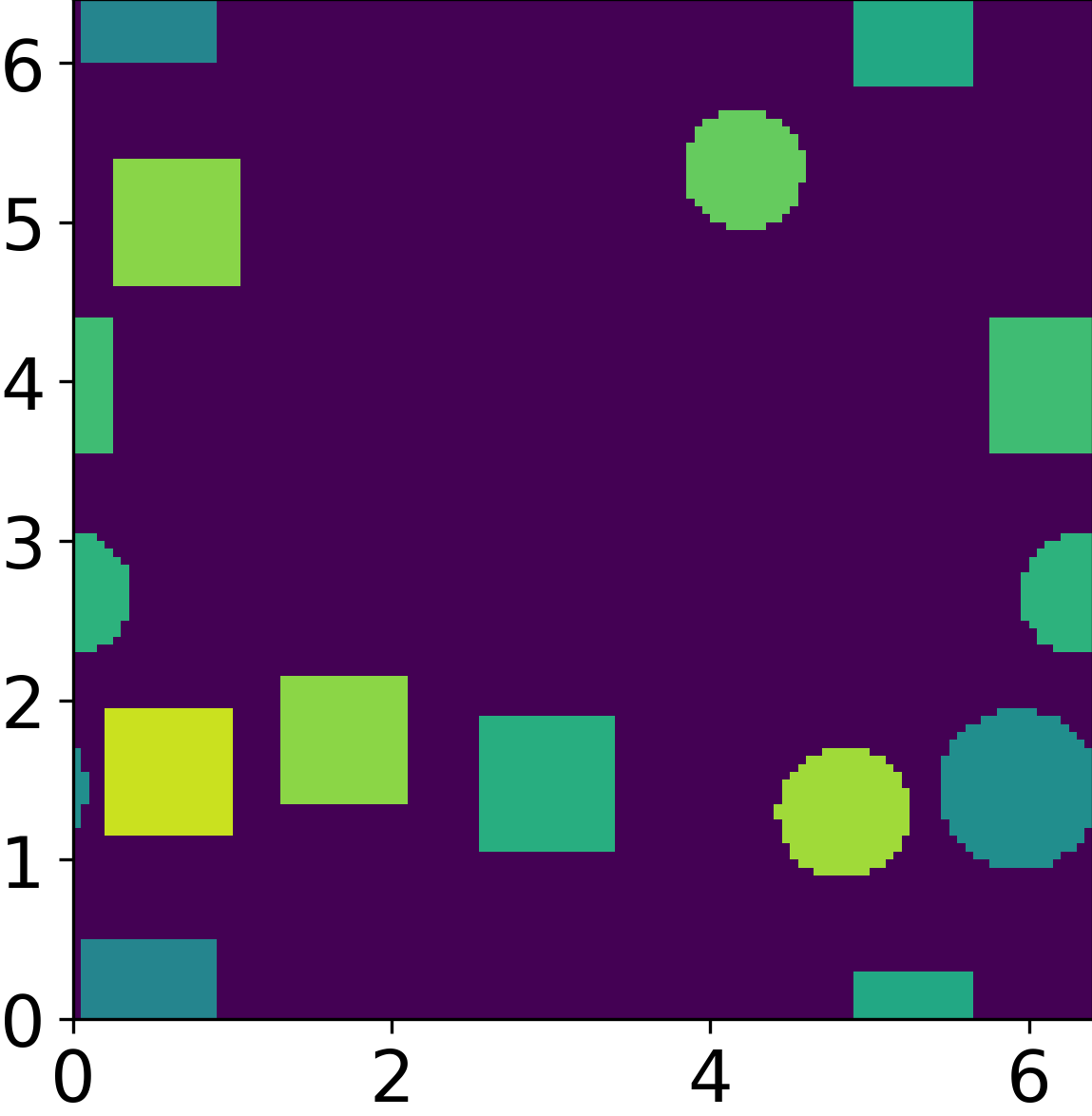}{$x$  [\si{\micro\meter}]}{$y$  [\si{\micro\meter}]}{0.29}{}{reforig}{inner sep=0pt}
	\end{tikzpicture}
}

\vspace{-1.15em}
\subfloat[]{
	\begin{tikzpicture}
		\tikzPlotAdvanced{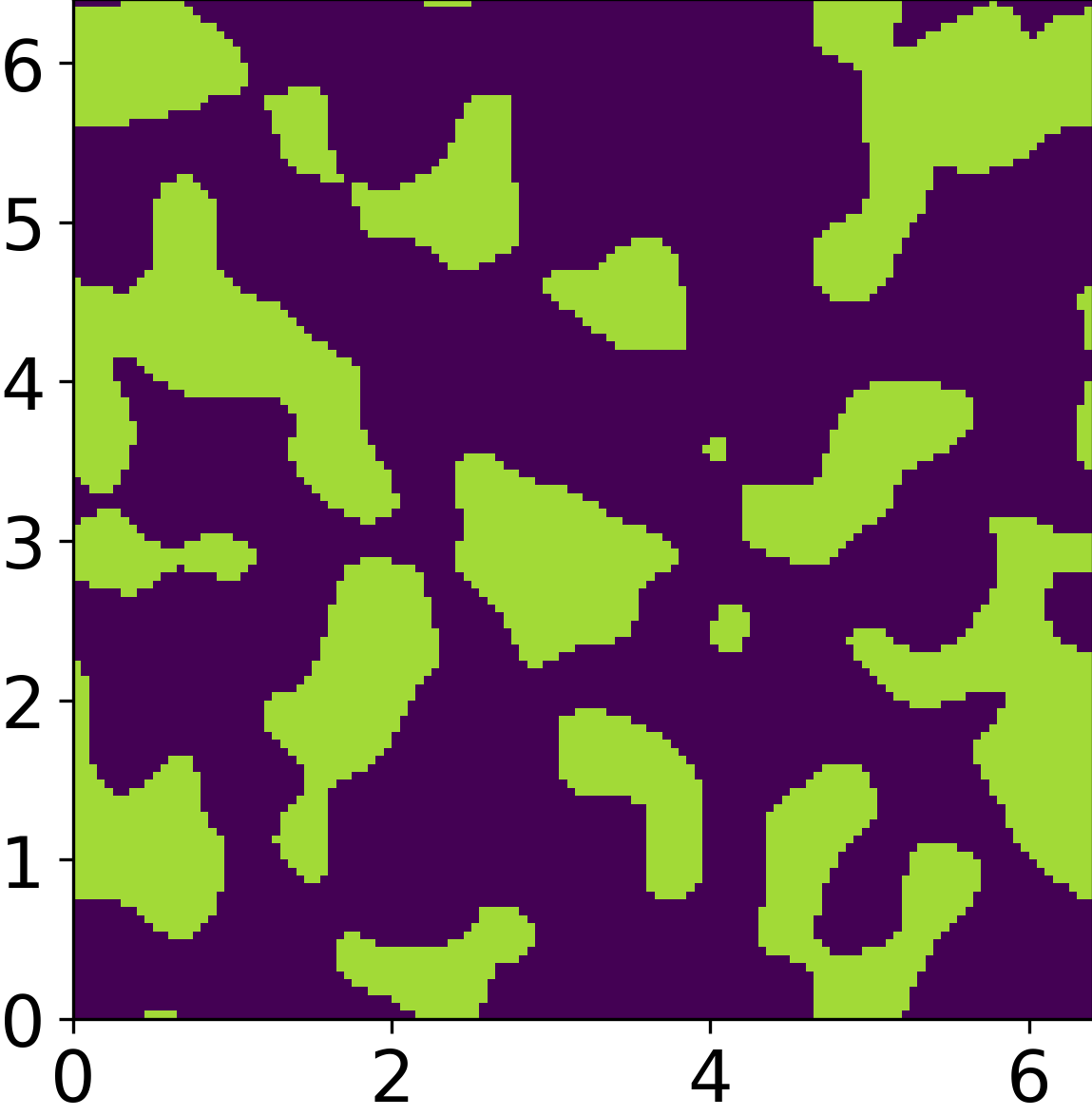}{$x$  [\si{\micro\meter}]}{$y$  [\si{\micro\meter}]}{0.29}{}{reforig}{inner sep=0pt}
	\end{tikzpicture}
}
\subfloat[]{
	\begin{tikzpicture}
		\tikzPlotAdvanced{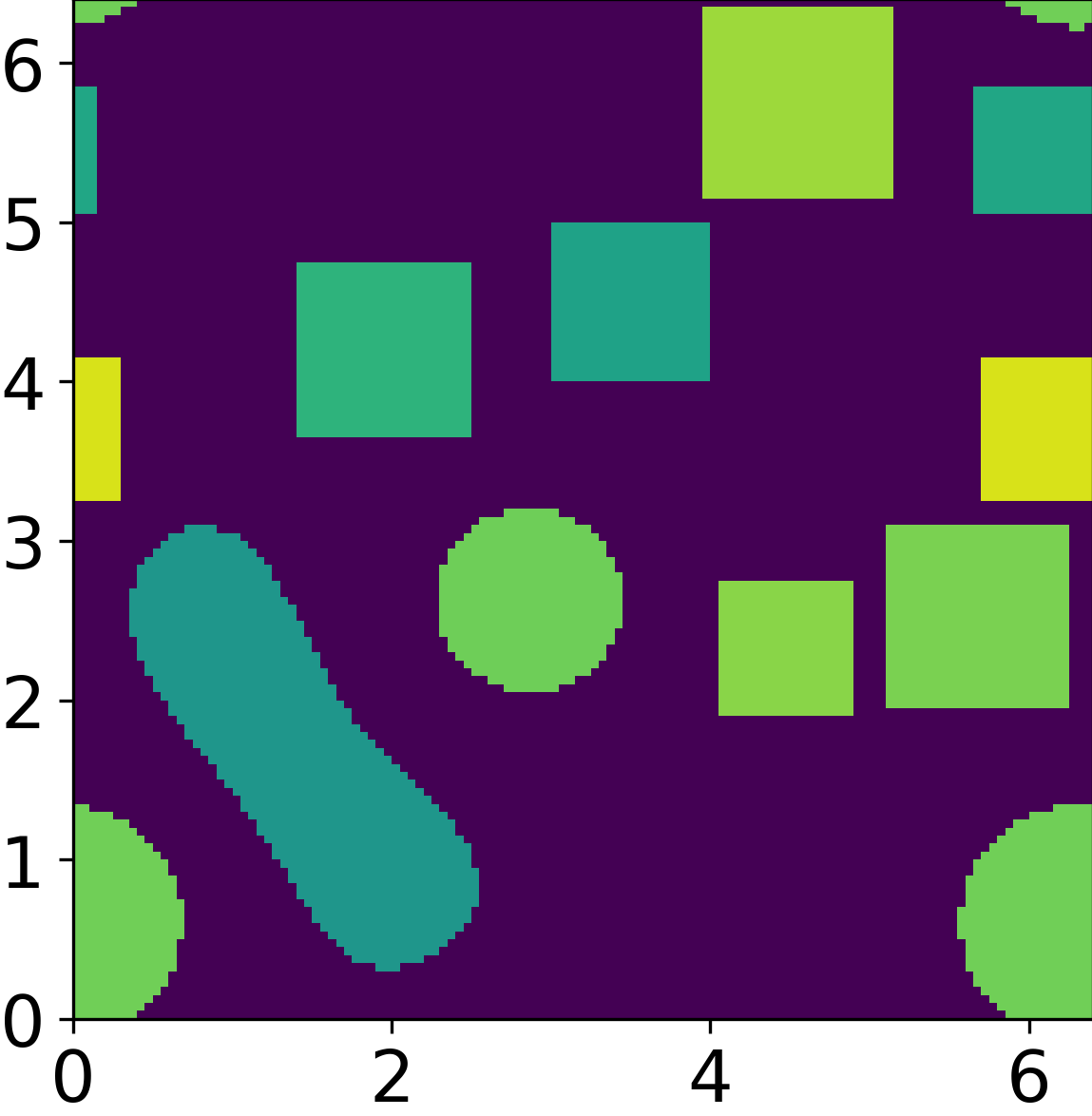}{$x$  [\si{\micro\meter}]}{$y$  [\si{\micro\meter}]}{0.29}{}{reforig}{inner sep=0pt}
	\end{tikzpicture}
}
\raisebox{2.5cm}{%
	\begin{tabular}{c}
		\subfloat[]{
			\begin{tikzpicture}
				\tikzPlotAdvanced{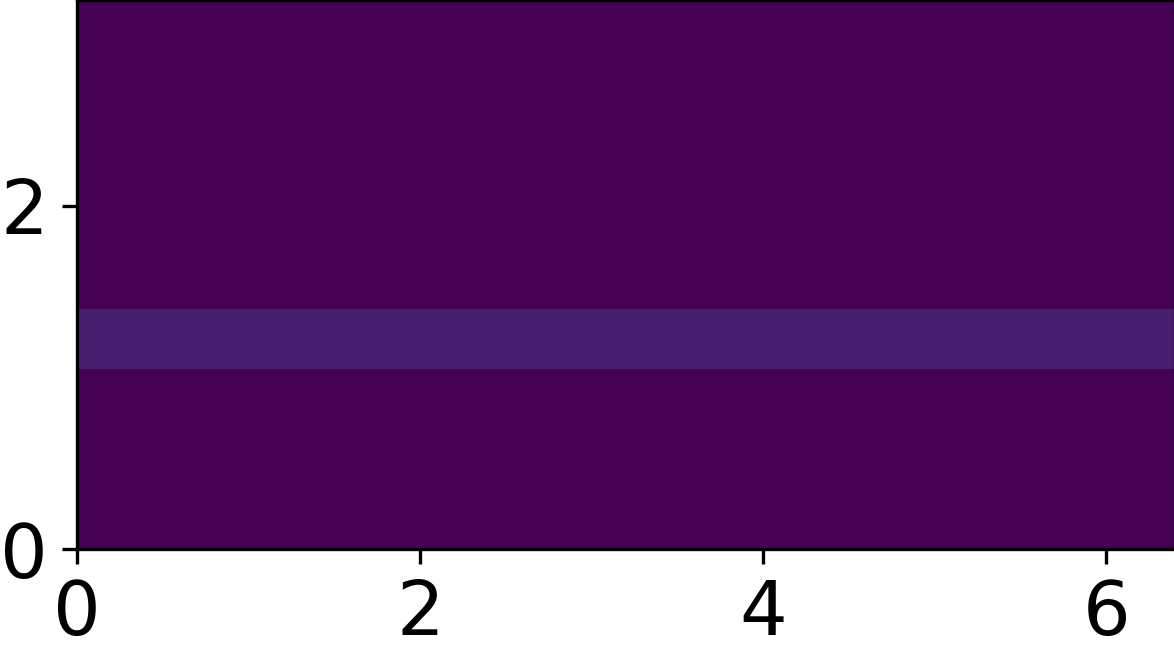}{$x$  [\si{\micro\meter}]}{$z$  [\si{\micro\meter}]}{0.2}{}{reforig}{inner sep=0pt}
			\end{tikzpicture}
		}
		\\[-1em]
		\subfloat[]{
			\begin{tikzpicture}
				\tikzPlotAdvanced{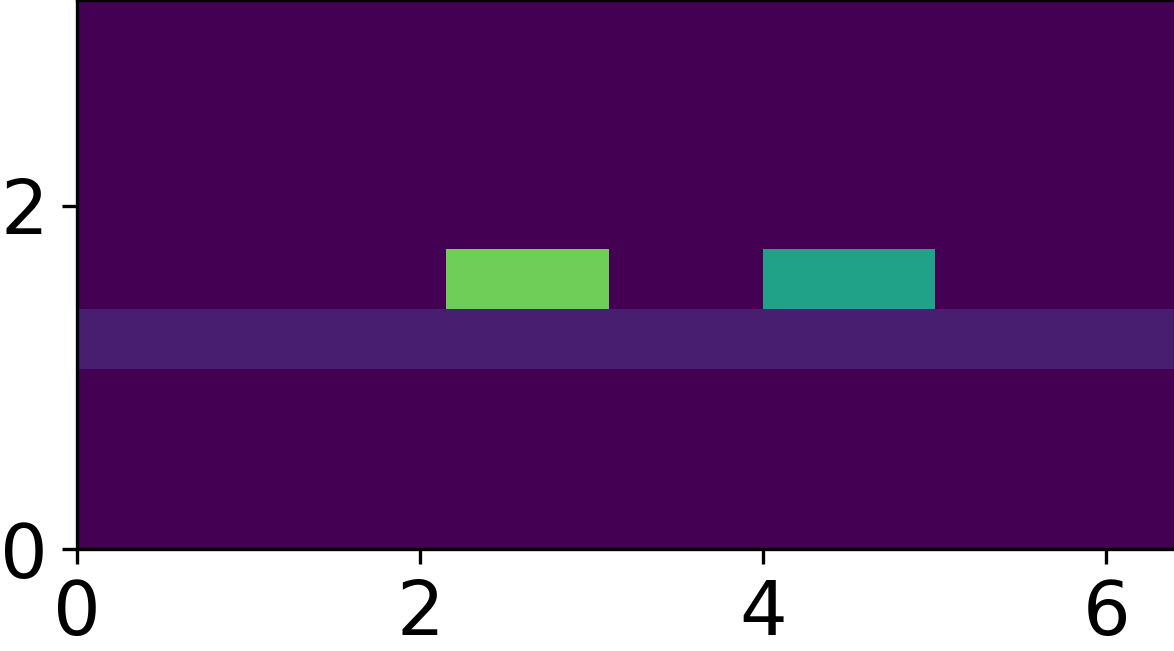}{$y$  [\si{\micro\meter}]}{$z$  [\si{\micro\meter}]}{0.2}{}{reforig}{inner sep=0pt}
			\end{tikzpicture}
		}
	\end{tabular}
}
\hspace{-1em}
	\begin{tikzpicture}	
		\node(ref){\phantom{a}};
		\node[above=0.4cm of ref, inner sep=0pt ] (colorbar) {
			\includegraphics[height=4.25cm]{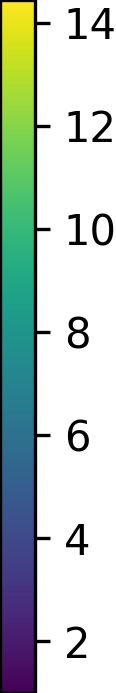}
		};
		\node[right=0.3cm of colorbar.south east, anchor=south west , inner sep=0pt,rotate=90,xshift=0.3cm ] {
			relative permittivity
		};
		
	\end{tikzpicture}	

	\caption{\textbf{Generated metasurfaces.}  Cross-sections parallel to the $x$-$y$-plane of generated metasurfaces associated with the (a) \emph{disks-only}, (b) \emph{squares-only}, (c)  \emph{disks-squares}, (d)  \emph{freeform-only} and (e) \emph{freeform-disks-squares} scenarios.
		Perpendicular cross-sections of (e) for the planes with $y=\SI{3200}{\nano\meter}$ and $x=\SI{3200}{\nano\meter}$ are visualized in (f) and (g), respectively.
	}
	\label{fig:visual:eps}
\end{figure}

In this section, we describe the methods from stochastic geometry  deployed in this paper to formulate stochastic 3D models that can be used to generate 3D metasurfaces.
To facilitate numerical simulations in Section~\ref{sec:simulation} with periodic boundary conditions in the $x$- and $y$-directions, all generated geometries are periodic in the lateral plane. The stochastic 3D models will generate metasurfaces in two steps: (i) first a 2D cross-section of the metasurface in the plane parallel to the $x$-$y$-plane is generated (see Figs.~\ref{fig:visual:eps}a-e for some examples), and (ii) this cross-section is extruded in the $z$-direction by assigning it with a height, followed by placing the resulting volumetric object on a SiO$_2$ substrate that is surrounded by vacuum, see Figs.~\ref{fig:visual:eps}e-g for visualizations of three orthogonal cross-sections through the same metasurface. In the following, we provide a summary of the stochastic 3D models used in this work; further details are given in Appendix~\ref{appendix:generation:metasurfaces}.

\subsubsection{Packings of disks and squares}
To stochastically model metasurfaces with relatively regularly shaped nanopillars, we consider 2D cross-sections that are generated as random packings of non-overlapping disks and/or squares in a rectangular (periodic) window $\windowPlane=[0,w_1)\times[0,w_2)$, where each disk or square represents the $x$-$y$-cross-section of a nanopillar. The sizes, shapes and relative permittivities of these cross-sections are drawn from prescribed distributions. Then, a packing algorithm is used to place them while approximately enforcing no overlap and a target area fraction of the placed objects. This procedure yields a 2D distribution $\epsPlane \colon \windowPlane \to [1,\epsMax]$ of relative permittivities representing a cross-section of the metasurface, where the value 1 corresponds to vacuum and $\epsMax=15$ is an upper bound for the relative permittivities.

In practice, due to computational constraints, $\epsPlane$ is evaluated not at all points in $\windowPlane$ but on a regular grid $\windowPlaneDiscrete=\{0\cdot \rho,1\cdot\rho,\dots,(n_\mathrm{x}-1)\rho\} \times \{0\cdot \rho,1\cdot\rho,\dots,(n_\mathrm{y}-1)\rho\}\subset \windowPlane$, where $n_\mathrm{x},n_\mathrm{y}>0$ denote the image resolution and $\rho>0$ is the pixel size. The resulting discretized version $\epsPlane \colon \windowPlaneDiscrete \to [1,\epsMax]$ will serve as the basis for constructing 3D distributions of relative permittivities. See Section~\ref{sec:database} for further details.

Overall, the model for generating disk/square packings is governed by several parameters that determine (i) the size distribution of disks/squares, (ii) the target area fraction they occupy, and (iii) the relative frequency of disks among the placed objects. By varying these parameters, we can generate cross-sections with different structural characteristics. In particular, depending on whether the frequency of disks is set to one, zero, or a value in between, we distinguish between three structural scenarios: a \textit{disks-only} scenario (all cross-sections are disks, see Fig.~\ref{fig:visual:eps}a), a \textit{squares-only} scenario (all cross-sections are squares, see Fig.~\ref{fig:visual:eps}b), and a \textit{disks-squares} scenario (both shapes appear, see Fig.~\ref{fig:visual:eps}c). A detailed description of the stochastic model for generating packings of disks and/or squares is provided in Appendix~\ref{appendix:generation:discs}.

\subsubsection{Excursion sets}
To generate freeform cross-sections, we introduce random masks $M\subset\windowPlane$ that are modeled by excursion sets of stationary Gaussian random fields (GRF) that are periodic on $\windowPlane$ \cite{adler2007random,LangPotthoff}. Structural statistics of generated masks can be controlled by two parameters: the area fraction  of the mask and a parameter that controls the correlation length of the GRF. The latter parameter influences the coarseness of generated masks. For further details on the deployed excursion set model, see Appendix~\ref{appendix:generation:excursion}.

By assigning  a random relative permittivity value (uniformly chosen in the interval $[1,\epsMax]$) to the generated mask $M$  and subsequently discretizing the result, the excursion set model produces 2D distributions $\epsPlane \colon \windowPlaneDiscrete \to [1,\epsMax]$  of relative permittivities. Note that
the relative permittivity in $\windowPlane\setminus M$ is set to 1, which corresponds to vacuum. Metasurfaces constructed solely from these masks form the \emph{freeform-only} scenario. See Fig.~\ref{fig:visual:validation}d for an example. Moreover, the packing procedure described above can also be used to place disks and squares inside $\windowPlane\setminus M$ without overlapping with the mask $M$, resulting in 2D distributions $\epsPlane$ that exhibit freeform regions together with disks and squares. We refer to this structural scenario as the \emph{freeform-disks-squares} class of metasurfaces. See Fig.~\ref{fig:visual:validation}e.

\subsubsection{Construction of a 3D metasurface database}\label{sec:database}
Generated 2D distributions $\epsPlane$ of relative permittivities can be used to construct 3D representations of metasurfaces which will be defined on a 3D grid $\windowVolumeDiscrete=\{0\cdot \rho,\dots,(n_\mathrm{x}-1)\cdot\rho\} \times \{0\cdot\rho,\dots,(n_\mathrm{y}-1)\cdot\rho\} \times \{0\cdot\rho,\dots,(n_\mathrm{z}-1)\cdot\rho\}$, where we set $\rho=\SI{50}{\nano\meter}$, $n_\mathrm{x}=n_\mathrm{y}=128$ and $n_\mathrm{z}=64$.
Then, a 3D distribution 
$\epsVol \colon \windowVolumeDiscrete \to [1, \epsMax]$ is constructed by
\begin{equation}\label{eq:epsVolume}
	\epsVol(x,y,z) =
	\left\{
	\begin{array}{ll}
		1, & z< z_\mathrm{substrate},\\
		\epssymbol_\text{SiO$_2$}, & z_\mathrm{substrate}\leq z < z_\mathrm{substrate}+h_\mathrm{substrate},\\
		\epsPlane(x,y), & z_\mathrm{substrate}+h_\mathrm{substrate}\leq z < z_\mathrm{substrate}+h_\mathrm{substrate}+h_\mathrm{metasurface},\\
		1, & \text{else,}
	\end{array}
	\right.
\end{equation}
for each grid point $(x,y,z)\in \windowVolumeDiscrete$.
Here, $\varepsilon_{\text{SiO}_2}=1.4585$ is the relative permittivity of SiO$_2$, $h_\mathrm{substrate}=\SI{350}{\nano\meter}$ is the thickness of the SiO$_2$ layer that is placed at height $z_\mathrm{substrate}=\SI{1050}{\nano\meter}$, and $h_\mathrm{metasurface}=\SI{350}{\nano\meter}$ specifies the layer thickness that is assigned to the planar section $\epsPlane$.
This construction yields 3D distributions $\epsVol$ of permittivity, see Fig.~\ref{fig:visual:eps}e-g, that can serve as input for numerical simulations of EM fields.

By varying the  parameters of model components associated with the generation of masks and packings, we generate differently structured 3D metasurfaces for each of the five considered structural scenarios (i.e., freeform-only, disks-only, squares-only, disks-squares and freeform-disks-squares). In total, we have generated 5296 3D distributions of permittivity, with approximately 1000 3D distributions per structural scenario. 
More details on how model parameters have been varied to generate this database are given in Appendix~\ref{appendix:generation:database}.

\subsection{Simulation of electromagnetic fields}\label{sec:simulation}
The generated 3D distributions $\epsVol$ of relative permittivities can be considered to be metasurfaces, which we will use as input for numerically 
simulating the transformation of EM fields induced by $\epsVol$. In particular, we will solve Maxwell's equations numerically using an FDTD solver \cite{mahlau2024flexibleframeworklargescalefdtd}, to compute discretized EM fields. Consequently,  
we will have pairs of metasurfaces and corresponding EM fields ready to train surrogate models using the methods described in Section~\ref{Sec:Surrogate}.

\subsubsection{Governing equations}
Let $\epsVol$ 
denote the (non-discretized) distribution of relative permittivities. In particular, we consider $\epsVol \colon \windowVolume \to [1, \epsMax]$, where $\windowVolume=[0,w_1] \times [0,w_2] \times [0,w_3]$ is a cuboidal observation window. 
We study Maxwell's equations that describe the resulting inhomogeneous EM fields induced by a monochromatic, linearly polarized plane wave (with the electric field in $x$-direction). The plane wave is propagating through a dielectric medium whose spatially varying relative permittivity is given by $\epsVol$, with electrical conductivity assumed to be negligible.
We denote electric and magnetic fields by $\EField, \HField \colon \windowVolume \times [0,\infty)\to \R^3$, where the vectors $\EField(\rvec,t)$ and $\HField(\rvec,t)$ are the electric and magnetic fields at position $\rvec \in \windowVolume$ at time $t\geq 0$, respectively. 
The  fields $\EField, \HField$ satisfy  Maxwell's equations, if
\begin{align}\label{eq:Maxwell}
	\nabla \times \HField &=\epssymbol_0 \epsVol \frac{	\partial \EField}{\partial t} +\JField,\\ 
	- \nabla \times \EField &=\mu_0\frac{	\partial \HField}{\partial t} , \nonumber\\
	\nabla \cdot (\epssymbol_0 \epsVol \EField) &=\rho_\mathrm{free},\quad
	\nabla \cdot \HField =0,\nonumber
\end{align}
in the interior of the domain $\windowVolume \times [0,\infty)$. Here, $\nabla \times$ and $\nabla \cdot$ denote the curl and divergence operations, respectively \cite{taflove}. The constants $\epssymbol_0$ and $\mu_0$ denote the permittivity and permeability of vacuum, respectively. 
At time $t=0$, the fields are set to an initial zero vector field, i.e.,
$\EField(\rvec,0)=0$ and $\HField(\rvec,0)=0$ for $\rvec\in \windowVolume$. 
The current density $\JField \colon \windowVolume \times [0,\infty) \to \R^3$ will be used to model the plane wave source.
Since the exciting plane will be placed in charge-free vacuum, 
$\JField$ is divergence free, i.e., $\nabla \cdot \JField=0$, which implies that the charge density satisfies $\rho_\mathrm{free}=0$. Further details on the plane-wave excitation and representation are provided below.

Periodic boundary conditions are imposed on the lateral facets of the cuboidal window $\windowVolume$ (i.e., points with $x=0$, $x=w_1$, $y=0$ or $y=w_2$) to model an infinitely large structure in the $x$-$y$-directions, e.g., for $\HField$ we have
\begin{equation}
	\HField(0,y,z,t) = \HField(w_1,y,z,t) \quad \text{and} \quad
	\HField(x,0,z,t) = \HField(x,w_2,z,t),
\end{equation}
for each $x\in [0,w_1],y\in [0,w_2],z\in [0,w_3], t\geq 0$. Analogously, these boundary conditions are applied for $\EField$. 
On the remaining facets of $\windowVolume$ (i.e., for the facets with $z=0$ and $z=w_3$), absorbing boundary conditions are imposed, that is, perfectly matched layers (PML), to minimize artificial reflections of outgoing waves \cite{BERENGER1994185,taflove}.

To avoid trivial solutions, following the approach described in \cite{Oskooi}, we introduce a plane wave source term that emits a plane incident wave by 
\begin{equation}\label{eq:source:term}
	\JField(\rvec,t) = \mathrm{Re}\!\left(
	\sourceJ \exp\!\left( \frac{2 \pi \mathrm{i}}{\lambda} (\propDir \cdot \rvec- c_0 t) \right)
	\right) \delta(z-z_\mathrm{source}),
\end{equation}
for $\rvec=(x,y,z)\in \windowVolume$ and $t\in [0,\infty)$, where $\sourceJ \in \C^3$ is the complex polarization vector the magnitude of which defines the amplitude, $\lambda>0$ is the excitation wavelength, $c_0$ the speed of light and the unit vector $\propDir \in \R^3$ defines the direction of propagation.
The Dirac delta $\delta(z -z_\mathrm{source})$ confines the source to the plane 
$z= z_\mathrm{source}$ for some $z_\mathrm{source} \in [0, w_3]$, representing a surface current sheet that launches the incident plane wave\footnote{Since the source term involves a Dirac delta distribution, the source current $\JField$ in Eq.~\eqref{eq:source:term}, as well as the system of partial differential equations corresponding to Maxwell's equations in Eq.~\eqref{eq:Maxwell}, are to be understood in the weak (distributional) sense.}. In the following, we set the polarization vector to $\sourceJ=(1,0,0)$ and the direction of propagation to $\propDir=(0,0,-1)$. Thus, in the source plane 
a linearly polarized wave with the electric field in the $x$-direction 
is excited that propagates in the $z$-direction towards the metasurface, which is placed parallel to the $x$-$y$-plane, see Section~\ref{sec:generation:metasurfaces}. Finally, the excitation wavelength is set to $\lambda=\SI{1050}{\nano\meter}$.

\subsubsection{Numerical implementation}
For a given discretized 3D distribution $\epsVol \colon \windowVolumeDiscrete \to [1,\epsMax]$ of permittivity, that represents a dielectric metasurface, the 
system of partial differential equations described above is numerically solved for $\HField$ and $\EField$, using an FDTD scheme implemented in the Python package \texttt{fdtdx} \cite{mahlau2024flexibleframeworklargescalefdtd}. In particular, the continuous domain 
$\windowVolume=[0,w_1]\times[0,w_2]\times[0,w_3]$  is discretized on a uniform Yee grid of size
$(n_\mathrm{x},n_\mathrm{y},n_\mathrm{z})=(128,128,64)$, with grid spacings of 
$\Delta x = \Delta y = \Delta z = \rho \SI{50}{\nano\meter}$.
The temporal resolution $\Delta t$ is chosen according to the Courant--Friedrichs--Lewy (CFL) stability condition \cite{5391985} with a Courant factor of $S=0.99$, i.e., 
\begin{equation}
	\Delta t = S \left(c_0 \left(\frac{1}{\Delta x} + \frac{1}{\Delta y}+\frac{1}{\Delta z}\right) \right)^{-1}
	=\frac{S \Delta x}{3 c_0}\approx  5.5\times 10^{-17}\,\si{s}.
\end{equation}
The discretized 3D distribution $\epsVol$ is used to define the local dielectric permittivities assigned to the Yee lattice cells \cite{1138693}. 
To accommodate the boundary conditions of the continuous partial differential equations described above, numerical implementations of periodic boundary conditions in the Python package \texttt{fdtdx} are applied along the $x$- and $y$-directions to model an infinitely extended periodic structure, while perfectly matched layers (PMLs) are used at the top and bottom boundaries ($z=0$ and $z=w_3$) to minimize reflections at the boundary of the computational domain.
In the numerical implementation, the PMLs are not planar, but they are assigned a thickness of $n_\mathrm{PML}=3$ cells, which corresponds to \SI{150}{\nano\meter} thick PML bands on both sides of the simulation domain.
The $z$-position of the plane that excites the source wave in Eq.~\eqref{eq:source:term} is set to $z_\mathrm{source}=\SI{3050}{\nano\meter}$. In this configuration, the source plane is located close to the upper domain boundary at $z=w_3$, but remains outside the PML region. 
Moreover, in this manner the source plane is placed above the metasurface, which occupies the positions with heights $z \in [\SI{1050}{\nano\meter},\SI{1750}{\nano\meter}]$.

The total simulation duration is set to $T_\mathrm{total}=
200\times 10^{-15}\,\si{s}$, which corresponds to roughly fifty optical periods for the excited wave with wavelength $\lambda=\SI{1050}{\nano\meter}$. This corresponds to $n_t=T/\Delta t= 3636$ simulated time steps. We expect that the chosen simulation duration reduces transient field contributions to a negligible level after which the steady-state EM field can be determined. 
After deploying the FDTD approach, we acquire the discretized version $\HFieldDisrete, \EFieldDisrete$ of the fields $\HField, \EField$ on the Yee grid. 
For brevity, we write $\EFieldDisrete\colon \windowVolumeDiscrete \times \{0,\Delta t, \dots , (n_t-1)\Delta t\}\to \R^3$ and $\HFieldDisrete\colon \windowVolumeDiscrete \times \{0.5\Delta t,1.5 \Delta t, \dots , (n_t-1.5)\Delta t\}\to \R^3$, even though the Yee grid actually discretizes $\EFieldDisrete$ and $\HFieldDisrete$ on staggered spatial grids, see Appendix~\ref{appendix:efields} for further details.

\subsubsection{Computation of complex field amplitudes}
As mentioned above, we assume that after a sufficiently long simulation time $T_\mathrm{total}$, the EM wave reaches a steady state.
For continuous solutions that satisfy the steady state solution, an efficient representation is given by utilizing complex amplitudes. For example, a magnetic field  $\HField$ that satisfies the steady state can be expressed by
\begin{equation}\label{eq:phasor}
	\HField(\rvec,t) = \mathrm{Re}\!\left(\widehat{\HField}(\rvec) e^{\mathrm{i}\omega t}
	\right),
\end{equation}
for $\rvec \in \windowVolume$ and for sufficiently large times $t$, where $\widehat{\HField}\colon \windowVolume \to \C^3$ is the (complex)  amplitude field and $\omega= 2\pi c_0 / \lambda$ is the angular frequency. Eq.~\eqref{eq:phasor} shows that in steady state, the magnetic field is completely determined by its  amplitude field for a given frequency $\omega$. Analogously, the electric field is described by its  amplitude field $\widehat{\EField}\colon \windowVolume \to \C^3$.

Substituting these representations for the electric and magnetic fields into Eq.~\eqref{eq:Maxwell}, followed by omitting the common factor $e^{\mathrm{i}\omega t}$, 
yields the corresponding frequency-domain system of equations 
\begin{align}\label{eq:Maxwell:curlH}
	\nabla \times \widehat{\HField} & = \mathrm{i} \omega \epssymbol_0 \epsVol   \widehat{\EField},  \\
	-\nabla \times  \widehat{\EField} &=  \mathrm{i} \omega \mu_0   \widehat{\HField}, \label{eq:Maxwell:curlE}\\
	\nabla \cdot (\epssymbol_0 \epsVol \widehat{\EField})&=0, \quad
	\nabla \cdot \widehat{\HField}=0. \nonumber
\end{align} 
By substituting $\widehat{\EField}$ in Eq.~\eqref{eq:Maxwell:curlE} with its representation in Eq.~\eqref{eq:Maxwell:curlH} yields
\begin{equation}\label{eq:WaveEquation}
	\nabla \times \left(\frac{1}{\epssymbol_0 \epsVol} \nabla \times  \widehat{\HField}\right) =\omega^2 \mu_0 \widehat{\HField}.
\end{equation}
This representation will enable us to quantify the physical consistency of  magnetic fields predicted by surrogate models, i.e., to define residuals that capture deviations from Maxwell’s equations (so-called Maxwell residuals), see Section~\ref{Sec:Surrogate}.

For a given continuous steady-state solution, the corresponding  amplitude field can be determined by integrating over a single oscillation period of length $\frac{2\pi}{\omega}$.
For example, the complex amplitude $\widehat{\HField}$ can be obtained by 
\begin{equation}\label{eq:phase:reconstruction}
	\widehat{\HField}(\mathbf{r}) = \frac{\omega}{\pi} \int_{T-\frac{2\pi}{\omega}}^{T} \HField(\mathbf{r},t)\, e^{-\mathrm{i}\omega t}\, \mathrm{d}t,
\end{equation}
for some time $T \in \mathbb{R}$.
The integral in Eq.~\eqref{eq:phase:reconstruction} can be approximated numerically using the discretized solution $\HField_\mathrm{d}$ obtained from the FDTD simulation.
Let $m$ denote the number of discrete time steps corresponding to one oscillation period, i.e., $m \cdot \Delta t \approx \tfrac{2\pi}{\omega}$. Then, assuming that the last $m$ simulation time steps fulfill the steady-state, the discretized  amplitude field $\HFieldPhasorDisrete \colon \windowVolumeDiscrete \to \C^3$ is given by
\begin{equation}
	\HFieldPhasorDisrete(\rvec)=\frac{\omega}{\pi}\sum_{\ell=n_t-m+1}^{n_t}\HFieldDisrete(\rvec,\ell \, \Delta t)\, e^{-\mathrm{i}\omega \ell \Delta t}.
\end{equation}
Analogously, we determine the  amplitude field $\EFieldPhasorDisrete\colon \windowVolumeDiscrete \to \C^3$ for the simulated  electric fields.

\begin{figure}
	\centering
\subfloat[]{
	\begin{tikzpicture}
		\tikzPlotAdvanced{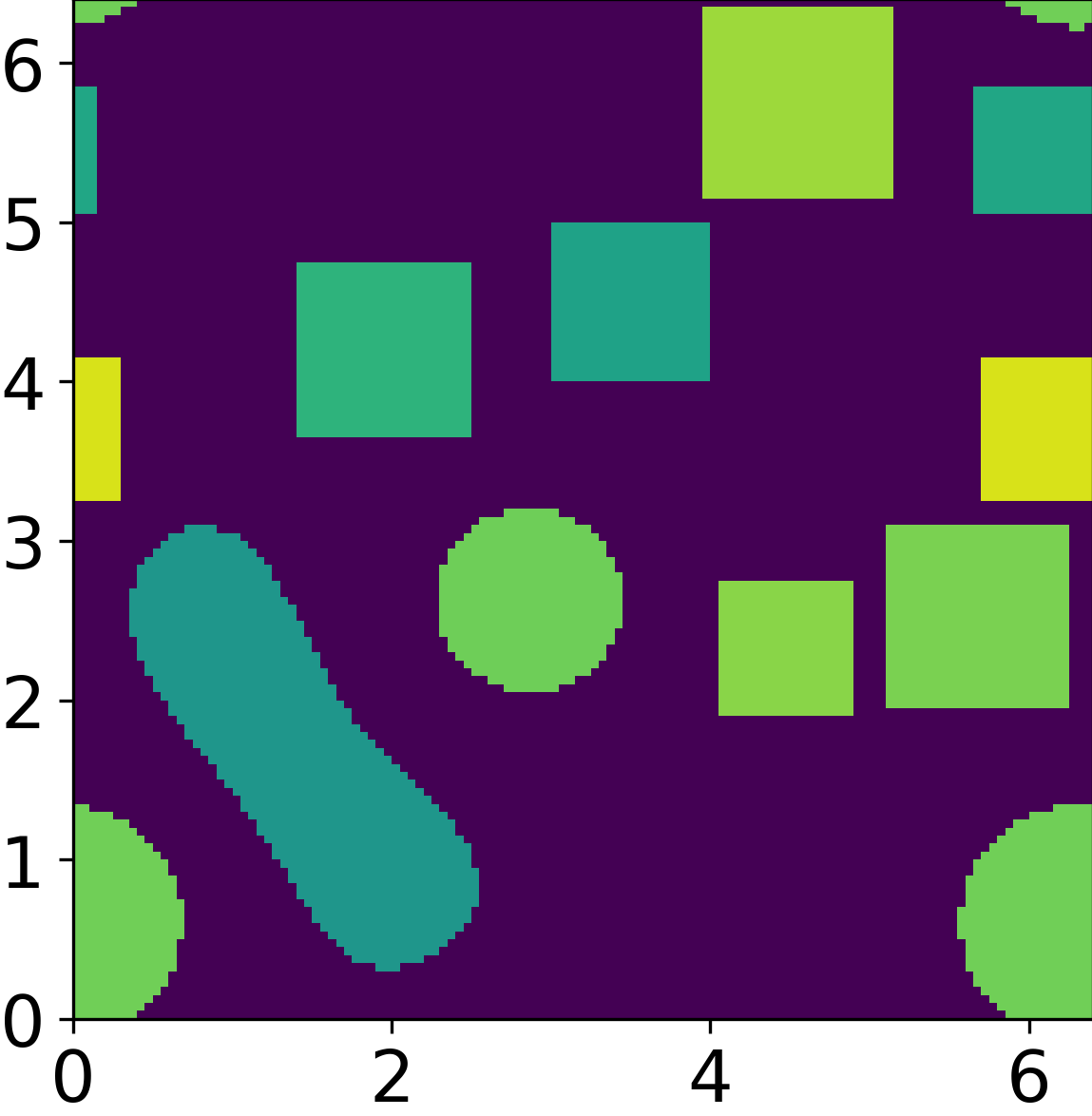}{$x$  [\si{\micro\meter}]}{$y$  [\si{\micro\meter}]}{0.3}{}{reforig}{inner sep=0pt}
	\end{tikzpicture}
}
\begin{tikzpicture}	
	\node(ref){\phantom{a}};
	\node[above=0.4cm of ref, inner sep=0pt ] (colorbar) {
		\includegraphics[height=4.25cm]{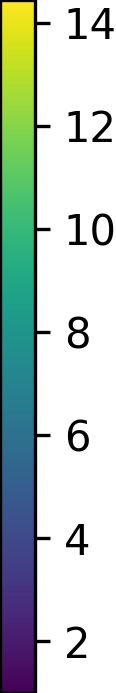}
	};
	\node[right=0.3cm of colorbar.south east, anchor=south west , inner sep=0pt,rotate=90,xshift=0.3cm ] {
		relative permittivity
	};
	
\end{tikzpicture}	
%
\subfloat[]{
	\begin{tikzpicture}
		\tikzPlotAdvanced{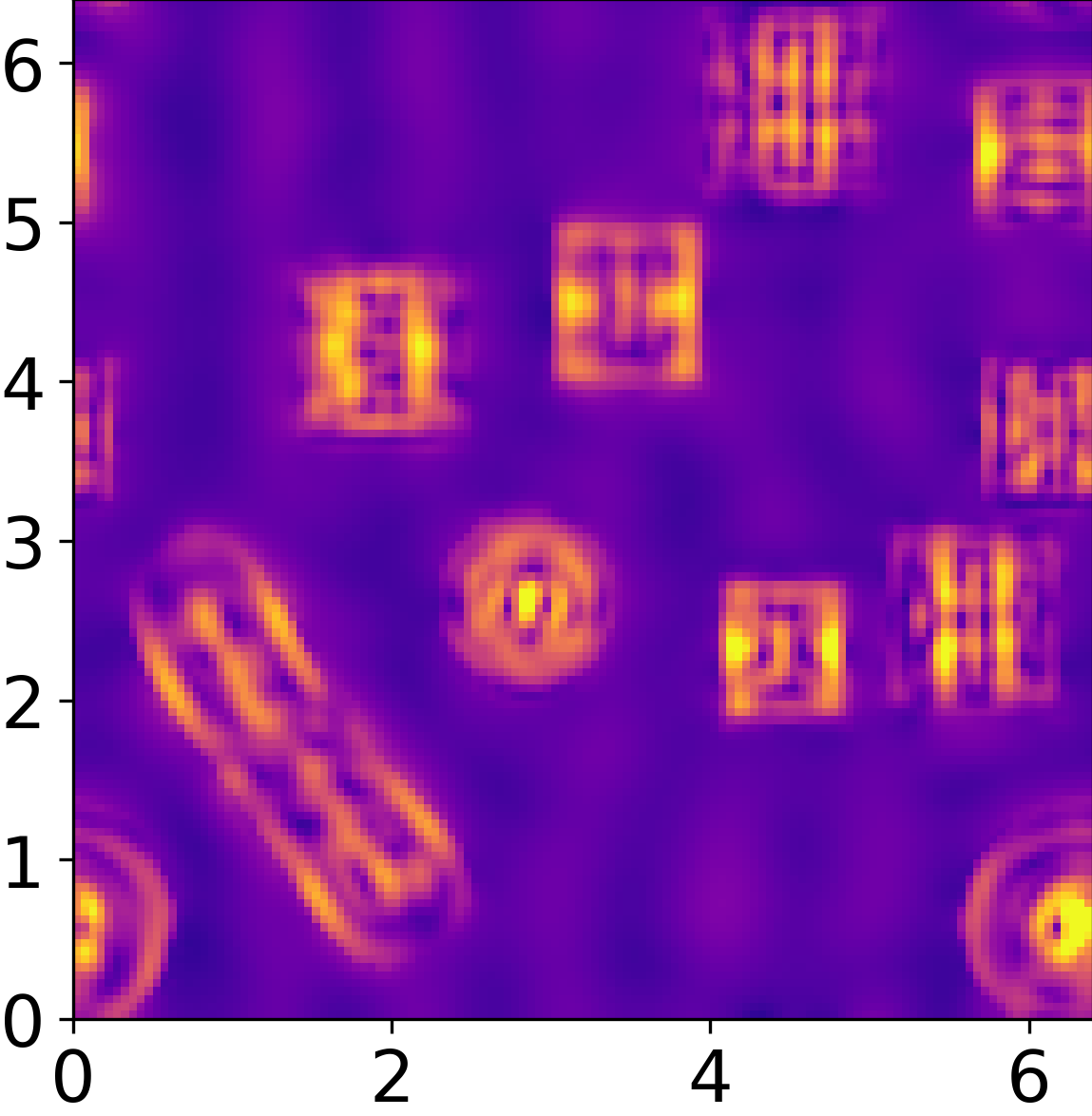}{$x$  [\si{\micro\meter}]}{$y$  [\si{\micro\meter}]}{0.3}{}{reforig}{inner sep=0pt}
	\end{tikzpicture}
}
	\begin{tikzpicture}	
		\node(ref){\phantom{a}};
		\node[above=0.4cm of ref, inner sep=0pt ] (colorbar) {
			\includegraphics[height=4.25cm]{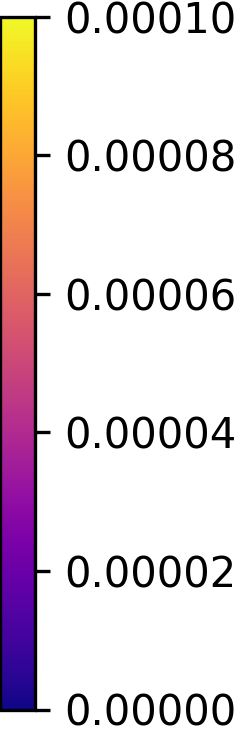}
		};
		\node[right=0.35cm of colorbar.south east, anchor=south west , inner sep=0pt,rotate=90,xshift=1.3cm ] {
			$|\HFieldDisrete|_2$
		};
		
	\end{tikzpicture}

	\caption{\textbf{Training data.}
		Cross-section of a 3D distribution $\epsVol$ of relative permittivities (a) and associated magnitude of the simulated magnetic field $\HFieldDisrete$ (b).
	}\label{fig:training:data}
\end{figure}

The resulting complex-valued amplitude fields $\EFieldPhasorDisrete$ and $\HFieldPhasorDisrete$ can be represented in a real-valued form by splitting the complex-valued vectors into vectors of real and imaginary parts, which will facilitate the training of neural networks as surrogate models in Section~\ref{Sec:Surrogate}.  More precisely, the real-valued representation $\HFieldPhasorDisreteReal\colon \windowVolumeDiscrete \to \R^6$ of $\HFieldPhasorDisrete$ is given by
\begin{equation}
	\HFieldPhasorDisreteReal(\rvec) =
	\left(
	\mathrm{Re}(\HFieldPhasorDisrete(\rvec)),
	\mathrm{Im}(\HFieldPhasorDisrete(\rvec))
	\right).
\end{equation}
Analogously, $\EFieldPhasorDisreteReal\colon \windowVolumeDiscrete \to \R^6$ is defined. For brevity and with mild abuse of notation, we denote the real-valued, discretized fields $\HFieldPhasorDisreteReal,\EFieldPhasorDisreteReal$, by $\HFieldPhasorDisreteRealAlt$ and $\EFieldPhasorDisreteRealAlt$ from now on. Throughout the remainder of this paper, we also refer to these amplitude fields $\HFieldPhasorDisreteRealAlt,\EFieldPhasorDisreteRealAlt$ simply as  \emph{magnetic fields} and \emph{electric fields}, respectively.

By computing $\HFieldPhasorDisreteRealAlt$  for each simulated 3D distribution $\epsVol$ of relative permittivities (see Section~\ref{sec:generation:metasurfaces}), we obtain a large dataset  consisting of pairs
$(\epsVol,\HFieldPhasorDisreteRealAlt)$. Visualizations of a simulated field  $\HFieldDisrete$ and the corresponding 3D distribution $\epsVol$ are shown in Fig.~\ref{fig:training:data}. 
The dataset of pairs $(\epsVol,\HFieldPhasorDisreteRealAlt)$ serves as the basis for training surrogate models that learn to predict magnetic fields directly from the distribution of permittivity; electric fields can be computed from predicted magnetic fields by means of Eq.~\eqref{eq:Maxwell:curlH}, see Appendix~\ref{appendix:efields} for further details. Once trained, the surrogate models can approximate the results of FDTD simulations while providing orders-of-magnitude speedups during inference.

\subsection{Surrogate models for predicting electromagnetic fields}\label{Sec:Surrogate}
The goal of this section is to train computational models that take  distributions of permittivity as input and approximate the associated magnetic fields, i.e., to train functions $f$ with $f(\epsVol)\approx \HFieldPhasorDisreteRealAlt$. 
In the literature~\cite{Chen2022Stanford,zhelyeznyakov2023large}, a well-established approach for learning such mappings is the deployment of CNNs \cite{goodfellow2016deep}, which, among other components, consist of convolutional layers. A typical 3D convolutional layer
$\mathcal{C}: \mathbb{R}^{C_\mathrm{in} \times H\times W\times D} \to \mathbb{R}^{C_\mathrm{out} \times H\times W\times D}$ 
with activation function $\sigma \colon \mathbb{R}^{C_\mathrm{out}}\to \mathbb{R}^{C_\mathrm{out}}$ is given by 
\begin{equation}
	\mathcal{C}(x) = \sigma \! \left(\left( \sum_{k=1}^{C_\mathrm{in}}
	W_{ki} \ast x_k + b_i
	\right)_{i=1}^{C_\mathrm{out}} \right),
\end{equation}
for an input tensor  $x = (x_1,\dots,x_{C_\mathrm{in}}) \in \mathbb{R}^{C_\mathrm{in} \times H\times W\times D}$ with $C_\mathrm{in}$ channels, height $H$, width $W$ and depth $D$
\cite{torch}. The trainable parameters of the convolutional layer $\mathcal{C}$ are the kernels $W_{ki} \in \R^{K_\mathrm{h} \times K_\mathrm{w} \times K_\mathrm{d}}$ with spatial dimensions $K_\mathrm{h}, K_\mathrm{w}, K_\mathrm{d}>0$ and the bias vector $b=(b_1,\dots,b_{C_\mathrm{out}}) \in \R^{C_\mathrm{out}}$.
For example, the discretized 3D distribution $\epsVol$ can be considered to be a tensor $x=(x_{1,i,j,k})_{i,j,k=1}^{n_\mathrm{x},n_\mathrm{y},n_\mathrm{z}} \in \R^{1\times  n_\mathrm{x} \times n_\mathrm{y} \times n_\mathrm{z} }$ with 1 channel, by 
$x_{1,i,j,k}=  \epsVol((i-1) \Delta x, (j-1) \Delta y,(k-1) \Delta z )$.  The convolutional layer processes such input tensors locally: its \emph{field of view} (FOV)---the spatial region of the input that contributes to the value of a component in the output voxel---is restricted by the kernel size $(K_\mathrm{h}, K_\mathrm{w}, K_\mathrm{d})$. 
In practice, CNNs increase their effective FOV by concatenating multiple convolutional layers and employing pooling operations. Nevertheless, the FOV remains finite and may still be a limiting factor for predicting EM fields, since the field strengths of any pair of spatially distant positions can be strongly correlated due to the non-local nature of Maxwell's equations. To remedy this, in this study, artificial neural networks are considered that deploy a different type of layer.

\subsubsection{Layers in neural operators} 
In contrast to convolutional layers that operate on tensors, the layers of neural operators define mappings between function spaces\footnote{Formally, neural operators map between spaces of square-integrable functions, i.e., $L^2$ function spaces.} \cite{azizzadenesheli2024neural}. 
In the present paper, we solely consider so-called Fourier neural operators \cite{FourierNeuralOperator} that we constrain on functions with domain $\windowVolume \subset \R^3$. Some of the layers used in neural operators utilize vector-valued convolutions. 
For an input function $v \colon \windowVolume \to \R^{C_\mathrm{in}}$ the convolution with the kernel $\kappa\colon \windowVolume \to \R^{C_\mathrm{out}\times C_\mathrm{in}}$ is given by 
\begin{equation}
	(\kappa \ast v)(\xvec) = \int_{\windowVolume} \kappa (\xvec-\yvec) v(\yvec) \,\mathrm{d}\yvec.
\end{equation}
where the kernel $\kappa$ is periodically extended in $\windowVolume$ if $\xvec - \yvec \notin \windowVolume$. 
According to the convolution theorem \cite{grafakos2008classical,grafakos2014Modern}, this operation can be expressed equivalently in the Fourier domain by
\begin{equation}
	\kappa \ast v=  \mathcal{F}^{-1}\left( \mathcal{F}(\kappa) \mathcal{F}(v) \right),
\end{equation}
where $\mathcal{F}$ denotes the Fourier transform in the periodic domain $\windowVolume$. In particular, $\mathcal{F}(v)$ and $\mathcal{F}(\kappa)$ map the mode vectors $\modeVector\in \Z^3$ to the amplitude vectors $(\mathcal{F}(v))(\modeVector) \in \C^\mathrm{C_\mathrm{in}}$ and the amplitude matrices  $(\mathcal{F}(\kappa))(\modeVector) \in \C^\mathrm{C_\mathrm{out}\times C_\mathrm{in}}$, respectively. 
The Fourier neural operator leverages this representation by substituting the Fourier transform $ \mathcal{F}(\kappa)$ of the kernel $\kappa$ with 
a function $\mathcal{R}_\theta \colon \Z^3 \to \C^\mathrm{C_\mathrm{out}\times C_\mathrm{in}}$ that is  parameterized by some parameter vector $\theta$.
In practice, $\theta$ describes a finite number  of frequency modes of $\mathcal{R}_\theta$.
More precisely, a common parametrization
$\theta \in 
\mathbb{C}^{C_\mathrm{out} \times C_\mathrm{in} \times (2m_\mathrm{x}+1) \times (2m_\mathrm{y}+1) \times (2m_\mathrm{z}+1)}
$
represents the complex matrix-valued amplitudes of the Fourier modes retained within some truncation limits 
$m_\mathrm{x}, m_\mathrm{y}, m_\mathrm{z} > 0$ in each spatial direction. 
The matrix-valued function $\mathcal{R}_\theta \colon \mathbb{Z}^3 \to \mathbb{C}^{C_\mathrm{out} \times C_\mathrm{in}}$ 
is then defined by
\begin{equation}\label{eq:FrequencyKernel}
	\mathcal{R}_\theta(i,j,k) =
	\begin{cases}
		\bigl(\theta_{p,q,\,i+m_\mathrm{x}+1,\,j+m_\mathrm{y}+1,\,k+m_\mathrm{z}+1}\bigr)_{p,q=1}^{C_\mathrm{out},\,C_\mathrm{in}}, 
		& \text{if } -m_\mathrm{x} \le i \le m_\mathrm{x},\; -m_\mathrm{y} \le j \le m_\mathrm{y},\; -m_\mathrm{z} \le k \le m_\mathrm{z}, \\[4pt]
		0, & \text{otherwise,}
	\end{cases}
\end{equation}
for $(i,j,k) \in \mathbb{Z}^3$. To ensure that the parameterization $\mathcal{R}_\theta$ preserves the symmetries that the Fourier transform of a real-valued kernel $\mathcal{F}(\kappa)$ would exhibit, 
the parameters $\theta$ are constrained such that
\[
\mathcal{R}_\theta(-i,-j,-k) = \mathcal{R}_\theta(i,j,k)^{\ast},
\]
where $(\cdot)^{\ast}$ denotes the complex conjugation. Finally, a so-called Fourier layer $\mathcal{L}$ with $C_\mathrm{in}$ input channels and $C_\mathrm{out}$ output channels is given by
\begin{equation}\label{eq:layer}
	\mathcal{L}(v) = \sigma\left(
	\mathcal{F}^{-1}\left( \mathcal{R}_\theta \mathcal{F}(v) \right) + A v +b
	\right),
\end{equation}
for  input functions $v\colon \windowVolume \to \R^{C_\mathrm{in}}$, 
where $A\in \R^{C_\mathrm{out}\times C_\mathrm{in}}$ and $b\in \R^{C_\mathrm{out}}$ are, in addition to $\theta$, trainable parameters. 
Effectively, the convolution performed by $\mathcal{L}$ has an ``infinitely large field of view'', since the bounded representation $\mathcal{R}_\theta$ of a kernel in the frequency domain has, in general, an unbounded support in the spatial domain. 
As a result, the values of the function $\mathcal{L}(v) \colon \windowVolume \to \R^{C_\mathrm{out}}$
can incorporate information from all spatial positions of the input $v$ simultaneously, allowing the Fourier layer to capture long-range and nonlocal dependencies that would be inaccessible to conventional convolutional architectures.

\subsubsection{Network architecture} 
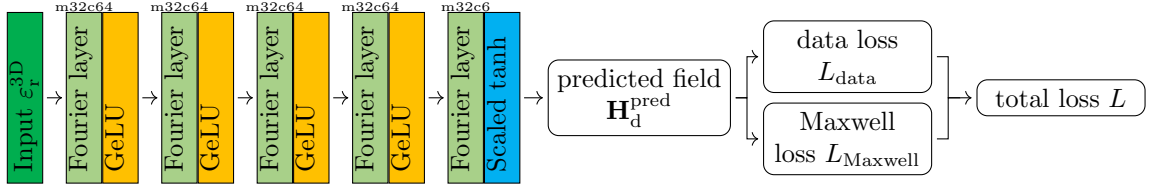
\begin{figure}[t!]
	\centering
\begin{tikzpicture}
	\inputLayer

	\Fourier{inputLayer}{fourier1}{\defaultDistance}{m32c64}{1}
	\GeLU{fourier1}{relu1}{0cm}{m32c64}{1}

	\Fourier{relu1}{fourier2}{\defaultDistance}{m32c64}{1}
	\GeLU{fourier2}{relu2}{0cm}{k32c64}{1}	
	
	\Fourier{relu2}{fourier3}{\defaultDistance}{m32c64}{1}
	\GeLU{fourier3}{relu3}{0cm}{k32c64}{1}	
	
	\Fourier{relu3}{fourier4}{\defaultDistance}{m32c64}{1}
	\GeLU{fourier4}{relu4}{0cm}{k32c64}{1}	
	
	\Fourier{relu4}{fourier5}{\defaultDistance}{m32c6}{1}

	\ScaledTanh{fourier5}{leaky}{0cm}{0}
	
	\node[ draw, thin, rounded corners=4pt, right=\defaultDistance of leaky.east,anchor=west, align=center, outer sep =2pt](output){predicted field\\
		$\HFieldPhasorDisreteRealAlt^\mathrm{pred}$
	};
	
	\draw[->] (leakyright) -- (output.west);

	\node[
	draw, thin, rounded corners=4pt,
	right=\defaultDistance of output.east,
	anchor=south west, align=center, outer sep=2pt, text width=2cm
	](boxTop){data loss $L_\mathrm{data}$};
	
	\node[
	draw, thin, rounded corners=4pt,
	right=\defaultDistance of output.east,
	anchor=north west, align=center, outer sep=2pt, text width=2cm
	](boxBottom){Maxwell loss $L_\mathrm{Maxwell}$};
	
	\draw[->]
	(output.east) -- ++(0.15cm,0)
	|- (boxTop.west);
	
	\draw[->]
	(output.east) -- ++(0.15cm,0)
	|- (boxBottom.west);

	\node[
	draw, thin, rounded corners=4pt,
	right=6cm of leaky.east,
	anchor=west, align=center, outer sep=2pt, text width=2cm
	](totalLoss){total loss $L$};

	\coordinate (merge) at ($(totalLoss.west)-(0.15cm,0)$);
	
	\draw[-] (boxTop.east) -- ++(0.15cm,0) |- (merge);
	\draw[-] (boxBottom.east) -- ++(0.15cm,0) |- (merge);
	\draw[->] (merge) -- (totalLoss.west);
	

	

\end{tikzpicture}
	\vspace{4mm}
	\caption{\textbf{Network architecture.}
		Input tensors are single-channel 3D distributions $\epsVol$ that are processed through five Fourier layers, the first four of which are deploying a GeLU activation function, whereas the last Fourier layer is followed by a scaled tanh activation function. 
		The labels above Fourier layers indicate the truncation limit (m) and the number of
		output channels (c). For example, the label m32c64 indicates a Fourier layer with truncation limits $m_\mathrm{x}=m_\mathrm{y}=m_\mathrm{z}=32$ and $c=64$ output channels.    	
		For predicted fields 	$\HFieldPhasorDisreteRealAlt^\mathrm{pred}$ the data and Maxwell losses loss $L_\mathrm{data}$, $L_\mathrm{Maxwell}$. The total loss $L$ is defined as a weighted sum of these contributions.
	}
	\label{fig:architecture}
\end{figure}

In our work, we employ an architecture consisting of five Fourier layers as defined in Eq.~\eqref{eq:layer}. 
More precisely, the network is given by
\begin{equation}\label{eq:neuralOperator}
	f(v) = \mathcal{L}_5 \circ \dots \circ \mathcal{L}_1 (v),
\end{equation}
where the layers $\mathcal{L}_1, \dots, \mathcal{L}_5$ have input--output channel pairs given by $(1,64)$, $(64,64)$, $\dots$, $(64,6)$, respectively, see Fig.~\ref{fig:architecture}. 
For the first four layers, the GeLU activation function \cite{hendrycks2016gelu} is used, while the final layer has a scaled tanh function as an activation function to ensure that the predicted values are in a reasonable interval, i.e., for $\mathcal{L}_5$ the activation function is given by
\begin{equation}\label{eq:activation}
	\sigma=\alpha \cdot \tanh,
\end{equation}
with $\alpha=0.0002$. 
The output dimension of $6$ is chosen to represent magnetic fields,  which are  three-dimensional vector fields with complex-valued components. 
Representing each complex component by its real and imaginary parts yields 
$2 \times 3 = 6$ real-valued quantities, which necessitates six real-valued channels.

Recall that the pairs $(\epsVol,\HFieldPhasorDisreteRealAlt)$  in the training data are discretized on the grid $\windowVolumeDiscrete$. However, the neural operator $f$ processes fields that are defined in continuous domains. To facilitate learning from discretized fields, the Fourier transform in Eq.~\eqref{eq:layer} can be substituted by the discrete Fourier transform, which can be efficiently computed via the fast Fourier transform (FFT) \cite{nussbaumer1981fast}. This discretization of the architecture $f$, which we denote again with abuse of notation by $f$, will facilitate learning with the training dataset consisting of pairs $(\epsVol,\HFieldPhasorDisreteRealAlt)$, i.e., learning $f(\epsVol)\approx \HFieldPhasorDisreteRealAlt$.

In order to increase computational efficiency and reduce the number of trainable parameters, in this study, we use Tucker-decomposed Fourier neural operator layers instead of the standard FNO layers \cite{kossaifi2023multigridtensorizedfourierneural}. This choice reduces the number of trainable parameters by an alternative parameterization of $\mathcal{R}_\theta$. 
Roughly speaking, in this parameterization, a so-called rank factor in the interval $[0,1]$ controls the degree of compression of $\mathcal{R}_\theta$. 
A lower rank factor leads to stronger compression and thus fewer parameters, while a rank factor close to~1 approaches the parametrization given in Eq.~\eqref{eq:FrequencyKernel}. 
In our implementation, we employ a rank factor of~0.3, resulting in approximately thirty percent of the parameters compared to the non-compressed implementation.

\subsubsection{Loss function} 
The loss function considered in this study measures (i) the discrepancy between $f(\epsVol)$ and $\HFieldPhasorDisreteRealAlt$, and (ii) the physical inconsistency with respect to the partial differential equation given in Eq.~\eqref{eq:Maxwell:curlH}.
To achieve this, the (total) loss function consists of a \emph{data loss} and a \emph{Maxwell loss}. Note that the total loss is an adapted 3D version of the loss introduced in~\cite{Chen2022Stanford}, where surrogate models were trained for two-dimensional EM field data.

The data loss  $L_\mathrm{data}$ is given by
\begin{equation}
	L_\mathrm{data}(\HFieldPhasorDisreteRealAlt^\mathrm{pred},\HFieldPhasorDisreteRealAlt)
	=\frac{1}{|\windowVolumeDiscrete|}\sum_{\rvec\in \windowVolumeDiscrete} |\HFieldPhasorDisreteRealAlt^\mathrm{pred}(\rvec) - \HFieldPhasorDisreteRealAlt(\rvec)|_p^p,
\end{equation}
for each predicted magnetic field $\HFieldPhasorDisreteRealAlt^\mathrm{pred} = f(\epsVol)$ and the corresponding ground truth $\HFieldPhasorDisreteRealAlt$, where $|\cdot|_p$ denotes the $L^p$ norm on $\R^6$ for $p\geq 1$ and 
$|\windowVolumeDiscrete|$ denotes the cardinality of the set $\windowVolumeDiscrete$ of grid points.

As mentioned above, the Maxwell loss  measures the deviation from Eq.~\eqref{eq:WaveEquation}. 
To avoid inaccuracies introduced by the PMLs, the evaluation of this loss function is constrained to regions within $\windowVolume$ away from the PMLs, i.e., this loss function will be constrained to
\begin{equation}
	W_\mathrm{d,const.}^\mathrm{3D}= \{
	\rvec=(x,y,z) \in \windowVolumeDiscrete \colon 3 \Delta z < z < (n_\mathrm{z}-3) \Delta z
	\}.
\end{equation}
Then, the Maxwell loss $L_\mathrm{Maxwell}$ is given by
\begin{equation}\label{eq:MaxwellLoss}
	L_\mathrm{Maxwell}(\epsVol,\HFieldPhasorDisreteRealAlt^\mathrm{pred},\HFieldPhasorDisreteRealAlt)
	=\frac{1}{|\windowVolumeDiscrete|}\sum_{\rvec\in W_\mathrm{d,const.}^\mathrm{3D} } \left| 
	\left(
	\nabla \times \left(\frac{1}{\epssymbol_0 \epsVol} \nabla \times  {\HFieldPhasorDisreteRealAlt^\mathrm{pred}}\right) 
	\right)\!(\rvec) - \omega^2 \mu_0\HFieldPhasorDisreteRealAlt(\rvec)
	\right|_p^p,
\end{equation}
for each predicted magnetic field $\HFieldPhasorDisreteRealAlt^\mathrm{pred} = f(\epsVol)$ and the corresponding ground truth $\HFieldPhasorDisreteRealAlt$. 
Note that the curl operator $\nabla \times$ in Eq.~\eqref{eq:MaxwellLoss} is defined in continuous space, whereas the field $\HFieldPhasorDisreteRealAlt^\mathrm{pred}$ is defined on the discrete grid $\windowVolumeDiscrete$. 
To overcome this issue, the differential curl operator in Eq.~\eqref{eq:MaxwellLoss} is approximated numerically using finite-difference quotients, 
see~\cite{Chen2022Stanford} for further details. 
Note that the ``double curl'' operator in Eq.~\eqref{eq:MaxwellLoss} does not consider PML boundary conditions nor is it influenced by any source term. However, the ground truth source $\HFieldPhasorDisreteRealAlt$ has been simulated using a planar wave source that has been placed in the vicinity of a PML. 
To ensure that the Maxwell operator in Eq.~\eqref{eq:MaxwellLoss} is evaluated under consistent source and boundary 
conditions, we enforce the known source/boundary values of the ground-truth field in the predicted field 
$\HFieldPhasorDisreteRealAlt^\mathrm{pred}$. 
More precisely, the first and last $(n_\mathrm{PML}+1)$ slices in the $z$-direction of 
$\HFieldPhasorDisreteRealAlt^\mathrm{pred}$ are replaced by the corresponding slices of 
$\HFieldPhasorDisreteRealAlt$. 
This substitution affects only the vicinity of the PMLs, including the plane in which the source wave is 
excited.

For a given dataset $\mathcal{X}$ that consists of  $n_\mathcal{X}\geq 1$ pairs $(\epsVol,\HFieldPhasorDisreteRealAlt)$ of permittivity distributions and magnetic fields, the overall loss $L$ achieved by the neural network $f$ is given by
\begin{equation}\label{eq:lossFunction}
	L= \frac{1}{n_\mathcal{X}} \sum_{(\epsVol,\HFieldPhasorDisreteRealAlt) \in \mathcal{X}}
	L_\mathrm{data}(f(\epsVol),\HFieldPhasorDisreteRealAlt) + \lambda_\mathrm{Maxwell}
	L_\mathrm{Maxwell}(\epsVol,f(\epsVol),\HFieldPhasorDisreteRealAlt),
\end{equation}
where $\lambda_\mathrm{Maxwell} > 0$ is a weighting factor that controls the relative influence of the Maxwell loss function during training.

\subsubsection{Training} 
The dataset consisting of 5296 pairs of permittivity distributions and magnetic fields 
was divided into training, validation and test subsets with ratios of $0.7$, $0.2$ and $0.1$, respectively. 
The  neural-operator-based surrogate model has been implemented using 
the Python package \texttt{neuralop} \cite{JMLR:v24:21-1524,kossaifi2024library}.
The surrogate model was trained using the AdamW optimizer with a learning rate of $2\times10^{-3}$ and a weight decay of $10^{-4}$ \cite{liang2024cautious}. 
A batch size of four samples was used and training was performed for $n_\mathrm{epochs}=200$~epochs. After training, the weights corresponding to the best performance on the validation set were restored to prevent overfitting and ensure good generalization.

To ensure a balanced contribution between the data and Maxwell loss terms during training, the weighting factor $\lambda_\mathrm{Maxwell}$ was adaptively updated  after each epoch based on the ratio of the data and Maxwell loss, measured on the validation set, see Appendix~\ref{appendix:weights} for further details. In order to avoid an overly strong influence of the Maxwell loss during the early stages of training, 
$\lambda_\mathrm{Maxwell}$ was set to zero for the first 20 epochs and only thereafter updated according to 
the adaptive strategy.

\section{Experimental results and discussion}\label{sec:results}

Using the data generated in Sections~\ref{sec:generation:metasurfaces} and~\ref{sec:simulation}, we perform a quantitative evaluation of different variants of the proposed surrogate modeling framework and compare them against state-of-the-art surrogate models. 
To this end, we train five surrogate models:
\begin{enumerate}[label=(\roman*)]
	\item A Fourier neural operator as defined in Eq.~\eqref{eq:neuralOperator}, trained on 70\,\% of the full dataset comprising 5296 samples (referred to as \emph{FNO}), where the $p$-values in $L_\mathrm{data}$ and $L_\mathrm{Maxwell}$ are both set to~1.
	\item The same network as in (i) with a loss function in which the $L^2$ norm is deployed in Eq.~\eqref{eq:MaxwellLoss} (referred to as \emph{\FNOL}), i.e., with the $p$-value of $L_\mathrm{Maxwell}$ set to~2 while keeping $p=1$ in $L_\mathrm{data}$.
	\item The same network as in (i) trained on sparse data, i.e., on 70\,\% of the subset corresponding to the \emph{freeform-only} structural scenario (referred to as \emph{\FNOConstr}), again using $p=1$ in both $L_\mathrm{data}$ and $L_\mathrm{Maxwell}$.
	\item A 3D extension of the WaveY-Net architecture~\cite{Chen2022Stanford} trained on 70\,\% of the full dataset (referred to as \emph{3D-WaveY-Net}),  with $p$-values in $L_\mathrm{data}$ and $L_\mathrm{Maxwell}$ set to~1. Details of the 3D-WaveY-Net architecture are provided in Appendix~\ref{appendix:wavey}.
	\item The same network as in (iv) trained on the \emph{freeform-only} subset (referred to as \emph{\WaveYConstr}),  with $p$-values in $L_\mathrm{data}$ and $L_\mathrm{Maxwell}$ set to~1. 
\end{enumerate} 
In addition, we explored further combinations for assigning $p \in {1,2}$ in the losses $L_\mathrm{data}$ and $L_\mathrm{Maxwell}$.  However, all experiments employing $p=2$ in $L_\mathrm{data}$  resulted in significantly degraded performance. Consequently, these configurations were not pursued further.
The models trained on the full dataset (FNO, \FNOL{}, and 3D-WaveY-Net) are trained for $n_\mathrm{epochs}=200$ epochs. To compensate for the reduced amount of training data in the \emph{freeform-only} scenario, the models \FNOConstr{} and \WaveYConstr{} are trained for $n_\mathrm{epochs}=1000$ epochs. Correspondingly, the Maxwell loss weight is set to $\lambda_\mathrm{Maxwell}=0$ for the first 100 epochs for these constrained models, compared to 20 epochs for the full-data models.

In Section~\ref{sec:results:training}, we first analyze the training dynamics, including the convergence behavior and the relative contributions of the data loss and the Maxwell loss. We then assess surrogate model performance on a holdout test set using quantitative error and similarity metrics.
Beyond the predictive performance investigated in Section~\ref{sec:quantitative}, we analyze the influence of metasurface geometry and heterogeneity on surrogate model performance in Section~\ref{sec:results:influence}. Then, in Section~\ref{sec:results:runtime}, we discuss the computational efficiency of the surrogate models, demonstrating their speedup over conventional solvers and outlining the implications for large-scale or real-time design studies. Finally, Section~\ref{sec:results:sr} demonstrates the capability of the proposed surrogate models to perform super-resolution. An overview of the comparative evaluation of FNO and 3D-WaveY-Net carried out in this section is given in Table~\ref{tab:head2head}.

\begin{table}[h]
	\centering
	\caption{Comparison of FNO and 3D-WaveY-Net surrogate models.}
	\label{tab:head2head}
	\begin{tabular}{p{7cm}ll}
		\toprule
		\textbf{Criterion} 
		& \textbf{FNO} 
		& \textbf{3D-WaveY-Net} \\
		\midrule
		Relative error in diffraction efficiency $e_{\mathrm{diffr}}$ (\%) 
		& $3.9$ 
		& $5.3$ \\
		
		Relative $L^1$ error in $\HFieldPhasorDisreteRealAlt$ ($e_{L^1}$) (\%) 
		& $6.1$ 
		& $10.09$ \\
		
		Performance on unseen geometries 
		& Robust performance 
		& Reduced performance \\
		
		Performance on finer grids 
		& Direct inference without retraining 
		& Not possible \\
		
		Inference efficiency (TFLOPS) 
		& 0.1696 
		& 0.1205 \\
		
		\bottomrule
	\end{tabular}
\end{table}

\subsection{Training dynamics}\label{sec:results:training}

\begin{figure}[H]
	\centering
	\subfloat[]{
		\tikzPlot{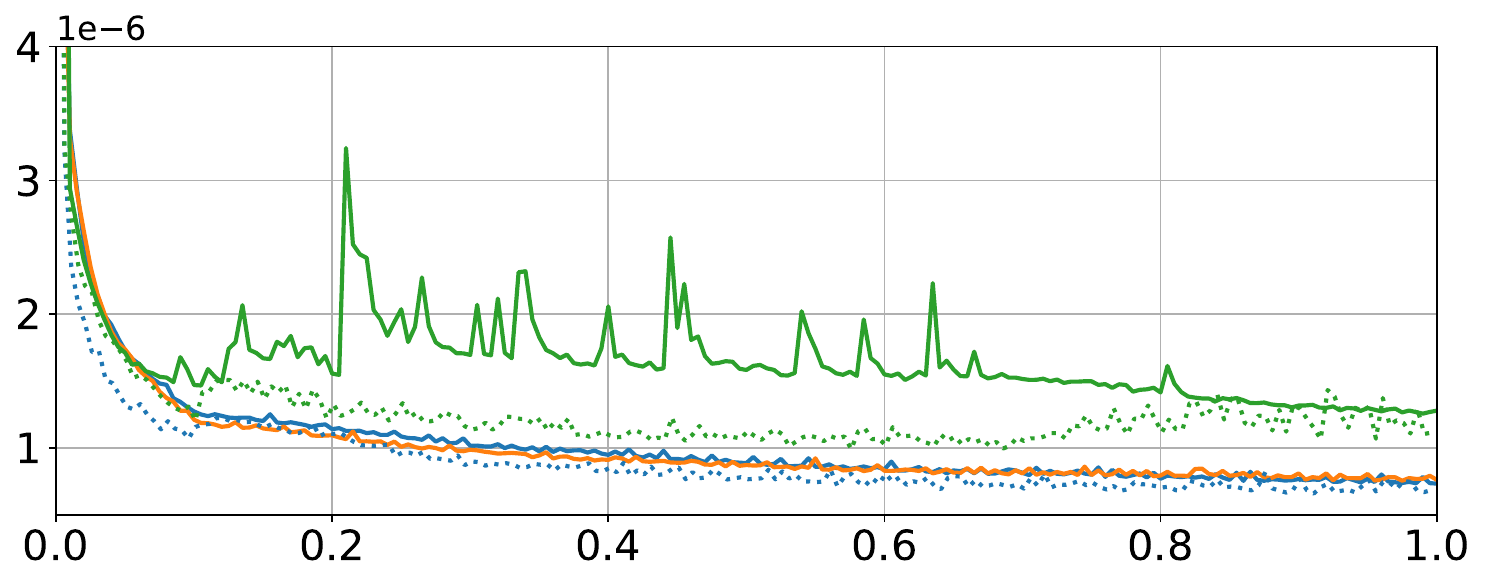}{relative epoch $\mathrm{epoch}/n_{\mathrm{epochs}}$}{training loss}{0.75}{}
	}
	\vspace{-1em}
	
	\subfloat[]{
		\begin{tabular}{c}
			\tikzPlot{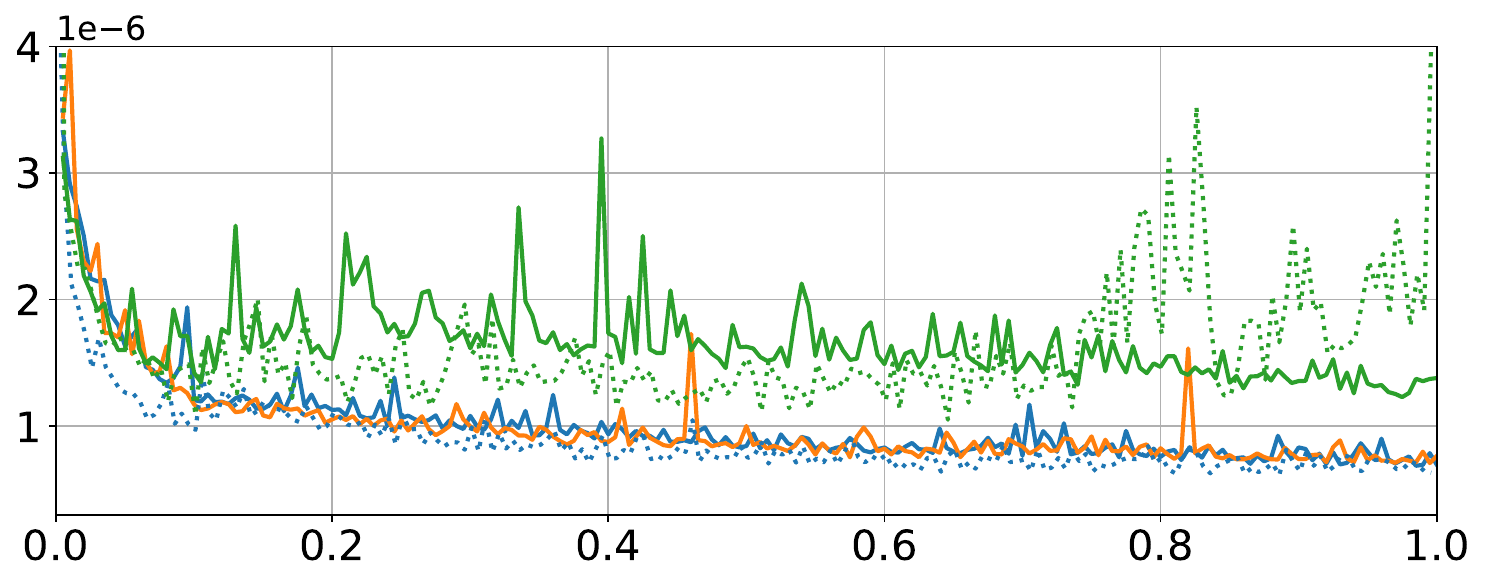}{relative epoch $\mathrm{epoch}/n_{\mathrm{epochs}}$}{validation loss }{0.75}{}\\[-1em]
			\tikzPlot{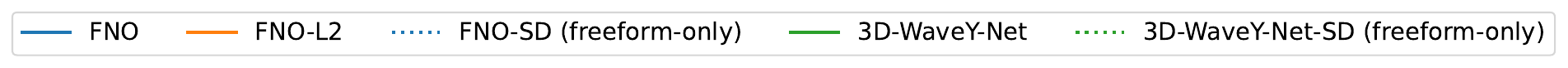}{}{}{0.75}{}
		\end{tabular}
	}
	\vspace{-1em}
	
	\subfloat[]{
		\tikzPlot{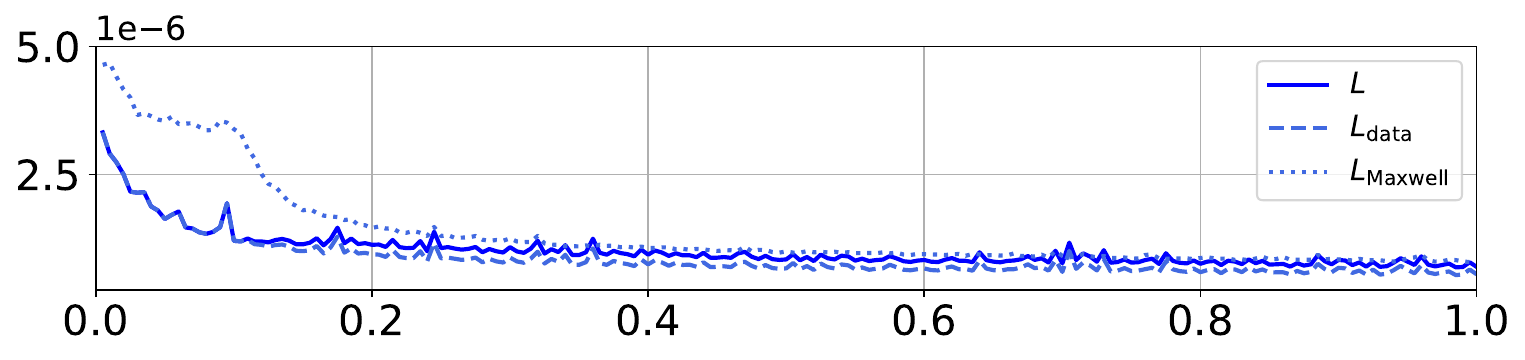}{relative epoch $\mathrm{epoch}/n_{\mathrm{epochs}}$}{validation loss}{0.75}{}
	}
	\caption{\textbf{Loss evolution.} Training (a) and validation (b) loss $L$ for \emph{FNO}, \emph{\FNOConstr}, \emph{3D-WaveY-Net} and \emph{\WaveYConstr}. Validation loss values for $L_\mathrm{data}, L_\mathrm{Maxwell}$ and $L$ during the training of 
		\emph{FNO} are shown in (c). }
	\label{fig:evolution}
\end{figure}

The evolution of training and validation losses for the five surrogate models is shown in Fig.~\ref{fig:evolution}a and b. 
Overall, the five surrogate models exhibit relatively fast convergence, although their behavior differs in stability and final performance. The FNO-type models converge the fastest and most robustly among the five. The {\FNOConstr} model also reaches low training loss values, and its final loss is even lower than that of the 3D-WaveY-Net. 
However, note that the evaluation of the {\FNOConstr} model is performed only on samples of the test set that correspond to the freeform-only scenario. Consequently, the loss values for the {\FNOConstr} model shown in Fig.~\ref{fig:evolution} may not be representative of its performance across all considered structural scenarios.  
Interestingly, the {\WaveYConstr} model appears to converge to a higher validation loss than its 
unconstrained counterpart 3D-WaveY-Net, which is trained on the full training dataset. This observation suggests 
that 3D-WaveY-Net may struggle to approximate the underlying differential operator when the diversity of 
training data is reduced to the freeform-only scenario.  
For a more objective comparison between the surrogate models, see Section~\ref{sec:quantitative}. 
When comparing the training and validation loss curves in Fig.~\ref{fig:evolution}a and b, we observe that FNO, \FNOL{} {\FNOConstr} and 3D-WaveY-Net exhibit a similar behavior between their respective training and validation losses, indicating that none of the models shows signs of overfitting within the structural scenarios for which they have been trained. In contrast, the validation loss of the {\WaveYConstr} model seems to diverge from its training loss, indicating overfitting and a reduced ability to generalize beyond the training set.

Further insights into the contribution of the loss components $L_\mathrm{data}$ and $L_\mathrm{Maxwell}$ to the total loss $L$  of the FNO model are given in  Fig.~\ref{fig:evolution}c.
Both the data $L_{\mathrm{data}}$ and the Maxwell loss $L_{\mathrm{Maxwell}}$ decrease
rather rapidly within the initial epochs, with $L_{\mathrm{data}}$ stabilizing earlier and at a lower magnitude. Recall that during the initial 20 epochs, the weight $\lambda_{\mathrm{Maxwell}}$
is kept at zero, so that the FNO model is trained solely with respect to the data loss term.
Consequently, in the interval $0 \leq \mathrm{epoch}/n_{\mathrm{epochs}} \leq 20/200=0.1$, we observe a more pronounced decrease of $L_{\mathrm{data}}$ in comparison with $L_{\mathrm{Maxwell}}$. Once the adaptive weighting strategy as described in Section~\ref{sec:simulation} and Appendix~\ref{appendix:weights} is switched on at epoch~20, the Maxwell loss starts to contribute to the total loss, which leads to a rapid drop in $L_{\mathrm{Maxwell}}$ and a slight increase in $L_{\mathrm{data}}$ at $20/n_{\mathrm{epochs}}=0.1.$
As training progresses, $L_{\mathrm{Maxwell}}$ decreases to 
a magnitude that is comparable to $L_{\mathrm{data}}$, indicating that the surrogate model increasingly satisfies Maxwell's equations, while its outputs continue to show good agreement with the simulated magnetic fields $\HFieldPhasorDisreteRealAlt$. Therefore, Fig.~\ref{fig:evolution}c indicates that a reasonable adaptive weighting strategy for $\lambda_{\mathrm{Maxwell}}$ has been chosen.

\begin{figure}[h]
	
\subfloat[ground truth]{
	\begin{tikzpicture}
		\tikzPlotAdvanced{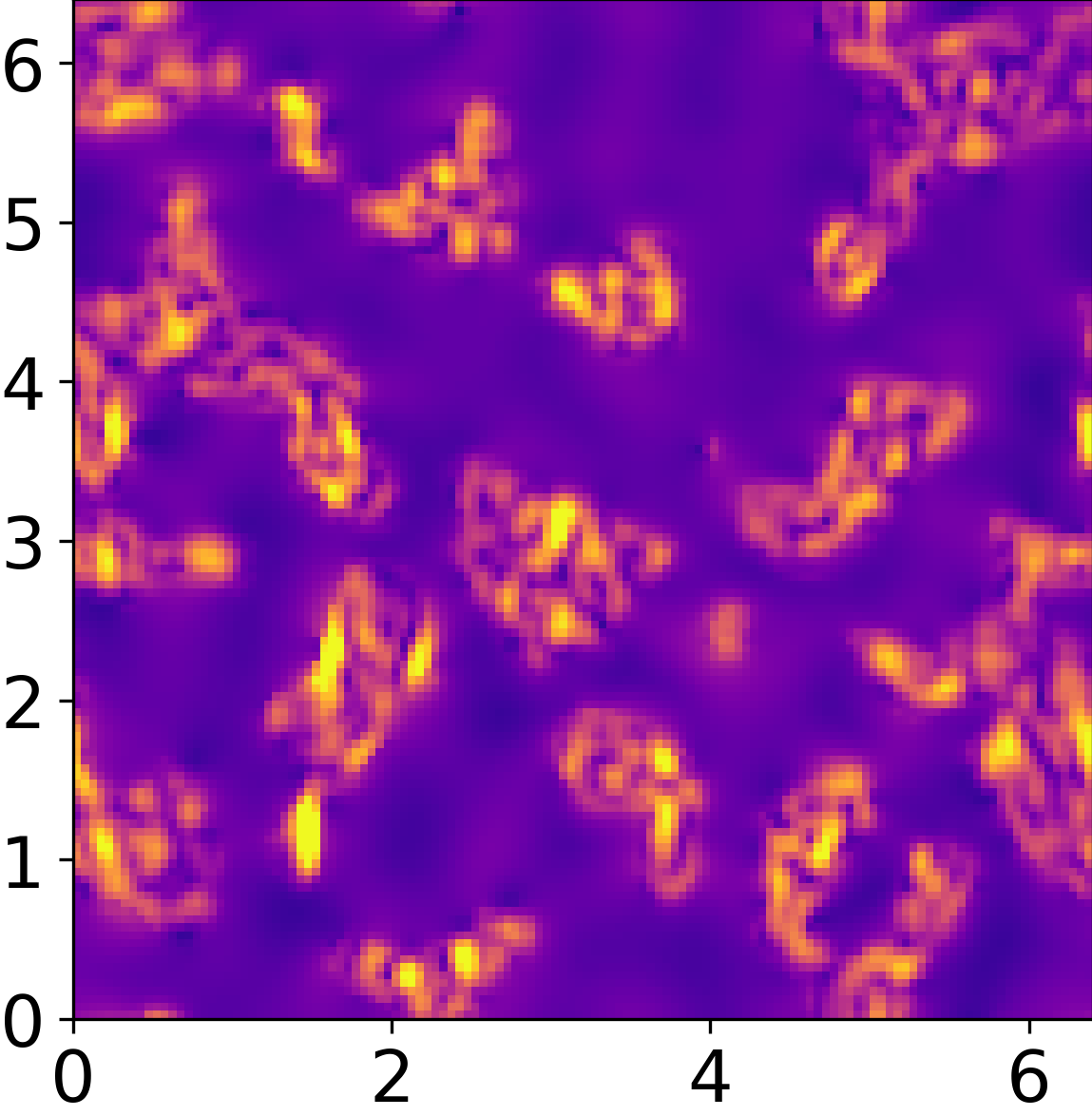}{$x$  [\si{\micro\meter}]}{$y$  [\si{\micro\meter}]}{0.25}{}{ref}{inner sep=0pt}
	\end{tikzpicture}
}
\subfloat[FNO]{
	\begin{tikzpicture}
		\tikzPlotAdvanced{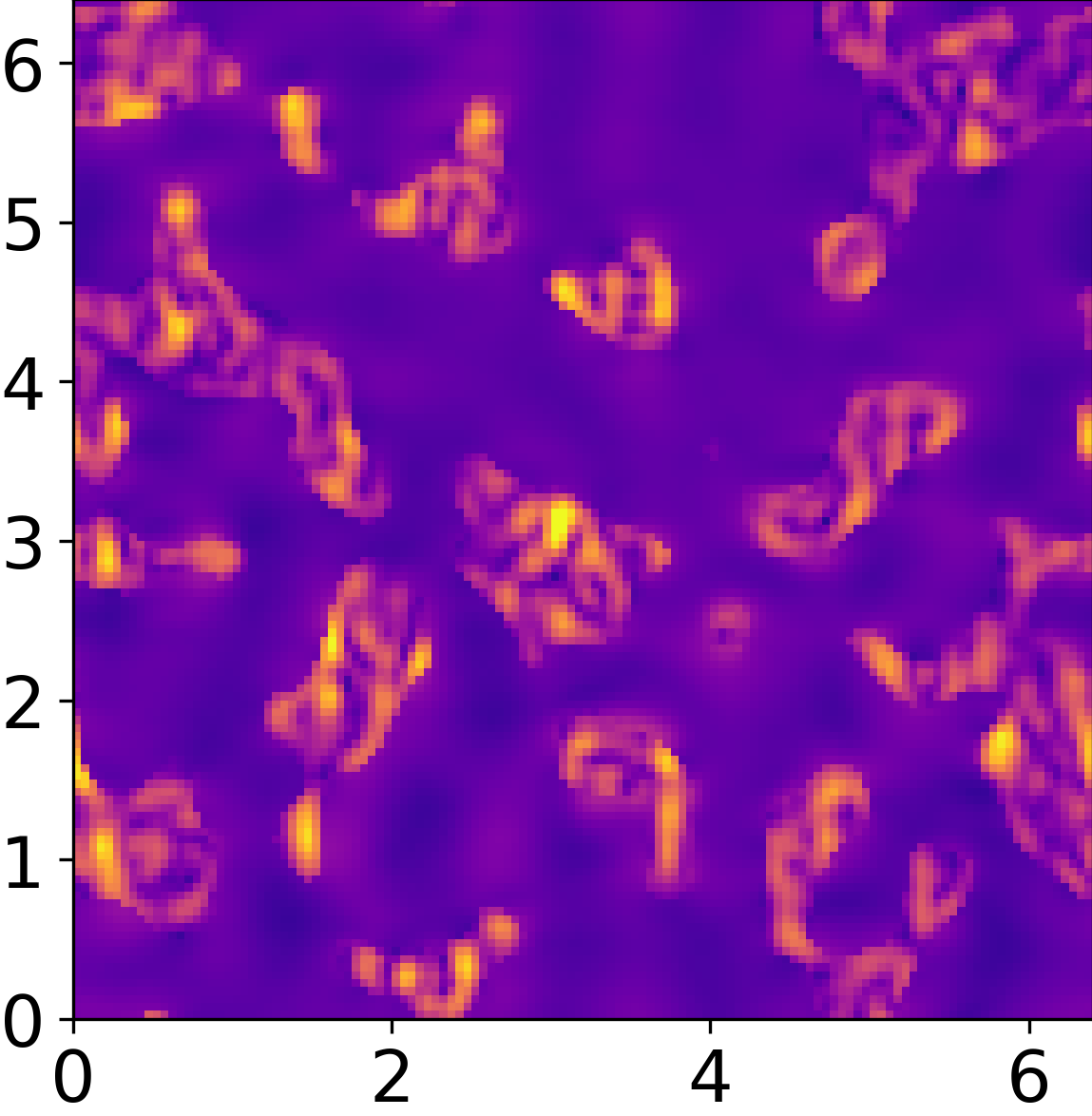}{$x$  [\si{\micro\meter}]}{$y$  [\si{\micro\meter}]}{0.25}{}{ref}{inner sep=0pt}
	\end{tikzpicture}
}
\subfloat[\FNOL]{
	\begin{tikzpicture}
		\tikzPlotAdvanced{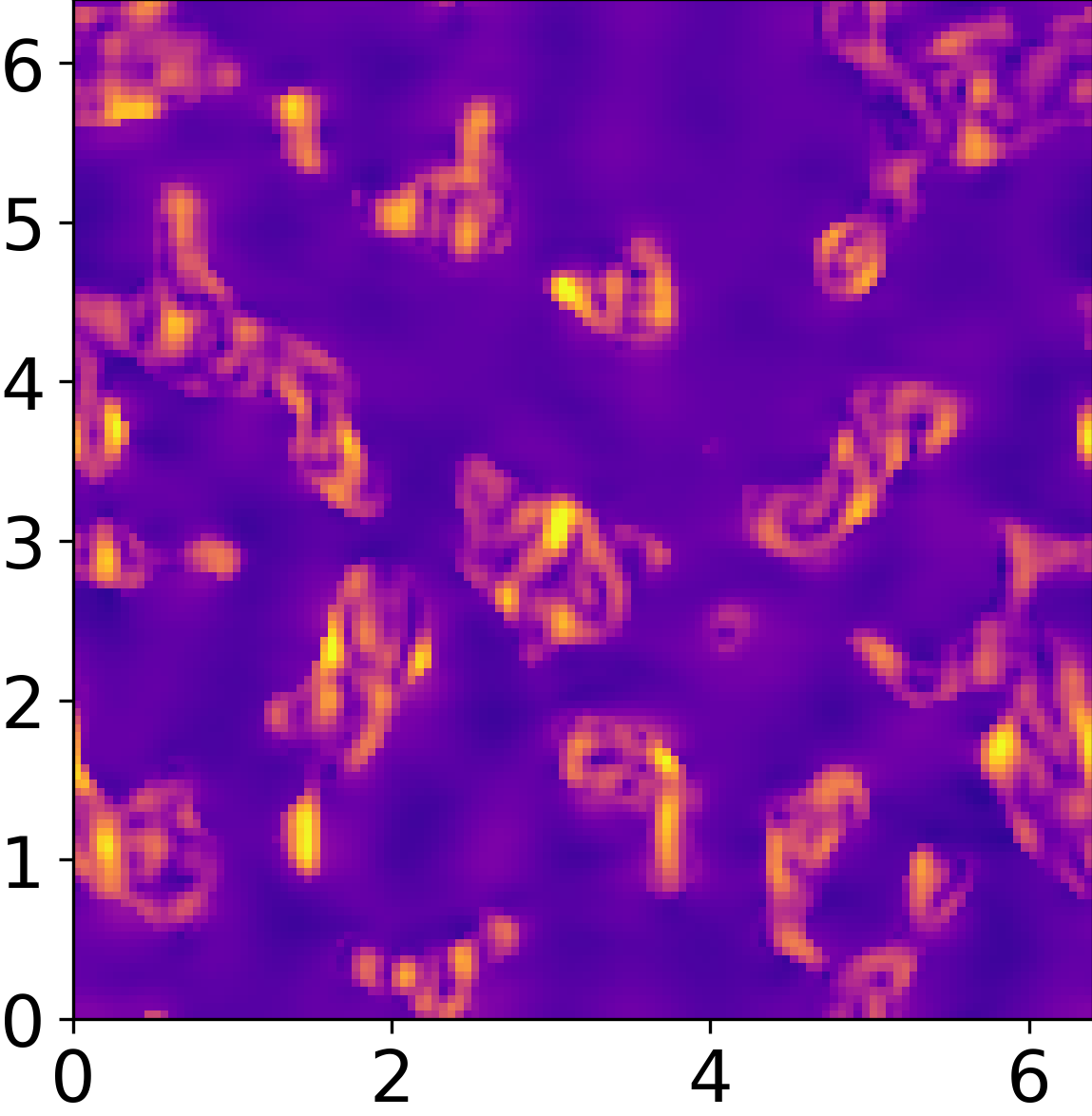}{$x$  [\si{\micro\meter}]}{$y$  [\si{\micro\meter}]}{0.25}{}{ref}{inner sep=0pt}
	\end{tikzpicture}
}
\vspace{-1em}

\subfloat[\FNOConstr]{
	\begin{tikzpicture}
		\tikzPlotAdvanced{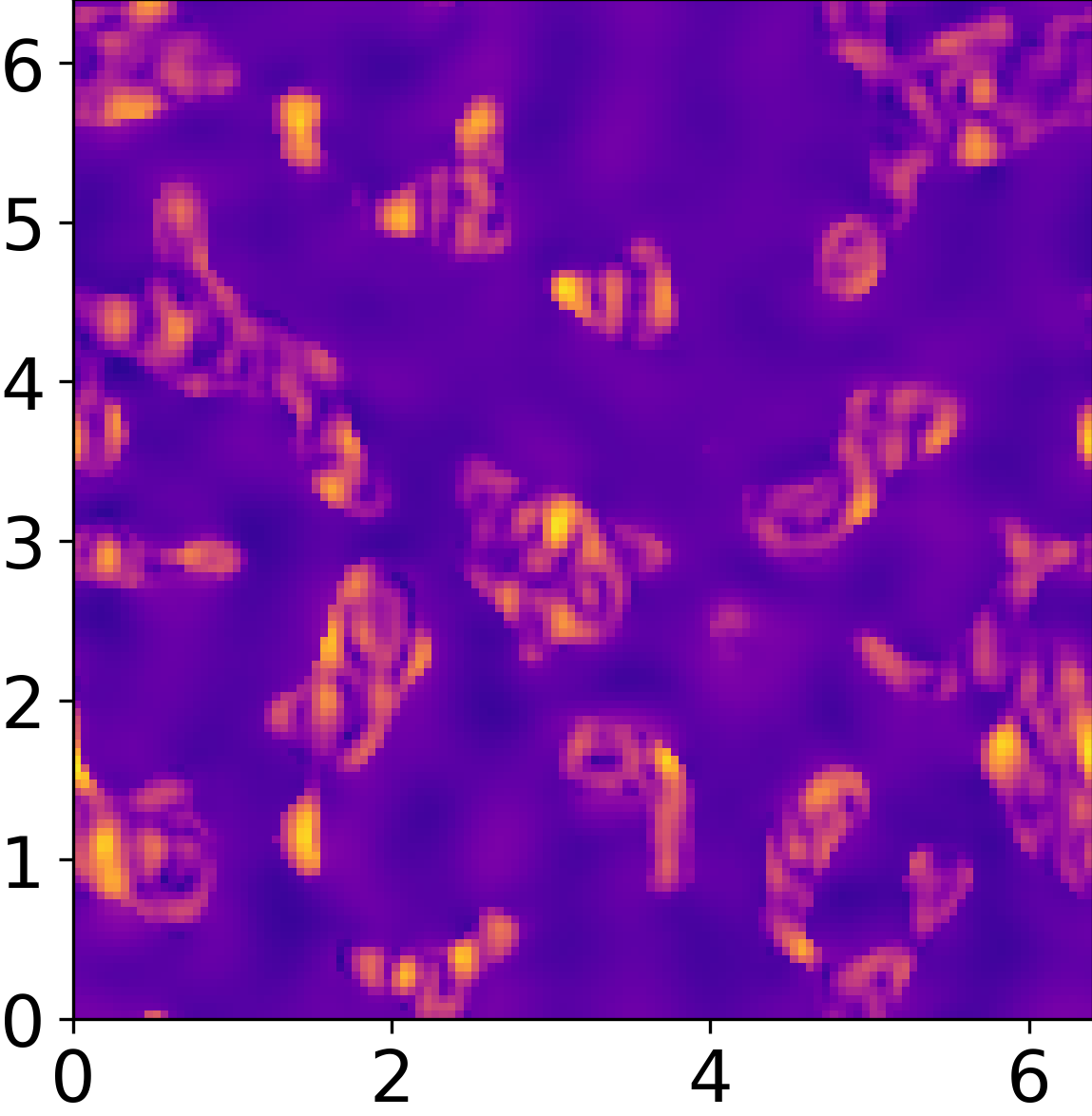}{$x$  [\si{\micro\meter}]}{$y$  [\si{\micro\meter}]}{0.25}{}{ref}{inner sep=0pt}
	\end{tikzpicture}
}
\subfloat[3D-WaveY-Net]{
	\begin{tikzpicture}
		\tikzPlotAdvanced{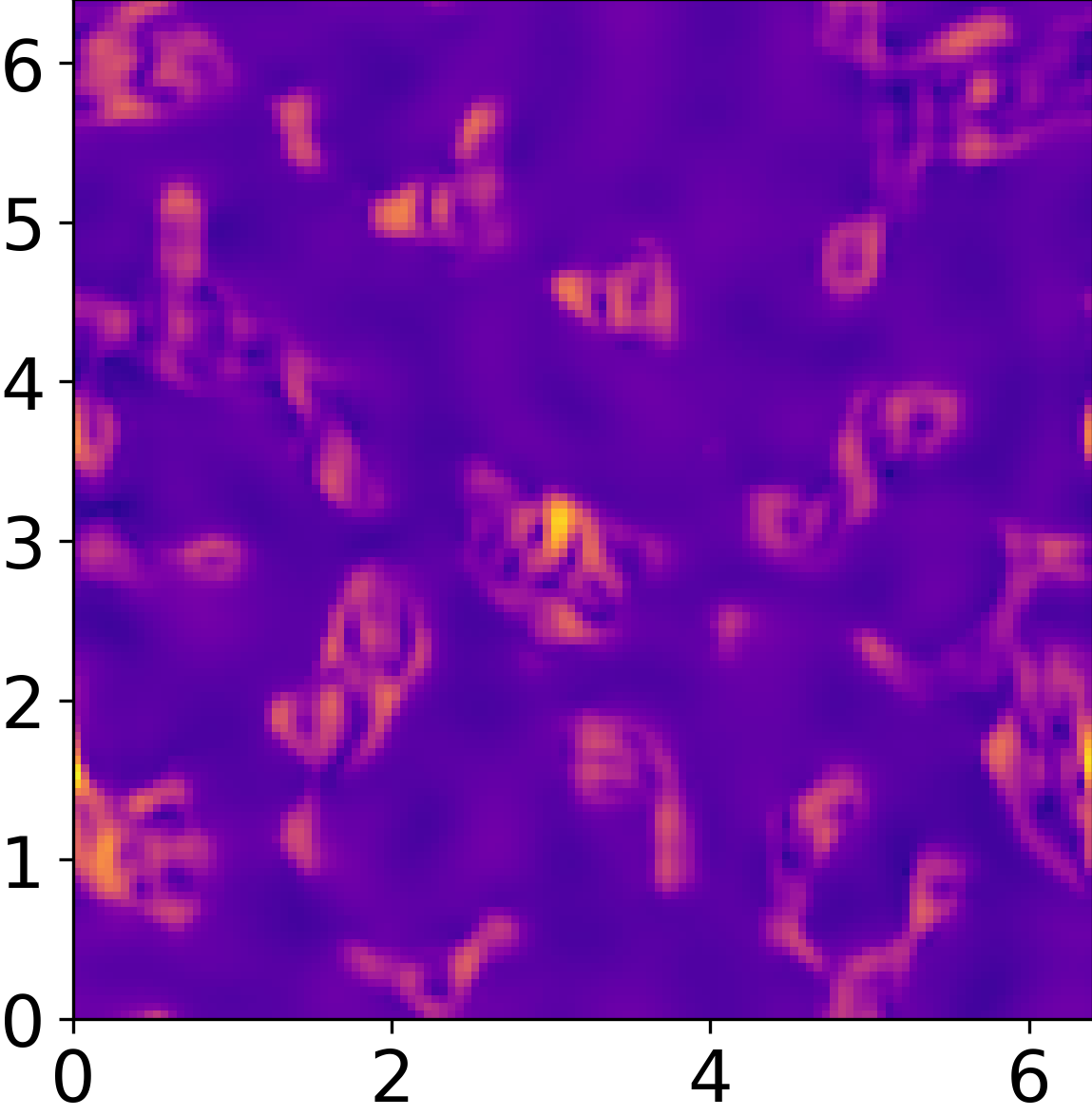}{$x$  [\si{\micro\meter}]}{$y$  [\si{\micro\meter}]}{0.25}{}{ref}{inner sep=0pt}
	\end{tikzpicture}
}
\subfloat[\WaveYConstr]{
	\begin{tikzpicture}
		\tikzPlotAdvanced{H_neural_oplimitted_data_xy.png}{$x$  [\si{\micro\meter}]}{$y$  [\si{\micro\meter}]}{0.25}{}{ref}{inner sep=0pt}
	\end{tikzpicture}
}
\raisebox{1cm}{
	\begin{tikzpicture}
		\node[inner sep=0pt ] (colorbar) {
			\includegraphics[width=1cm]{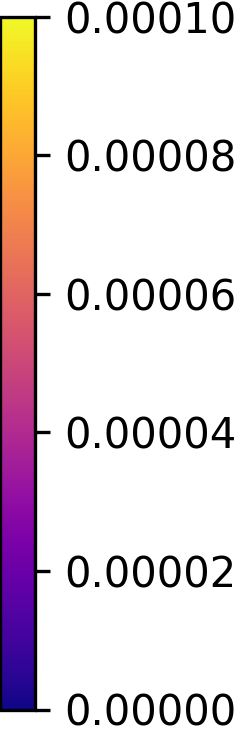}
		};
		\node[right=0.2cm of colorbar.east, anchor=west , inner sep=0pt,rotate=90 ] {
			$|\HFieldPhasorDisreteRealAlt|_2$
		};
	\end{tikzpicture}
}

	\caption{\textbf{Visual validation.}  Ground truth of the absolute values of $\HFieldPhasorDisreteRealAlt$ in a planar section (a), and the corresponding absolute values of $\HFieldPhasorDisreteRealAlt^\mathrm{pred}$predicted by FNO (b), \FNOL{} (c), {\FNOConstr} (d), 3D-WaveY-Net (e) and \WaveYConstr{} (f).  }
	\label{fig:visual:validation}	
\end{figure}

A visual comparison between the ground truth $\HFieldPhasorDisreteRealAlt$ and the predictions $\HFieldPhasorDisreteRealAlt^\mathrm{pred}$ achieved by the surrogate models 
is given in Fig.~\ref{fig:visual:validation}. At first glance, the predictions computed by the FNO, \FNOL, {\FNOConstr} and 3D-WaveY-Net models are qualitatively similar to the ground truth shown in Fig.~\ref{fig:visual:validation}a, whereas {\WaveYConstr} seems to struggle to reproduce the (amplitudes in the) ground truth. On closer inspection, the predictions obtained with the FNO and \FNOL{} models (Fig.~\ref{fig:visual:validation}b and c) most closely resemble the ground truth. In particular, regions in the ground truth that exhibit large magnetic field magnitudes  are 
reproduced more faithfully by  both models in comparison to the other surrogate models. These ``high-magnitude'' regions appear slightly smoothed in the predictions by the {\FNOConstr} and the 3D-WaveY-Net models. A detailed quantitative analysis of these differences is provided in Section~\ref{sec:quantitative}.
An analogous visual impression of predicted electric fields is provided in Fig.~\ref{fig:visual:validationE} of Appendix~\ref{appendix:efields}.

\begin{figure}[h]
	\centering
\subfloat[FNO]{
	\begin{tikzpicture}
		\tikzPlotAdvanced{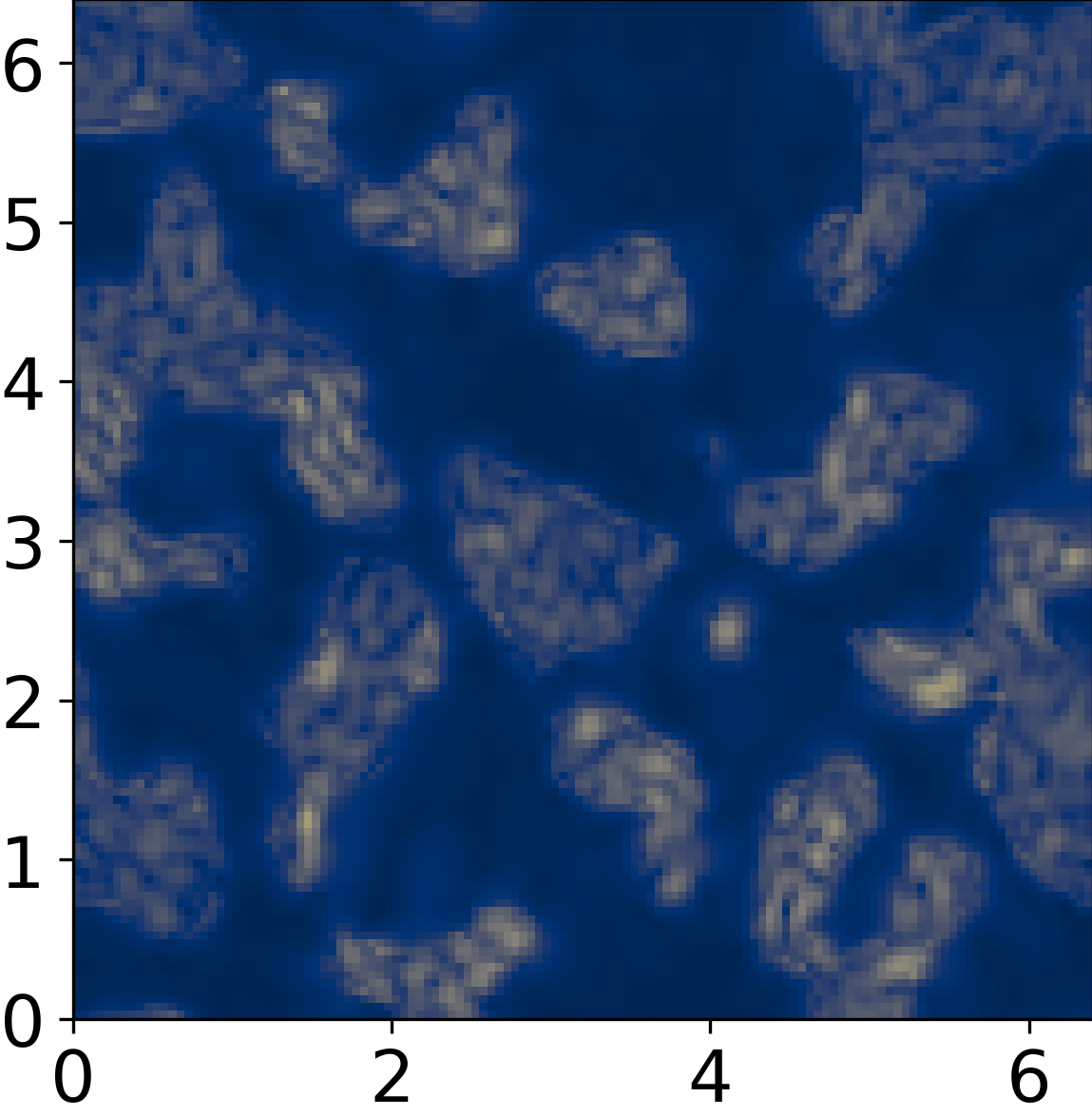}{$x$  [\si{\micro\meter}]}{$y$  [\si{\micro\meter}]}{0.25}{}{ref}{inner sep=0pt}
	\end{tikzpicture}
}
\subfloat[\FNOL]{
	\begin{tikzpicture}
		\tikzPlotAdvanced{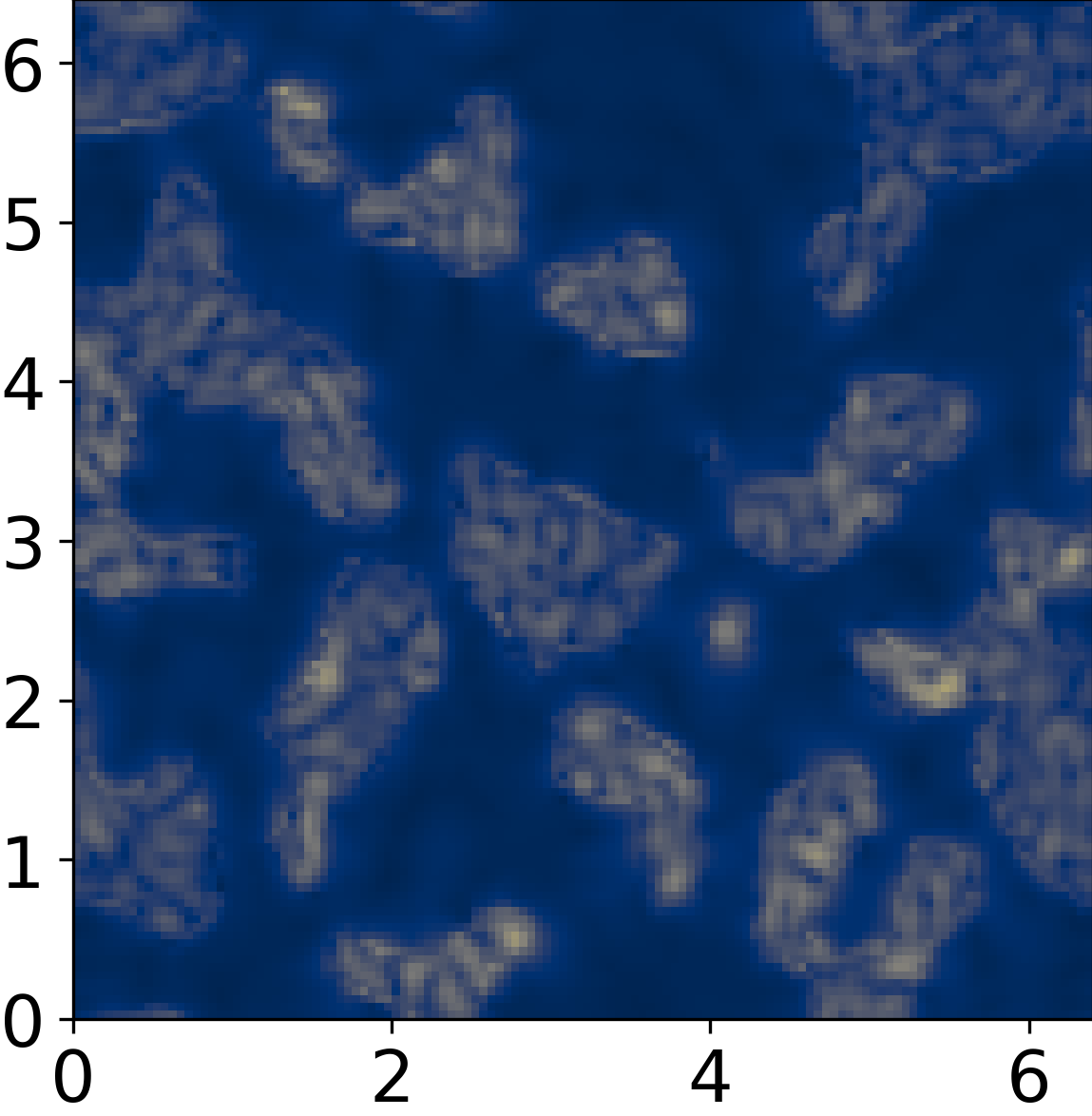}{$x$  [\si{\micro\meter}]}{$y$  [\si{\micro\meter}]}{0.25}{}{ref}{inner sep=0pt}
	\end{tikzpicture}
}
\subfloat[\FNOConstr]{
	\begin{tikzpicture}
		\tikzPlotAdvanced{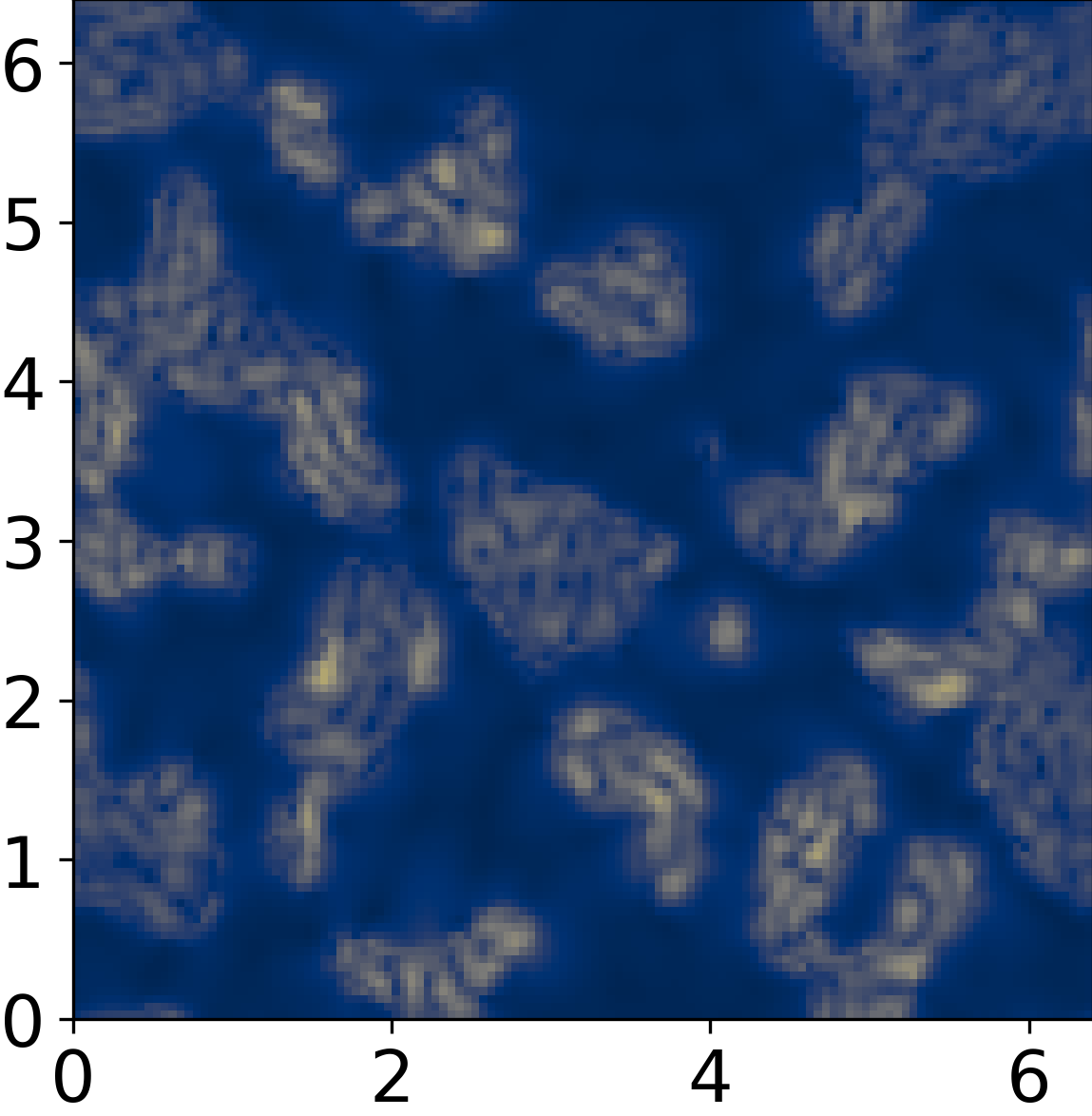}{$x$  [\si{\micro\meter}]}{$y$  [\si{\micro\meter}]}{0.25}{}{ref}{inner sep=0pt}
	\end{tikzpicture}
}
\vspace{-1em}

\hspace{5em}
\subfloat[3D-WaveY-Net]{
	\begin{tikzpicture}
		\tikzPlotAdvanced{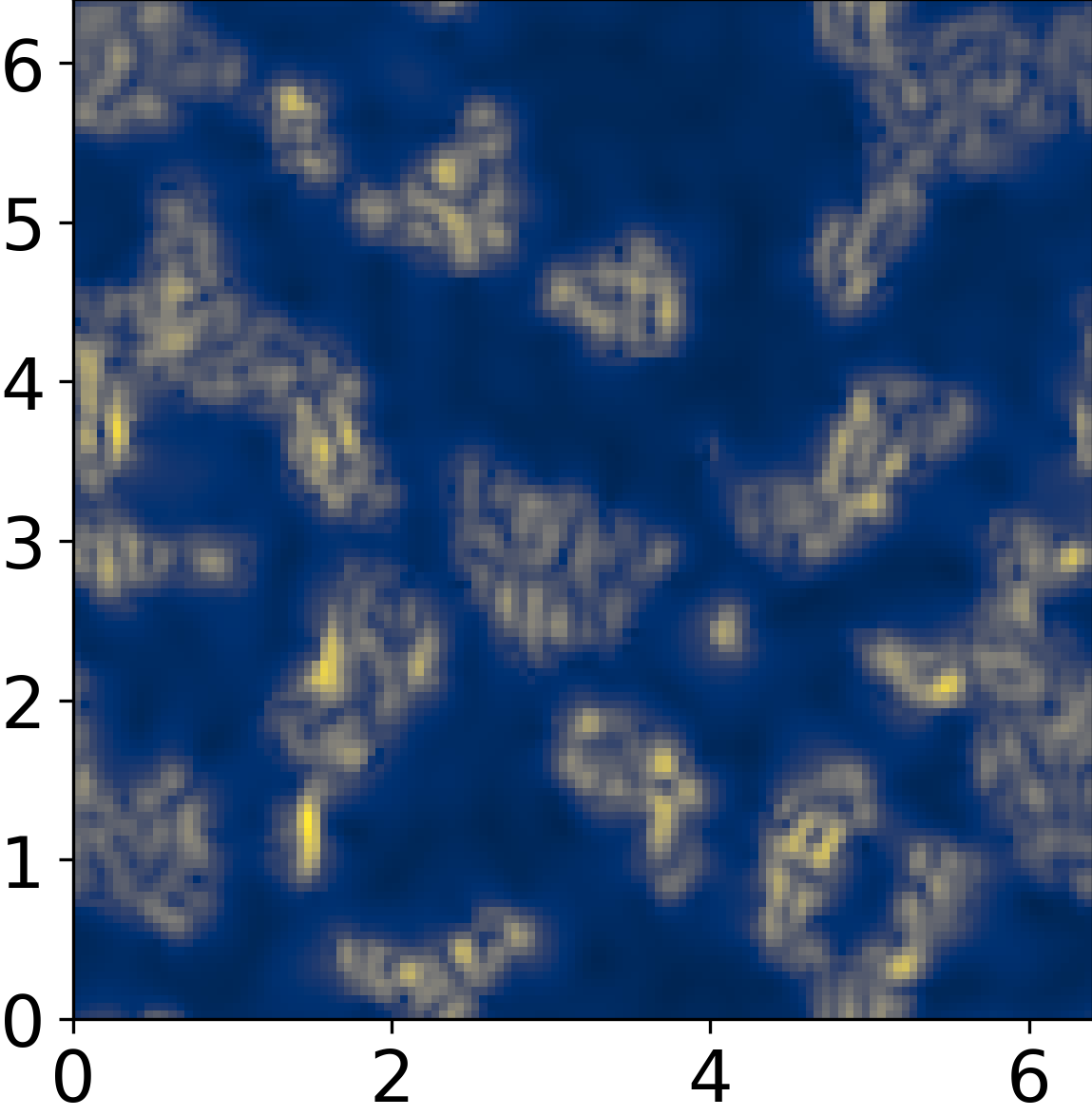}{$x$  [\si{\micro\meter}]}{$y$  [\si{\micro\meter}]}{0.25}{}{ref}{inner sep=0pt}
	\end{tikzpicture}
}
\subfloat[\WaveYConstr]{
	\begin{tikzpicture}
		\tikzPlotAdvanced{H_error_H_neural_oplimitted_data_xy.png}{$x$  [\si{\micro\meter}]}{$y$  [\si{\micro\meter}]}{0.25}{}{ref}{inner sep=0pt}
	\end{tikzpicture}
}
\raisebox{1cm}{
	\begin{tikzpicture}
		\node[inner sep=0pt ] (colorbar) {
			\includegraphics[width=1cm]{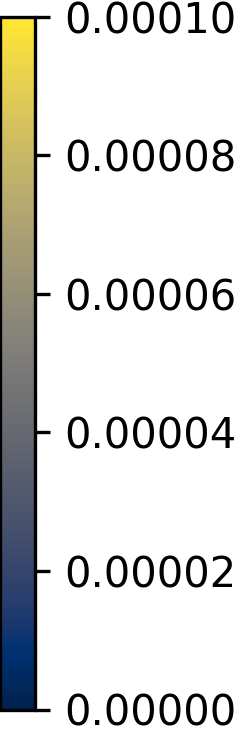}
		};
		\node[right=0.2cm of colorbar.east, anchor=west , inner sep=0pt,rotate=90,xshift=-1cm ] {
			$|\HFieldPhasorDisreteRealAlt-\HFieldPhasorDisreteRealAlt^{\mathrm{pred}}|_2$
		};
	\end{tikzpicture}
}
	\caption{\textbf{Point-wise errors.} Magnitudes of residuals $\HFieldPhasorDisreteRealAlt-\HFieldPhasorDisreteRealAlt^\mathrm{pred}$   in a planar section for FNO (a), \FNOL{} (b), {\FNOConstr} (c), 3D-WaveY-Net (d) and \WaveYConstr{} (e).  }
	\label{fig:visual:pointwise}	
\end{figure}

\subsection{Quantitative performance analysis}\label{sec:quantitative}
To complement the visual comparison shown in Fig.~\ref{fig:visual:validation}, we provide a quantitative comparison between the predicted magnetic fields $\HFieldPhasorDisreteRealAlt^{\mathrm{pred}}$ of the surrogate models and the corresponding ground truth $\HFieldPhasorDisreteRealAlt$, e.g., by aggregating point-wise prediction errors as shown in Fig.~\ref{fig:visual:pointwise}. To achieve this, for each 3D distribution $\epsVol$ of relative permittivity in the test set, the trained surrogate models are  deployed to compute their respective predictions $\HFieldPhasorDisreteRealAlt^{\mathrm{pred}}$, which can be compared to the ground truth $\HFieldPhasorDisreteRealAlt$ of the test set that has been computed by FDTD simulations.
To perform a quantitative comparison between such pairs $(\HFieldPhasorDisreteRealAlt^{\mathrm{pred}}, \HFieldPhasorDisreteRealAlt)$, we introduce several metrics that measure  discrepancies or similarities between the predicted and ground truth fields. Note that in this section we again consider magnetic fields as complex-valued phasor fields. 
Accordingly, we revert from the real-valued representation in $\R^6$ 
to the complex-valued representation
$
\HFieldPhasorDisreteRealAlt^{\mathrm{pred}}, \HFieldPhasorDisreteRealAlt
\colon \windowVolumeDiscrete \to \C^3.
$
The complex-valued component functions are denoted by $H_\mathrm{x},H_\mathrm{y},H_\mathrm{z} \colon \windowVolumeDiscrete \to \C$, i.e., we have $\HFieldPhasorDisreteRealAlt(\rvec)=(
H_\mathrm{x}(\rvec),H_\mathrm{y}(\rvec),H_\mathrm{z}(\rvec))$. Analogously, the component functions of $\HFieldPhasorDisreteRealAlt^{\mathrm{pred}}$ are denoted by $H_\mathrm{x}^\mathrm{pred},H_\mathrm{y}^\mathrm{pred},H_\mathrm{z}^\mathrm{pred}$.

\subsubsection{Metrics}

To quantify the discrepancy between two fields $Y,Y^\mathrm{pred} \colon \windowVolumeDiscrete \to \R$ we consider normalized mean absolute and normalized root mean squared errors denoted by $\mathrm{nMAE}$ and $\mathrm{nRMSE}$ which are given by
\begin{equation}
	\mathrm{nMAE}(Y^\mathrm{pred},Y)
	=
	\frac{
		\sum_{\rvec\in W_\mathrm{d,const.}^\mathrm{3D} } |Y^\mathrm{pred}(\rvec) - Y(\rvec)|
	}{
		\sum_{\rvec\in W_\mathrm{d,const.}^\mathrm{3D} } | Y(\rvec)|
	},
\end{equation}
and by
\begin{equation}
	\mathrm{nRMSE}(Y^\mathrm{pred},Y)
	=
	\frac{
		\left(
		\sum_{\rvec\in W_\mathrm{d,const.}^\mathrm{3D} } |Y^\mathrm{pred}(\rvec) - Y(\rvec)|^2
		\right)^{1/2}
	}{
		\left(
		\sum_{\rvec\in W_\mathrm{d,const.}^\mathrm{3D} } | Y(\rvec)|^2
		\right)^{1/2}
	},
\end{equation}

To quantify the discrepancy between the predicted magnetic field 
$\HFieldPhasorDisreteRealAlt^{\mathrm{pred}}$ and the ground truth field 
$\HFieldPhasorDisreteRealAlt$, we use $\mathrm{nMAE}$ and $\mathrm{nRMSE}$ to quantify relative 
$L^p$ errors. More precisely, following the approach outlined in \cite{Chen2022Stanford} we consider
\begin{equation}
	e_{L^1}
	(\HFieldPhasorDisreteRealAlt^\mathrm{pred},\HFieldPhasorDisreteRealAlt)
	= \frac{1}{2} \left(
	\mathrm{nMAE}(\mathrm{Re}(H_\mathrm{y}^\mathrm{pred}), \mathrm{Re}(H_\mathrm{y}))
	+
	\mathrm{nMAE}(\mathrm{Im}(H_\mathrm{y}^\mathrm{pred}), \mathrm{Im}(H_\mathrm{y}))
	\right),
\end{equation}
which we refer to as relative $L^1$ error. The relative $L^2$ error is given by
\begin{equation}
	e_{L^2}
	(\HFieldPhasorDisreteRealAlt^\mathrm{pred},\HFieldPhasorDisreteRealAlt)
	= \frac{1}{2} \left(
	\mathrm{nRMSE}(\mathrm{Re}(H_\mathrm{y}^\mathrm{pred}), \mathrm{Re}(H_\mathrm{y}))
	+
	\mathrm{nRMSE}(\mathrm{Im}(H_\mathrm{y}^\mathrm{pred}), \mathrm{Im}(H_\mathrm{y}))
	\right).
\end{equation}
These metrics quantify the global deviation between 
prediction and ground truth with respect to the $L^p$ norm and normalize it by the 
corresponding $L^p$ magnitude of the ground truth field, thereby providing a 
scale-invariant measure of overall prediction performance.
Note that, the metrics are computed for the dominant component of the magnetic field, namely the $y$-component.

To evaluate how accurately the surrogate models reproduce the magnitudes of 
the magnetic field, we consider the relative error $e_{\mathrm{amp}}$ of field amplitudes which is given by
\begin{equation}
	e_{\mathrm{amp}}
	=
	\frac{1}{|W_\mathrm{d,const.}^\mathrm{3D}|}
	\sum_{\rvec \in W_{\mathrm{d,const.}}^{\mathrm{3D}}}
	\left|
	\frac{
		|\HFieldPhasorDisreteRealAlt^{\mathrm{pred}}(\rvec)|_2 - 	|\HFieldPhasorDisreteRealAlt(\rvec)|_2
	}{
		|\HFieldPhasorDisreteRealAlt(\rvec)|_2 + \tau
	}
	\right|,
\end{equation}
where $|W_\mathrm{d,const.}^\mathrm{3D}|$ denotes the cardinality of the set $W_\mathrm{d,const.}^\mathrm{3D}$ of grid points and 
$\tau = 10^{-10}$  is a constant that is introduced in the denominator to avoid numerical instability. 

Analogously, these point-wise metrics can be computed for electric fields. More precisely,  for ground truth and predicted magnetic fields $\HFieldPhasorDisreteRealAlt, \HFieldPhasorDisreteRealAlt^\mathrm{pred}$, the corresponding electric fields
$\EFieldPhasorDisreteRealAlt, \EFieldPhasorDisreteRealAlt^\mathrm{pred}$ can be computed in order to quantify prediction errors for electric fields, see Appendix~\ref{appendix:efields} for further details.

To complement the metrics introduced above, we evaluate the performance of the surrogate model in terms of the \emph{diffraction efficiencies} it predicts.
In contrast to point-wise field errors, which quantify local discrepancies between predicted and ground-truth fields, diffraction efficiencies are physically meaningful quantities that measure how optical power is distributed among individual propagating diffraction orders.  
Following the approach in \cite{liu2009rigorous}, the diffraction efficiency
$\eta_{+1}$ of the transmitted $(+1)$ diffraction order is obtained, which 
corresponds to the ratio between the optical power carried by the transmitted $(+1)$ diffraction order and the total incident optical power.
Analogously, the predicted diffraction efficiency $\eta_{+1}^{\mathrm{pred}}$ is computed from
$\HFieldPhasorDisreteRealAlt^{\mathrm{pred}}$.
Finally, the relative diffraction efficiency error is defined as
\begin{equation}
	e_{\mathrm{diffr}}
	=
	\left|\frac{\eta_{+1}^{\mathrm{pred}} - \eta_{+1}}
	{\eta_{+1}}\right|.
\end{equation}
This performance metric quantifies the performance of surrogate models to reproduce the functional optical response of metasurfaces.

\subsubsection{Quantitative comparison of surrogate models}
\begin{figure}[H]
	\centering
	\subfloat[]{
		\begin{tikzpicture}
			\tikzPlotAdvanced{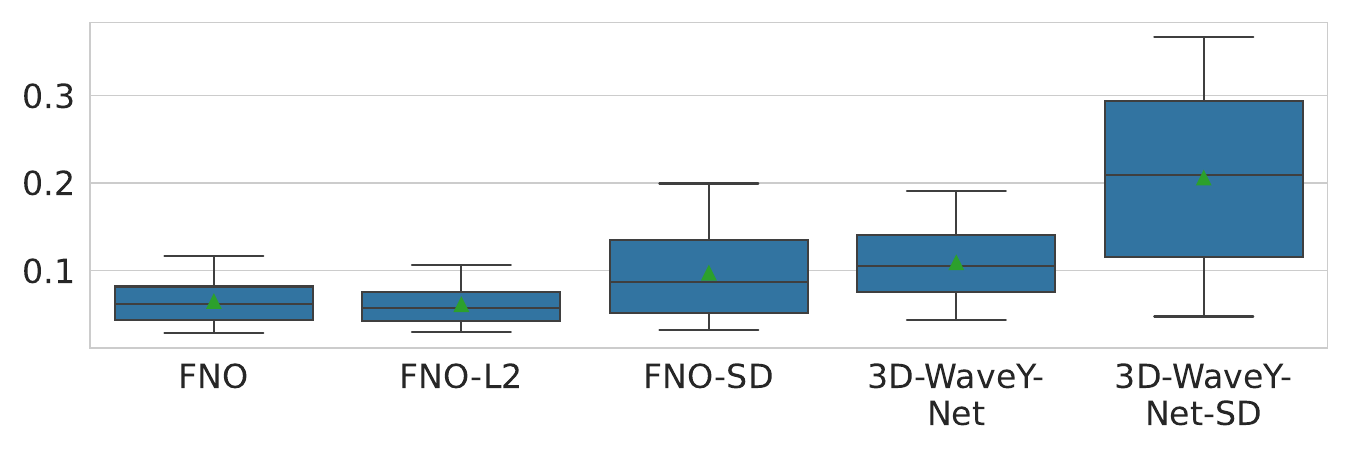}{}{$e_{L^1}$}{0.45}{}{ref}{inner sep=0pt}
		\end{tikzpicture}
	}
	\subfloat[]{
		\begin{tikzpicture}
			\tikzPlotAdvanced{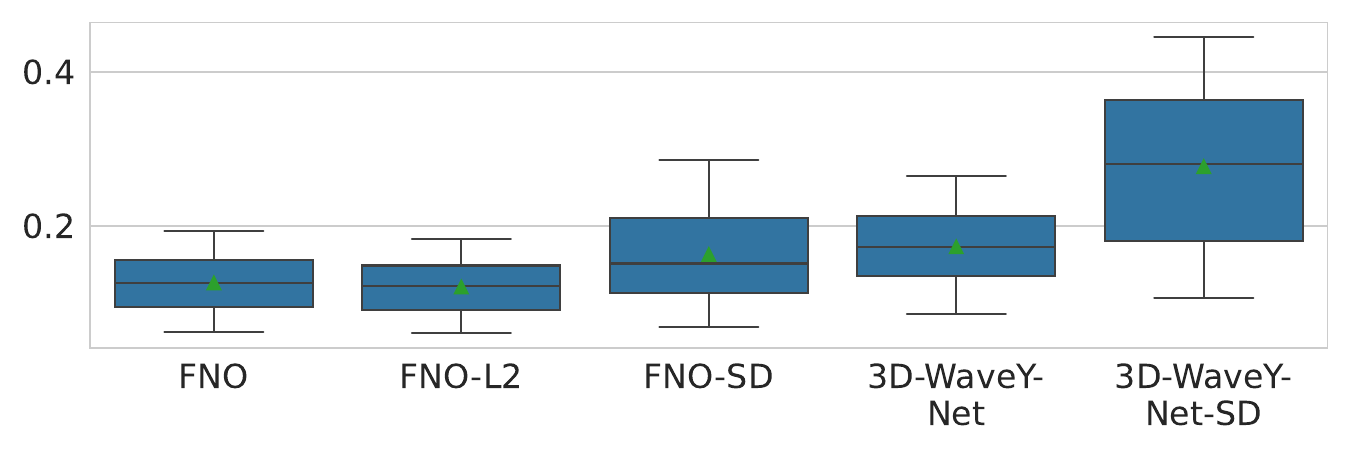}{}{$e_{L^2}$}{0.45}{}{ref}{inner sep=0pt}
		\end{tikzpicture}
	}
	\vspace{-1em}
	
	\subfloat[]{
		\begin{tikzpicture}
			\tikzPlotAdvanced{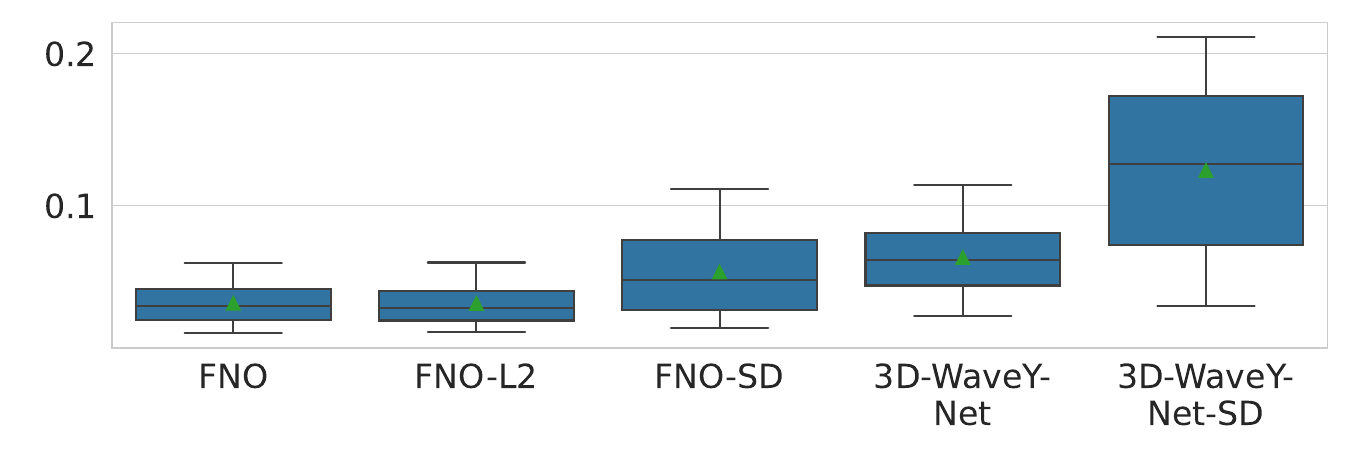}{}{$e_{\mathrm{amp}}$}{0.45}{}{ref}{inner sep=0pt}
		\end{tikzpicture}
	}
	\subfloat[]{
		\begin{tikzpicture}
			\tikzPlotAdvanced{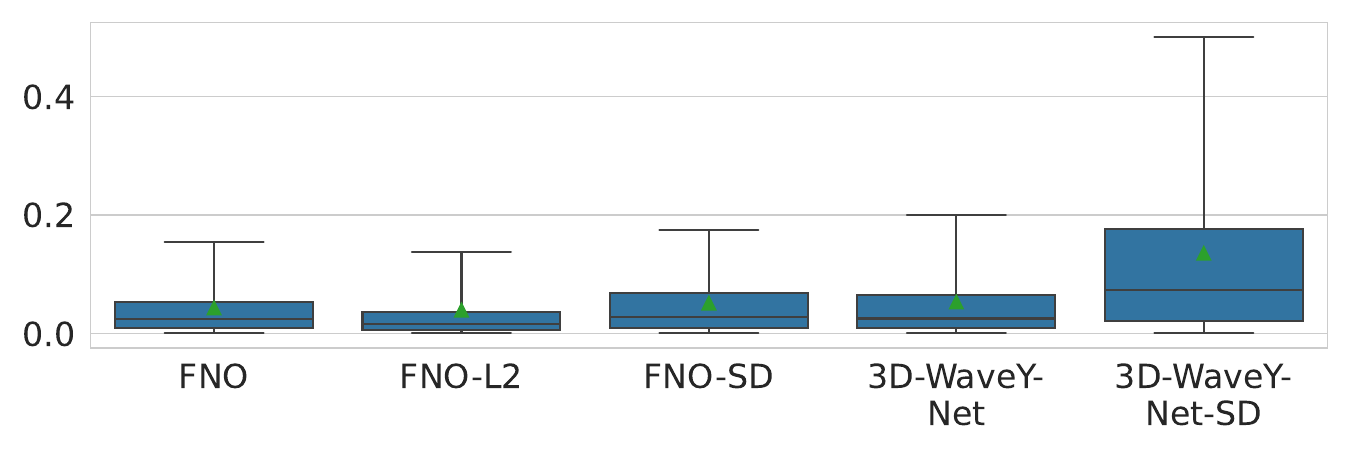}{}{$e_{\mathrm{diffr}}$}{0.45}{}{ref}{inner sep=0pt}
		\end{tikzpicture}
	}

	\caption{\textbf{Quantitative evaluation of surrogate model performance.}
		Box plots of metrics computed over the test data for  
		(a)  relative $L^1$ error $e_{L^1}$, 
		(b)  relative $L^2$ error $e_{L^2}$, 
		(c) amplitude error $e_{\mathrm{amp}}$, and
		(d) diffraction efficiency error $e_{\mathrm{diffr}}$. 
		Box heights represent the interquartile range; horizontal lines within boxes mark medians; the whiskers extend to the 5th and 95th percentiles; the green triangles indicate the mean.
	}
	\label{fig:quantitative}
\end{figure}

The metrics described above are computed for each individual sample in the test dataset and for each of the trained surrogate models. 
Statistics of the metrics computed in this manner are visualized in Fig.~\ref{fig:quantitative}, which summarizes the resulting distributions over test sets. Mean values and standard deviations of these metrics are listed in Table~\ref{table:performance}.

When comparing the  
relative $L^1$,  relative $L^2$ and relative amplitude errors for the five surrogate models (Figs.~\ref{fig:quantitative}a-c), 
we observe that the relative amplitude errors are consistently smaller than the corresponding 
$L^1$ and $L^2$ errors. This indicates that the surrogate models are more 
accurate in estimating the overall magnitudes of the magnetic fields than in 
reproducing their directional components.

\begin{table}[h]
	\caption{Mean and standard deviation of all performance metrics of surrogate models computed over the test set. The lowest mean error for each metric is highlighted in bold.}\label{table:performance}
	\centering
	\begin{tabular}{lcccc}
		\toprule
		& $e_{L^1}$ & $e_{L^2}$ & $e_{\mathrm{amp}}$ & $e_{\mathrm{diffr}}$ \\
		\midrule
		FNO & 0.065 $\pm$ 0.027 & 0.126 $\pm$ 0.043 & \textbf{0.036 $\pm$ 0.015} & 0.043 $\pm$ 0.059 \\
		FNO-L2 & \textbf{0.061 $\pm$ 0.025} & \textbf{0.121 $\pm$ 0.040} & \textbf{0.036 $\pm$ 0.015} & \textbf{0.039 $\pm$ 0.094} \\
		FNO-SD & 0.097 $\pm$ 0.053 & 0.163 $\pm$ 0.068 & 0.056 $\pm$ 0.029 & 0.051 $\pm$ 0.066 \\
		3D-WaveY-Net & 0.109 $\pm$ 0.045 & 0.173 $\pm$ 0.057 & 0.066 $\pm$ 0.029 & 0.053 $\pm$ 0.080 \\
		3D-WaveY-Net-SD & 0.206 $\pm$ 0.105 & 0.277 $\pm$ 0.108 & 0.123 $\pm$ 0.058 & 0.135 $\pm$ 0.178 \\
		\bottomrule
	\end{tabular}
\end{table}

Among the considered error metrics (Figs.~\ref{fig:quantitative}a-c), the \FNOL{} model achieves the lowest mean values together with the smallest interquartile ranges (closely followed by FNO), indicating both high predictive performance and low variability of performance on the test dataset. In contrast, the {\FNOConstr} model exhibits larger mean errors, and a larger interquartile range. This lowered performance reflects the model's restricted generalization capability, since it has been solely trained on the freeform-only scenario. The 3D-WaveY-Net exhibits an intermediate performance: its mean errors listed in Table~\ref{table:performance} are  higher than those of all FNO-type models.
Notably, the {\WaveYConstr} model shows the overall worst performance among all models, despite 
being trained on the same freeform-only dataset as \FNOConstr.
Interestingly, the performance of \FNOConstr{} is better than that of  3D-WaveY-Net, even 
though 3D-WaveY-Net has been trained on data from all structural scenarios. This observation indicates that 
neural operators may be more effective than conventional CNN architectures at learning the 
underlying differential operator involved in Maxwell's equations.

The relative errors of the diffraction efficiency shown in Fig.~\ref{fig:quantitative}d confirm the observations made with respect to the point-wise error metrics. In particular, FNO-type models exhibit consistently low relative errors, demonstrating their ability to accurately predict the diffraction efficiencies. Notably, the relative errors for \FNOL{} are at \SI{3.9}{\percent}. The 3D-WaveY-Net yields slightly larger errors. Again, the {\WaveYConstr} shows the worst performance. Overall, these results confirm that the FNO-type models provide the most accurate predictions among the considered surrogate models.

\subsection{Influence of metasurface design}\label{sec:results:influence}
In Section~\ref{sec:quantitative}, we evaluated the overall performance of the surrogate models. However, this analysis does not reveal how characteristics of the inputs influence the performance of surrogate models. In other words, how the performance of the trained surrogate models depends on the metasurface design encoded in the relative permittivity distributions $\epsVol$. 
To investigate these effects, we  introduce  descriptors that characterize the metasurface design encoded in the permittivity distributions $\epsVol$.  By correlating the resulting descriptors with the corresponding error and similarity metrics between $\HFieldPhasorDisreteRealAlt^{\mathrm{pred}}$ and $\HFieldPhasorDisreteRealAlt$, as discussed in Section~\ref{sec:quantitative}, we can assess how different aspects of the metasurface design influence the predictive performance of the surrogate models.

\subsubsection{Metasurface descriptors} Recall that the 3D distributions $\epsVol$ of permittivity were constructed by generating a 2D distribution $\epsPlane$ that represents a planar cross section of the metasurface, which is deposited on a virtual SiO$_2$ film. In particular, $\epsVol$ is uniquely determined by $\epsPlane$, so it suffices to consider descriptors computed directly from the cross-sectional distribution $\epsPlane$.

The first descriptor we consider is the mean relative permittivity $\varphi_\mathrm{mean}$ of $\epsPlane$, which is given by
\begin{equation}\label{eq:descriptor:mean}
	\varphi_\mathrm{mean}
	=\frac{1}{|B|} \sum_{\rvec\in B} \epsPlane(\rvec),
\end{equation}
where $B=\{\rvec \in \windowPlaneDiscrete \colon \epsPlane(\rvec)>1 \}.$
Note that by constraining the average in Eq.~\eqref{eq:descriptor:mean} on the set
$B$, we omit permittivity values of 1 in the cross-section that are associated with the vacuum.

To quantify the variability of the relativity permittivity values, we consider their empirical standard deviation $\varphi_\mathrm{std}$ which is given by
\begin{equation}
	\varphi_\mathrm{std}
	= \left(
	\frac{1}{|B|-1} \sum_{\rvec\in B} (\epsPlane(\rvec)-\varphi_\mathrm{mean})^2
	\right)^{1/2}.
\end{equation}

The area fraction $\varphi_{\mathrm{area}}=\frac{|B|}{|\windowPlaneDiscrete|}$ serves as a further descriptor that 
quantifies the proportion of the domain that is occupied by the high-permittivity 
material phase.

Finally, we introduce a descriptor that quantifies the ``coarseness'' of the 
metasurface structure. For this purpose, we compute the Euclidean distance 
transform $D \colon \windowPlaneDiscrete \to [0,\infty)$ with respect to the 
complement $B^{\complement}$ of the high-permittivity region \cite{DigitalImageProcessing}. It is defined by
$
D(\rvec) = \mathrm{dist}\bigl(\rvec,\, B^{\complement}\bigr), 
$
so that $D(\rvec)$ gives the distance from the point $\rvec$ to the closest point 
outside of $B$. An equivalent interpretation is that $D(\rvec)$ represents the 
radius $r\geq 0$ of the largest disk centered at $\rvec$ that is still fully contained in 
$B$. This provides a  measure of the local thickness or coarseness of 
the high-permittivity phase $B$. Then, the mean radius $\varphi_\mathrm{radius}$ given by
\begin{equation}
	\varphi_\mathrm{radius}
	=
	\frac{1}{|B|}
	\sum_{\rvec \in B }
	D(\rvec),
\end{equation}
is a descriptor with which we quantify the overall coarseness of a metasurface.

\subsubsection{Correlation analysis}
\begin{figure}[H]
	\centering
	\subfloat[FNO]{
		\begin{tikzpicture}
			\tikzPlotAdvanced{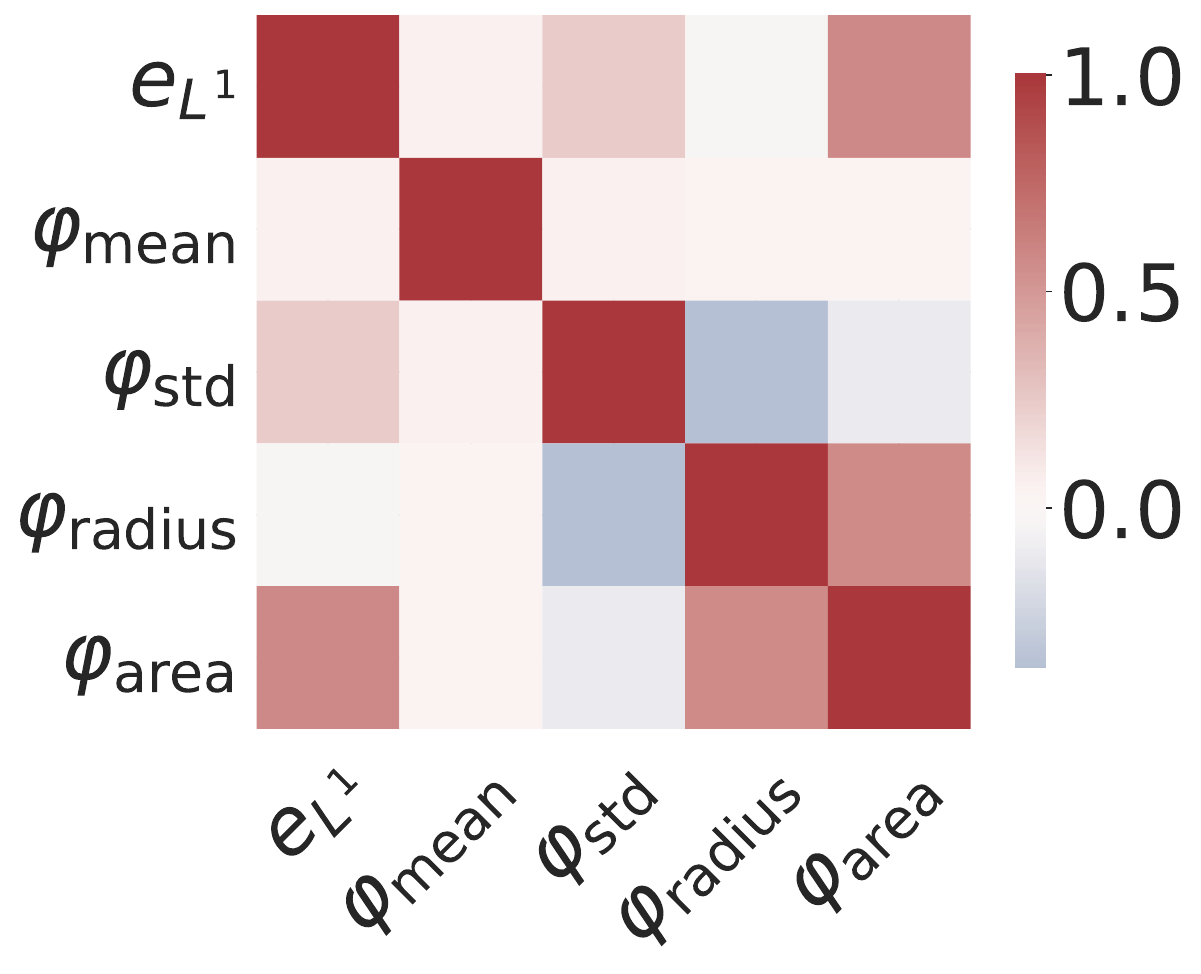}{}{}{0.3}{}{ref}{inner sep=0pt}
		\end{tikzpicture}
	}
	\subfloat[\FNOL]{
		\begin{tikzpicture}
			\tikzPlotAdvanced{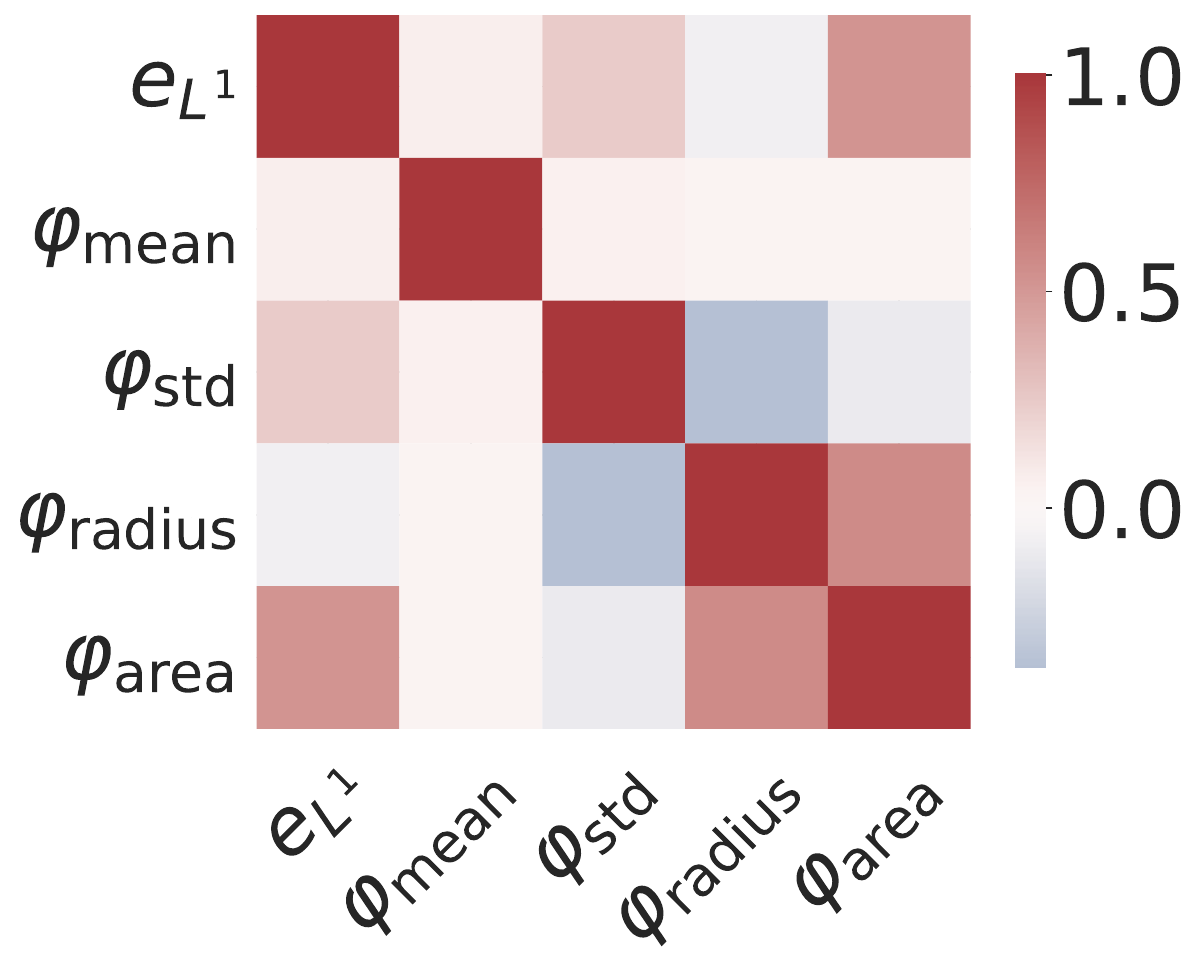}{}{}{0.3}{}{ref}{inner sep=0pt}
		\end{tikzpicture}
	}
	\subfloat[\FNOConstr]{
		\begin{tikzpicture}
			\tikzPlotAdvanced{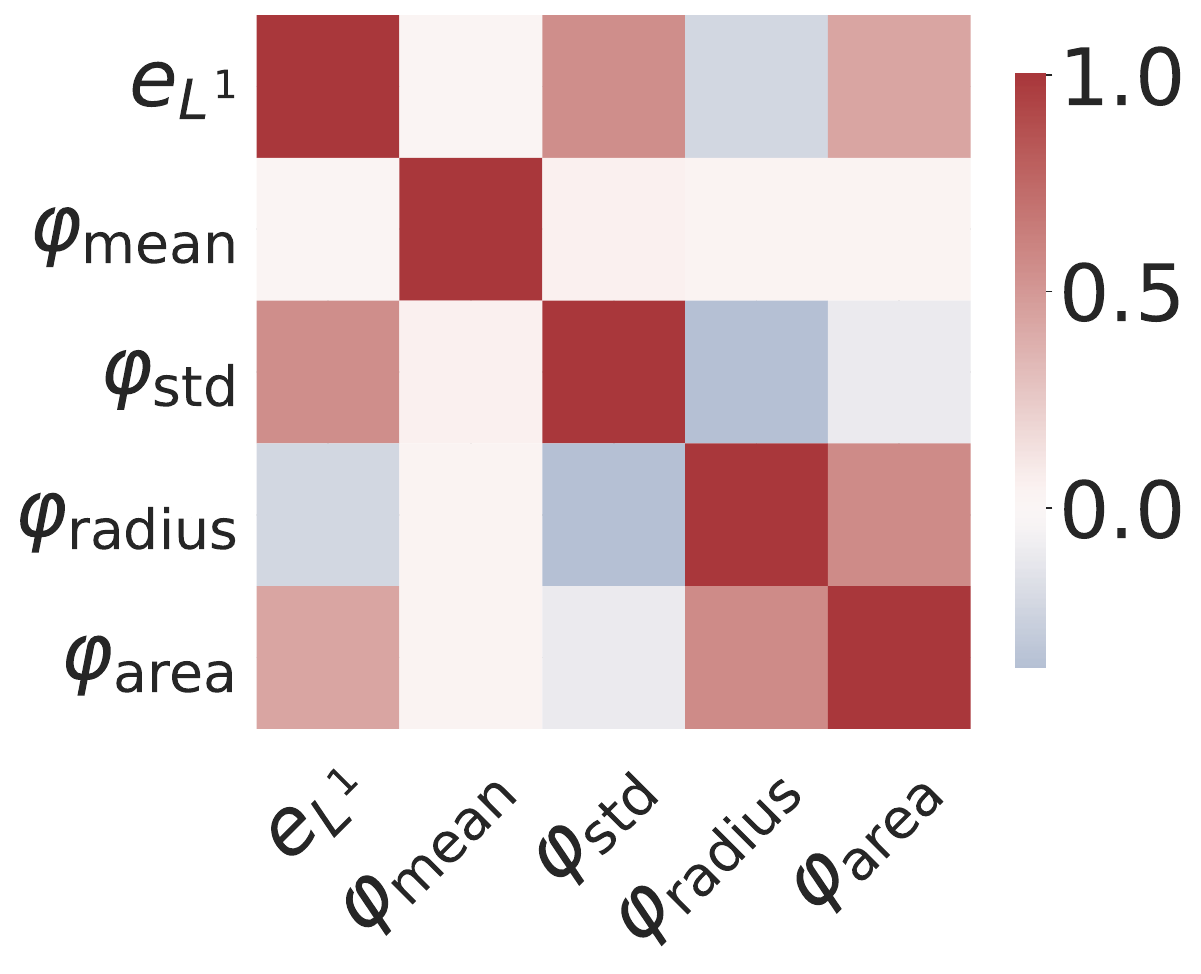}{}{}{0.3}{}{ref}{inner sep=0pt}
		\end{tikzpicture}
	}
	\vspace{-1em}
	
	\subfloat[3D-WaveY-Net]{
		\begin{tikzpicture}
			\tikzPlotAdvanced{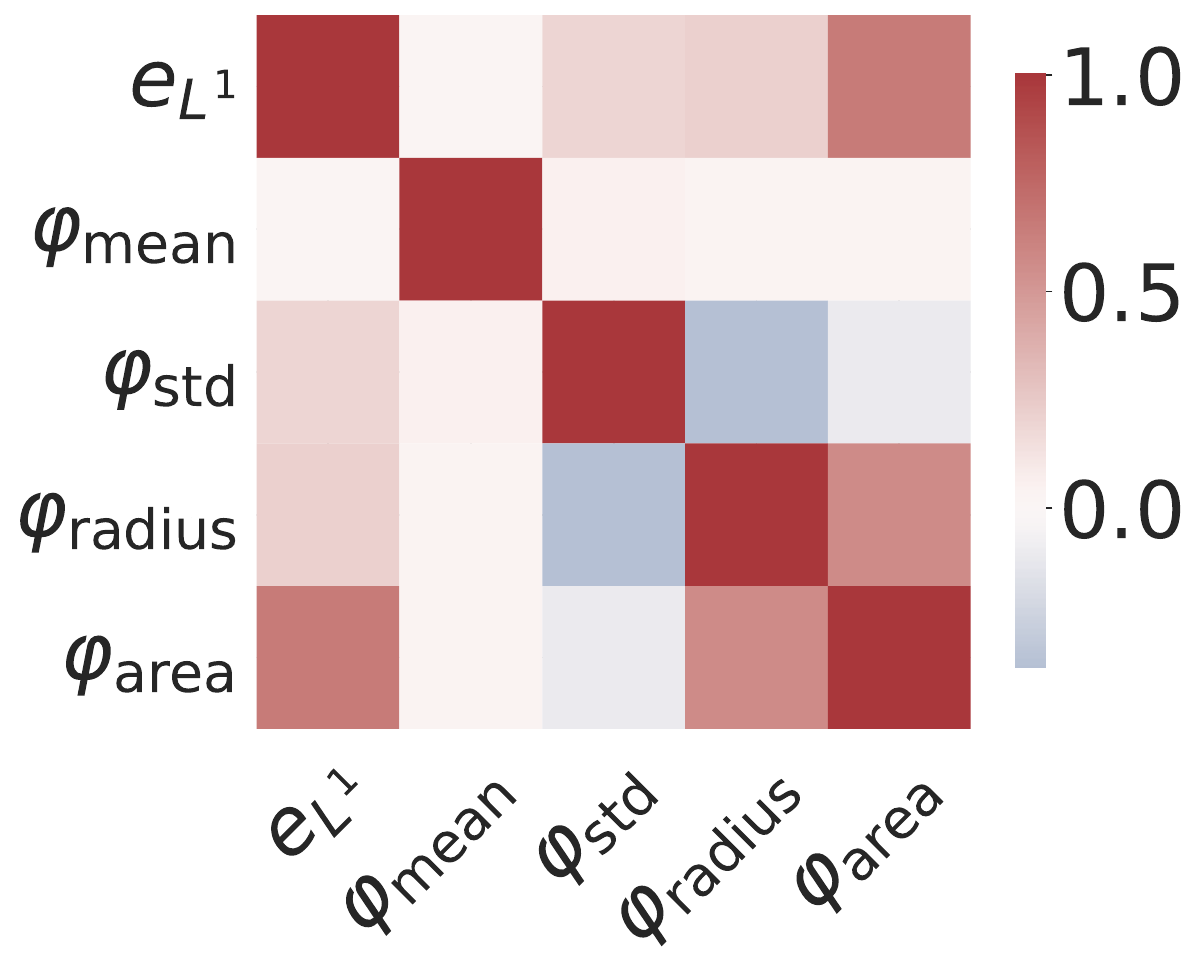}{}{}{0.3}{}{ref}{inner sep=0pt}
		\end{tikzpicture}
	}
	\subfloat[\WaveYConstr]{
		\begin{tikzpicture}
			\tikzPlotAdvanced{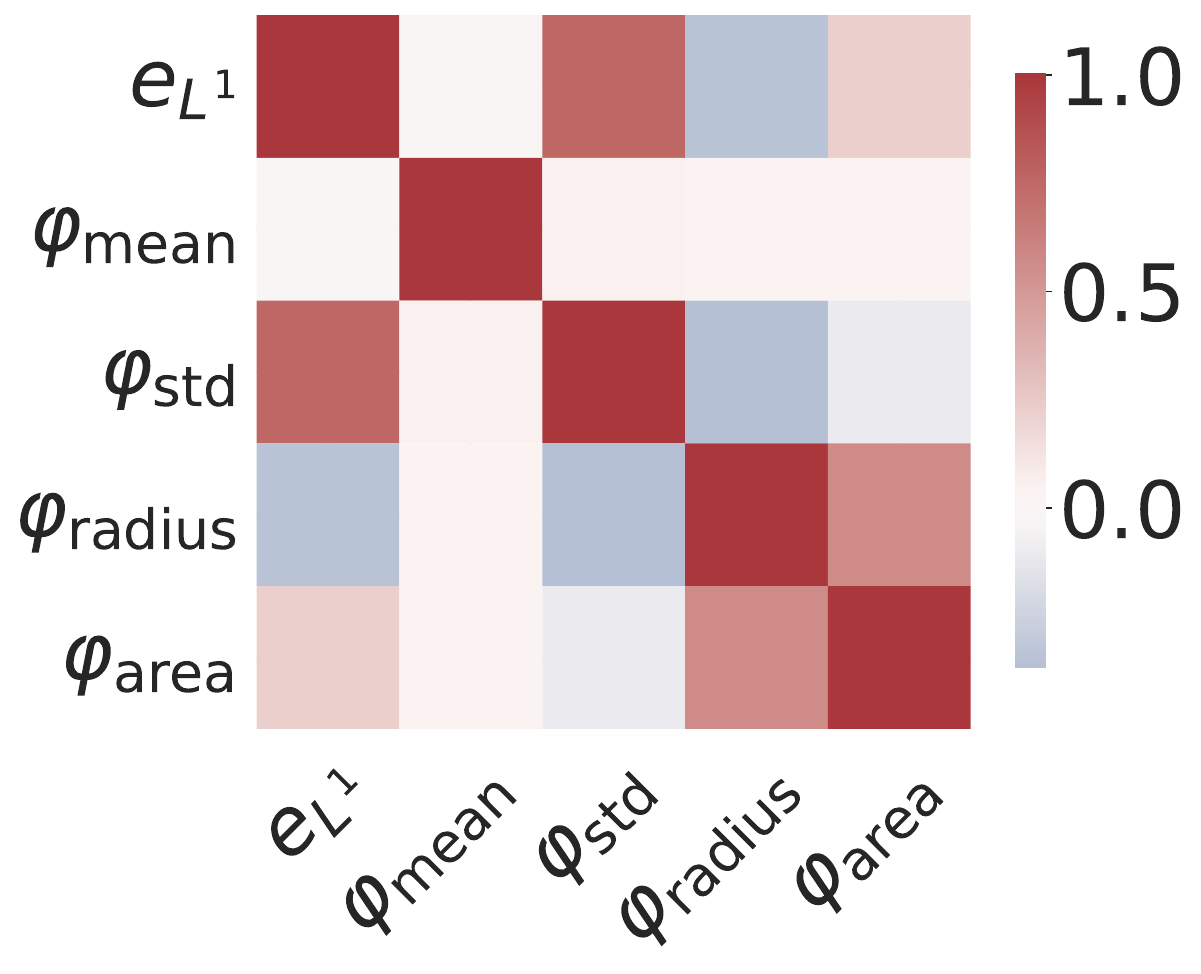}{}{}{0.3}{}{ref}{inner sep=0pt}
		\end{tikzpicture}
	}	
	\caption{\textbf{Correlations between metasurface descriptors and error metrics.}
		Pearson correlation coefficients computed on the test set for (a) FNO, (b) \FNOL, (c) \FNOConstr, (d) 3D-WaveY-Net and (e) \WaveYConstr.
	}
	\label{fig:correlation}
\end{figure}

\begin{figure}[H]
	\centering
	\subfloat[FNO]{
		\begin{tikzpicture}
			\tikzPlotAdvanced{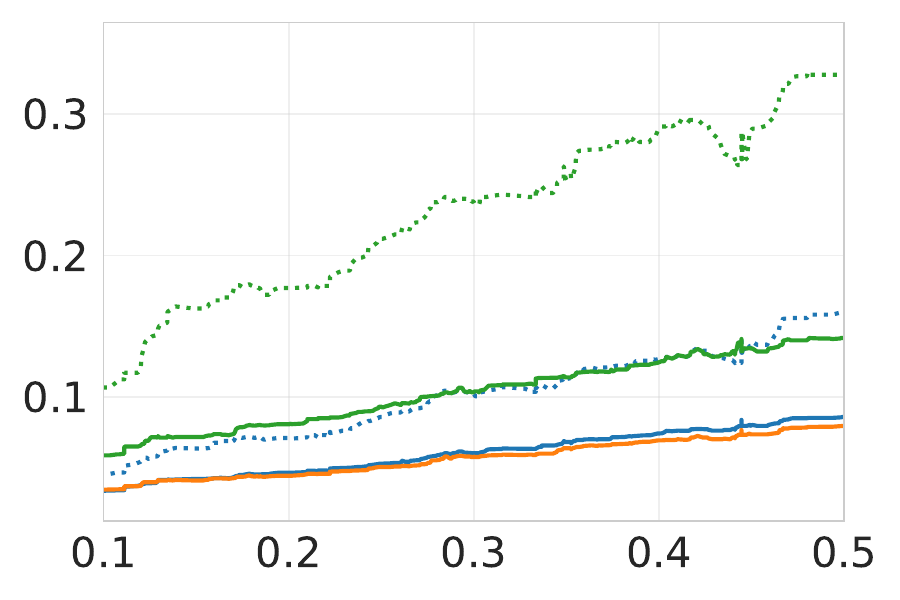}{$\varphi_\mathrm{area}$}{$e_{L^1}$}{0.4}{}{ref}{inner sep=0pt}
		\end{tikzpicture}
	}
	\subfloat[\FNOL]{
		\begin{tikzpicture}
			\tikzPlotAdvanced{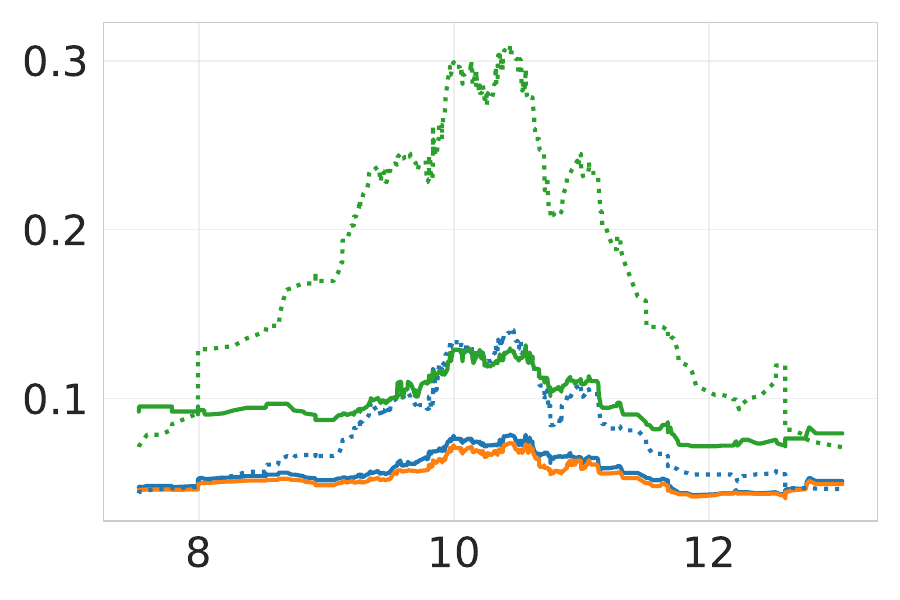}{$\varphi_\mathrm{mean}$}{$e_{L^1}$}{0.4}{}{ref}{inner sep=0pt}
		\end{tikzpicture}
	}
	
	\begin{tikzpicture}
		\node[inner sep=0pt]{\includegraphics[width=0.5\textwidth]{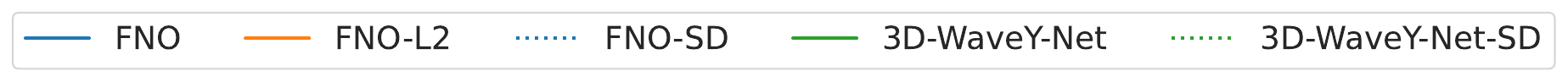}};
	\end{tikzpicture}
	\vspace{-1em}
	
	\subfloat[\FNOL]{
		\begin{tikzpicture}
			\tikzPlotAdvanced{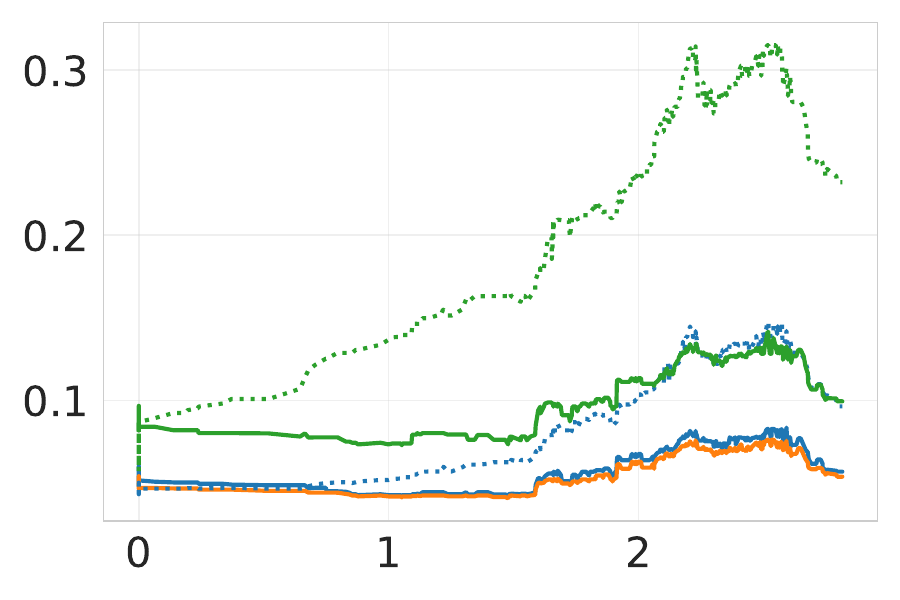}{$\varphi_\mathrm{std}$}{$e_{L^1}$}{0.4}{}{ref}{inner sep=0pt}
		\end{tikzpicture}
	}	
	\subfloat[\FNOConstr]{
		\begin{tikzpicture}
			\tikzPlotAdvanced{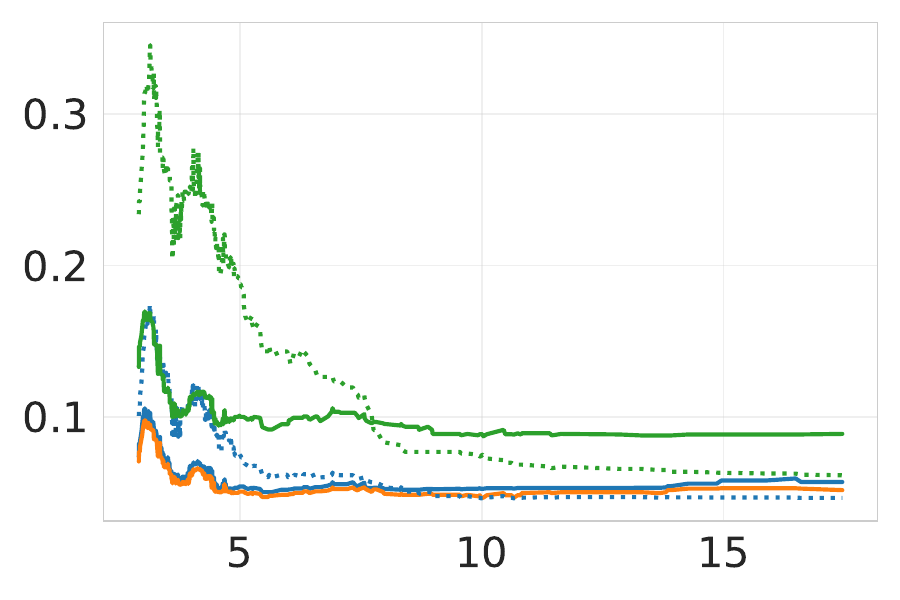}{$\varphi_\mathrm{radius}$}{$e_{L^1}$}{0.4}{}{ref}{inner sep=0pt}
		\end{tikzpicture}
	}

	\caption{
		\textbf{Dependence of prediction error on metasurface descriptors.} Dependence of prediction error on metasurface descriptors. Moving-median curves showing the relative $L^1$ error $e_{L^1}$ as a function of (a) area fraction $\varphi_{\mathrm{area}}$, (a) mean relative permittivity $\varphi_{\mathrm{mean}}$, (c) standard deviation $\varphi_{\mathrm{std}}$, and (d) mean radius $\varphi_{\mathrm{radius}}$ for the surrogate models. For each subfigure, the considered descriptor and the $e_{L^1}$  have been determined on the test dataset.  A moving median with a window size of 50 samples was then applied to the sorted error values to obtain smooth trends.
	}
	\label{fig:correlation:dependence}
\end{figure}

To investigate the (linear) relationship between metasurface descriptors and the error metric $e_{L^1}$, Pearson correlation coefficients were computed on the test set, see Fig.~\ref{fig:correlation}. To complement this analysis, a visual impression on the influence of  $\varphi_{\mathrm{area}}, \varphi_{\mathrm{std}}$ and $\varphi_{\mathrm{radius}}$ on $e_{L^1}$ is given in Fig.~\ref{fig:correlation:dependence}. 
Across the considered surrogate models, the correlations between the descriptors and $e_{L^1}$ reveal several trends. Overall, the mean relative permittivity $\varphi_{\mathrm{mean}}$ shows only weak linear correlations with the error. 
This observation is further reflected in Fig.~\ref{fig:correlation:dependence}b, where the dependence of $e_{L^1}$ on $\varphi_{\mathrm{mean}}$ exhibits a clearly non-linear behavior. While no strong linear trend is present, the error shows a noticeable increase for intermediate values of the mean relative permittivity and is most pronounced for \WaveYConstr followed by 3D-WaveY-Net. 
In contrast, the empirical standard deviation $\varphi_{\mathrm{std}}$ exhibits a stronger linear relationship, particularly for the {\FNOConstr} model (Fig.~\ref{fig:correlation}c). Here, higher values of $\varphi_{\mathrm{std}}$ are associated with higher errors. This behavior is expected: {\FNOConstr} was trained solely on the freeform-only scenario, for which $\varphi_{\mathrm{std}} = 0$ holds, because the permittivity within each mask is spatially constant. Consequently, a metasurface that exhibits large variations of relative permittivity values, as it is the case for the other structural scenarios, corresponds to a ``unseen'' example for \FNOConstr, leading to a lower predictive performance, see also Fig.~\ref{fig:correlation:dependence}c. Similar trends can be observed for {\WaveYConstr} which has been trained on the same dataset as \FNOConstr, see Figs.~\ref{fig:correlation}e and \ref{fig:correlation:dependence}c.

Although {\FNOConstr} and 3D-WaveY-Net exhibited similar mean and median error values in Fig.~\ref{fig:quantitative}, their behavior with respect to $\varphi_{\mathrm{std}}$ differs substantially. 
When comparing Figs.~\ref{fig:correlation}c and d, we observe that $\varphi_{\mathrm{std}}$ has a significantly more pronounced influence on the errors of {\FNOConstr} than on those of 3D-WaveY-Net. This could mean that the freeform-only scenario lacks samples with sufficient variability in relative permittivity to train surrogate models that generalize well to metasurfaces exhibiting strong  heterogeneity in permittivity.

Beyond the influence of $\varphi_{\mathrm{std}}$, several additional trends can be observed in Figs.~\ref{fig:correlation} and \ref{fig:correlation:dependence}. The area fraction $\varphi_{\mathrm{area}}$ shows a positive correlation with $e_{L^1}$ for all surrogate models. This effect is most pronounced for 3D-WaveY-Net, suggesting that metasurfaces with large material coverage on the SiO$_2$ film induce more complex EM interactions the prediction of which can become more difficult for surrogate models, see also Fig.~\ref{fig:correlation:dependence}a.

At first glance, the mean radius  $\varphi_{\mathrm{radius}}$ seems to have rather different influences on error metrics.
Fig.~\ref{fig:correlation} indicates a negative correlation between $\varphi_{\mathrm{radius}}$ and error metrics for predictions determined with FNO, \FNOL, {\FNOConstr} and \WaveYConstr, whereas the correlation coefficients are slightly positive for 3D-WaveY-Net. The trends observed for FNO, \FNOL,  {\FNOConstr} and {\WaveYConstr} are confirmed by Fig.~\ref{fig:correlation:dependence}d. For these three surrogate models, the error decreases as $\varphi_{\mathrm{radius}}$ increases, indicating that EM fields are easier to predict for metasurfaces with coarser structures. In contrast, 
the curve in Fig.~\ref{fig:correlation:dependence}d associated with 3D-WaveY-Net remains relatively flat and even slightly increases, showing that the 3D-WaveY-Net model does not seem to benefit from coarser structures to the same extent. The reason for this behavior is not yet clear and will require further investigation to be fully understood.

Overall, the models FNO and \FNOL{} seem to be more robust when processing metasurfaces with large heterogeneities and geometric complexity than conventional CNN-based architectures, whereas the {\FNOConstr} model that has been trained on limited data  seems to exhibit a similar performance as 3D-WaveY-Net.
Finally, we point out that the analysis performed in this section can offer guidance for improving future surrogate models: by examining how performance depends on specific structural descriptors, one can strategically adjust or enrich the training data, for example, by including samples that better represent scenarios in which the surrogate models currently perform comparatively poorly.

\subsection{Runtime analysis}\label{sec:results:runtime}
To assess the computational efficiency of the surrogate models, we compare their inference times after training with the time required for performing  FDTD simulations with the Python package \texttt{fdtdx} \cite{mahlau2024flexibleframeworklargescalefdtd}. To achieve this, we measure the time to process a set of 100 3D distributions $\epsVol$ of permittivity  
with each considered method, to compute average runtimes per 3D distribution, see Table~\ref{tab:runtime}. 
Each call of the surrogate models is performed with a batch size of $1$, ensuring that inference times are comparable with simulation times for performing FDTD.  Note that we omit the {\FNOConstr} and {\WaveYConstr} models in the runtime analysis, as they have the same architectures as FNO and 3D-WaveY-Net, respectively. 
All computations were performed on an HPC cluster equipped with a NVIDIA~H100 SXM5 (80\,GB) GPU and two AMD EPYC~9454 48-core processors running at 3.79\,GHz.

\begin{table}[ht]
	\centering
	\caption{
		Average inference/runtime for processing 3D distributions of permittivity 
		(batch size of 1) on an HPC node equipped with a  NVIDIA H100 SXM5 (80\,GB) GPU 
		and two AMD EPYC 9454 48-core CPUs. In addition, the number of floating point operations per second of surrogate models is provided.}
	\label{tab:runtime}
	\begin{tabular}{lccc}
		\toprule
		\textbf{Method} & 
		\textbf{Mean Runtime} & 
		\textbf{Speedup vs.\ FDTDX} &
		\textbf{TFLOPS}	 \\
		\midrule
		\texttt{fdtdx}  & 2.5877\,s  & $1\times$ & - \\
		FNO & 0.03829\,s & 67$\times$ & 0.1696\\
		3D-WaveY-Net &  0.0123 & 210$\times$ & 0.1205 \\
		\bottomrule
	\end{tabular}
\end{table}

The results reveal significant differences in runtimes between the numerical FDTD solver and the surrogate models. As expected, the FDTD solver \texttt{fdtdx} has the longest runtimes, as it must numerically solve Maxwell's equations on a fine spatial and temporal grid. In contrast, both surrogate models offer significant speedups, with 3D-WaveY-Net achieving the highest acceleration with a speedup factor of 210. FNO provides a speedup with a factor of 67. Although FNO is slower than 3D-WaveY-Net in terms of inference time, it still offers a dramatic reduction in computational cost compared to the full-wave solver. 
As a more objective, hardware-independent measure, Table~\ref{tab:runtime} additionally reports the number of floating-point operations per second (FLOPS), measured in ``TerraFLOPS'' (TFLOPS), during inference for both models.
The TFLOPS values in Table~\ref{tab:runtime} are comparable for both surrogate models, with FNO requiring $0.17$\,TFLOPS and 3D-WaveY-Net $0.12$\,TFLOPS during inference.

When the runtime results are interpreted alongside the  performance metrics reported in Section~\ref{sec:quantitative}, a trade-off emerges: 3D-WaveY-Net is computationally more efficient but delivers lower predictive performance, whereas FNO provides  more accurate predictions at the cost of moderately increased inference times. This balance between performance and speed suggests that the choice of surrogate model may depend on  intended applications.

\subsection{Super-resolving using neural operators}\label{sec:results:sr}
Recall that the FNO model is a neural operator that is a mapping between infinite dimensional function spaces. In contrast, conventional CNNs are  mappings between finite dimensional vector spaces, e.g., in the present paper 3D-WaveY-Net maps discretized 3D images onto discretized 3D images. After training, CNNs are typically tied to the resolution of the training data. Consequently, a model like 3D-WaveY-Net cannot be directly applied to 3D input images sampled on a finer spatial grid.

\begin{figure}[h]
	\centering
	\subfloat[]{
		\begin{tikzpicture}
			\tikzPlotAdvanced{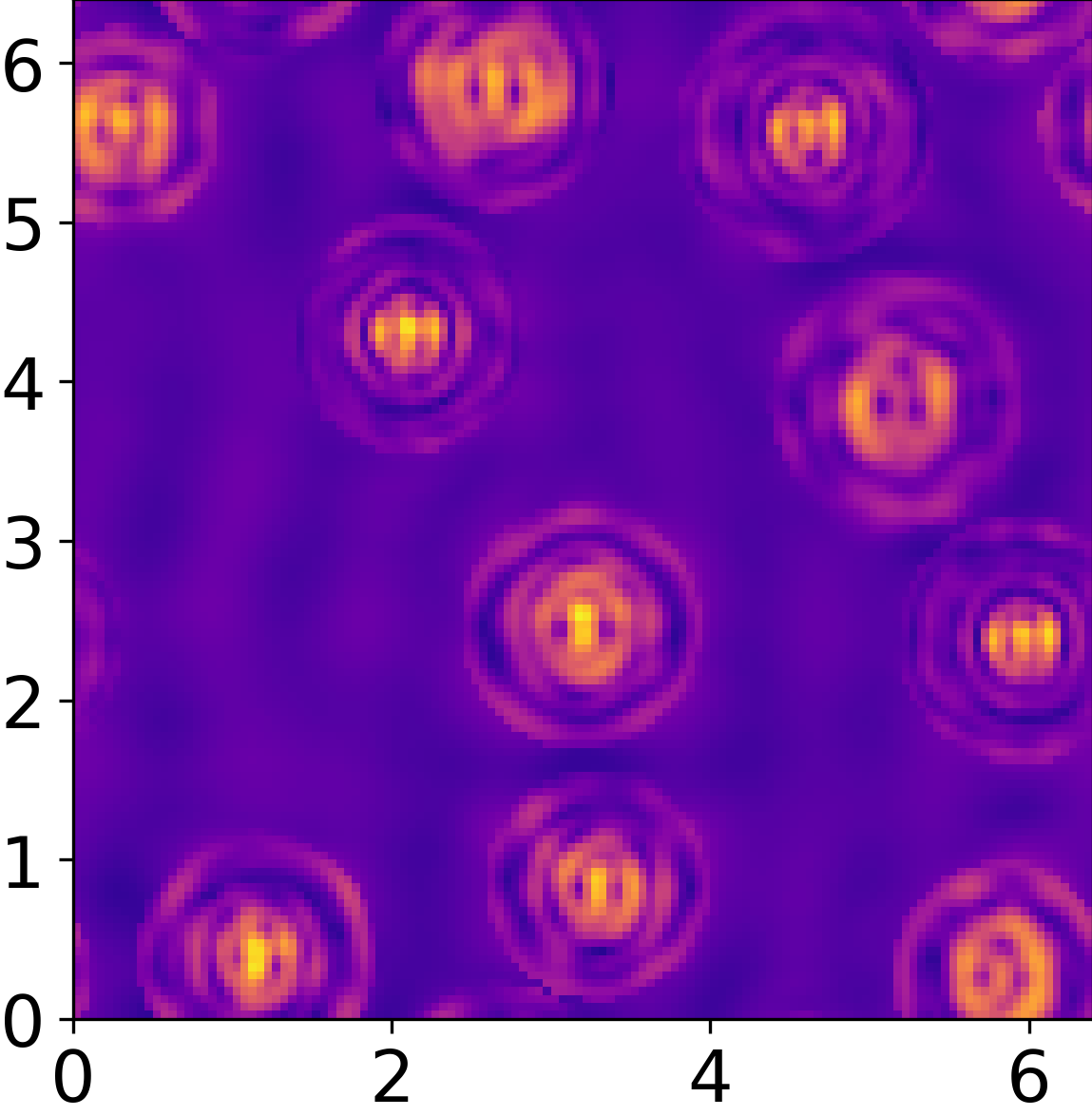}{\large $x$ in [\si{\micro\meter}]}{\large $y$ in [\si{\micro\meter}]}{0.4}{a}{ref}{inner sep=0pt}
		\end{tikzpicture}
	}
	\subfloat[]{
		\begin{tikzpicture}
			\tikzPlotAdvanced{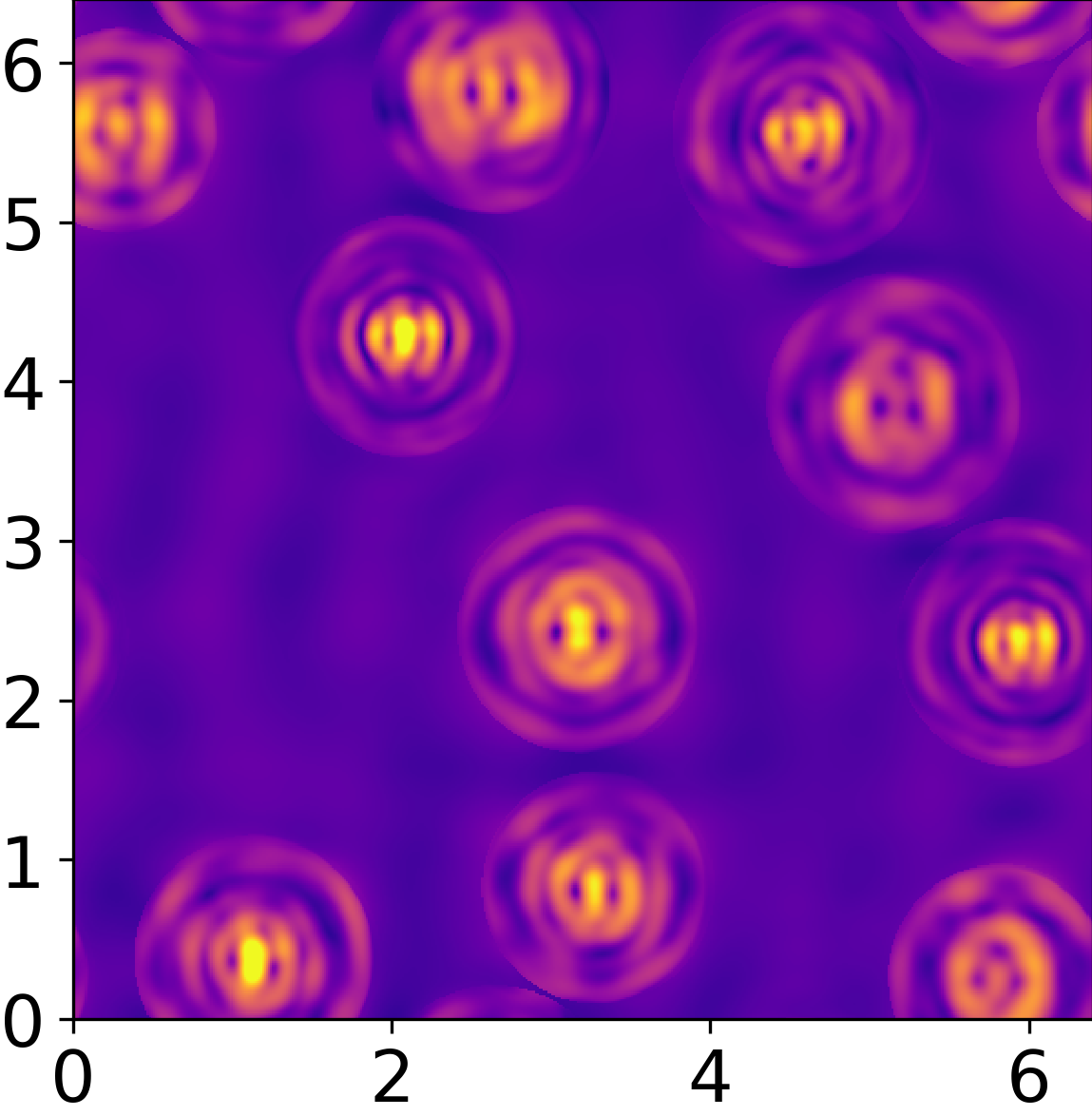}{\large $x$ in [\si{\micro\meter}]}{\large $y$ in [\si{\micro\meter}]}{0.4}{a}{ref}{inner sep=0pt}
		\end{tikzpicture}
	}
	\raisebox{1cm}{
		\begin{tikzpicture}

			\node[inner sep=0pt ] (colorbar) {
				\includegraphics[width=1.5cm]{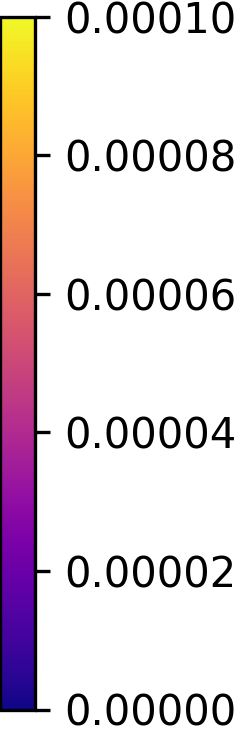}
			};
			\node[right=0.2cm of colorbar.east, anchor=west , inner sep=0pt,rotate=90 ] {
				\large	$|\HFieldPhasorDisreteRealAlt|_2$
			};
		\end{tikzpicture}
	}
	\caption{\textbf{Super-resolution.}  a: FNO prediction of the magnetic field amplitude at the original resolution used during training.  
		b: FNO prediction obtained by evaluating the same trained model on a permittivity field re-discretized at four times the resolution in each spatial dimension.  }
	\label{fig:fno_superresolution}
\end{figure}

Neural operators such as FNO have the advantage that, despite being trained on discretized samples, the  mapping learned by the neural operator can be evaluated on arbitrary discretizations. In our setting, this means that once the FNO model has learned a mapping from training data consisting of  relative permittivity  and magnetic fields on the grid 
$\windowVolumeDiscrete$, it can be applied to distributions of permittivity that are discretized with a finer resolution without any retraining. 

This resolution independence follows from the definition of a Fourier layer in Eq.~\eqref{eq:layer}.  
For discrete single-channel inputs $v=(v_{1,i,j,k})_{i,j,k=1}^{n_\mathrm{x},n_\mathrm{y},n_\mathrm{z}}
\in \mathbb{R}^{1\times n_\mathrm{x}\times n_\mathrm{y}\times n_\mathrm{z}}$, obtained by sampling the
3D distribution of permittivity via $v_{1,i,j,k}=\varepsilon\!\bigl((i-1)\Delta x,(j-1)\Delta y,(k-1)\Delta z\bigr)$,
the operator $\mathcal{L}$ applies an FFT to $v$, multiplies the resulting Fourier coefficients by the  
learned spectral kernel $\mathcal{R}_\theta$ on a fixed set of low-frequency modes. After some further operations in Fourier space an inverse FFT is applied.  
If the grid resolution is increased, the FFT is simply evaluated on a larger discrete frequency grid, but the
low-frequency modes of $\mathcal{R}_\theta$ remain unchanged.  
Thus, the multiplication in Fourier space with  $\mathcal{R}_\theta$ can still be performed for higher-resolved inputs.
As a result, the layer $\mathcal{L}$, and hence the full FNO, can be evaluated on inputs of higher spatial resolution
without any modification of the learned parameters.

To illustrate this, Fig.~\ref{fig:fno_superresolution} compares the FNO prediction at the  resolution of the training data with its prediction on a grid refined by a factor of four in each spatial dimension. For the refined case, the 3D distribution 
$\epsVol$ of relative permittivities was discretized on the finer grid and passed directly through the trained FNO. As shown in Fig.~\ref{fig:fno_superresolution} the resulting ``super-resolved'' field remains visually consistent with the corresponding prediction at the original resolution.
This capability highlights an additional advantage of neural operator architectures for learning differential operators, in particular, when flexibility in spatial discretization is required.

\section{Conclusions}\label{sec:conclusions}
We presented a framework for training neural operators in a physics-informed manner to predict electromagnetic fields transformed by three-dimensional metasurface geometries. The presented approach learns a continuous  mapping from 3D distributions of permittivity to the corresponding electromagnetic field solutions. Using a dataset of only $\approx 5000$ simulated 3D distributions and associated fields, we demonstrated that neural operators can generalize effectively across a broad family of  metasurfaces with irregularly-shaped nanopillars, without requiring simplifying assumptions on metasurface geometries.

Quantitative performance analysis indicates that the proposed surrogate model achieves high predictive accuracy across diverse geometries while offering substantial reductions in computational cost compared to full-wave solvers. This makes neural operators suitable for integration into iterative design workflows. Importantly, even the \FNOConstr{} model, which has been trained with only 1000 examples, exhibited reasonably good performance. This highlights the data efficiency gained by deploying neural operators as surrogate models for the prediction of electromagnetic fields.
Furthermore, we demonstrated that the neural operator supports resolution-independent inference, enabling super-resolved electromagnetic field predictions on spatial grids finer than those used in training data.

Moreover, we have observed that the methods from stochastic geometry that have been deployed for the 
generation of 3D metasurfaces provide a promising source of training data. In particular, this is 
evidenced by the \FNOConstr{} model, which generalizes rather effectively in all considered structural scenarios,  despite being trained exclusively on synthetically generated samples associated with one specific structural scenario.

We further observed that enforcing the Maxwell equations using an $L^2$ norm for the physics-based Maxwell loss yields slightly improved predictive performance compared to an $L^1$ formulation of the Maxwell loss, indicating that squared residuals could provide a more effective constraint for training surrogate models to predict EM fields transformed by metasurfaces.

Overall, the results indicate the potential of combining spatial stochastic modeling with  neural operators to derive scalable, data-efficient and physically consistent surrogate models for 3D  simulations. Future work may extend this framework to broadband simulations, allowing the surrogate models to predict electromagnetic fields consistently across a broad range of wavelengths. Furthermore, we will exploit the differentiability of the neural operator for gradient-based inverse design of large-area metasurfaces.

\newpage
\appendix

\titleformat{\section}
{\normalfont\Large\bfseries}
{Appendix \thesection}
{1em}{}

\renewcommand{\thesection}{\Alph{section}}
\renewcommand{\thesubsection}{\thesection.\arabic{subsection}}

\section{Generation of metasurfaces}\label{appendix:generation:metasurfaces}
In this section, we describe the stochastic 3D models that are deployed for generating a large database of metasurfaces that will be used as geometry input for numerical simulations of EM waves.
To facilitate simulations with periodic boundary conditions in $x-y$-directions, the generated geometries will be periodic as well. 
To simulate metasurfaces with the considered stochastic models the following steps are performed: (i) First, the 2D cross section of nanopillars is generated, (ii) then, the cross sections are cylindrically extended in 3D and placed on a virtual substrate. In the following, we describe three different stochastic models for the cross-sections of nanopillars.

\subsection{Packings of disks and squares}\label{appendix:generation:discs}
To stochastically model metasurfaces with relatively regular nanopillars, we consider cross section models that can generate packings of non-overlapping disks and squares.
Therefore, let $\windowPlane=[0,w_1) \times [0,w_2)$  be a rectangular observation window and consider the subset $B(\centroid,r,p)\subset \windowPlane$
which is given by 
\begin{equation}
	B(\centroid,r,p) = \{ \yvec \in \windowPlane \colon |\centroid- \yvec|_{p,\mathrm{per}} \leq 1 \},
\end{equation}
for $\centroid\in \windowPlane$, $r>0$ and $p\in [1,\infty]$, 
where $|\cdot|_{p,\mathrm{per}}$ denotes the periodic $L^p$ norm on $\windowPlane$. Particularly, for $p=2$ or $p=\infty$, the set $B(\centroid,r,p)$ is a disk with radius $r$ or a square with side length $2r$; both of which are (periodically) centered at $\centroid$. Consequently, a packing of $n \in \mathbb{N}$ disks and squares can be effectively represented by the sets $\{(\centroid_i,r_i,p_i)\colon i=1,\dots,n\}\subset \windowPlane \times [0,\infty) \times \{2,\infty\}$. 

The pseudocode outlined in Algorithm~\ref{alg:training_procedure} allows for the generation of differently structured disk/square packings. To facilitate the generation cross-sections that exhibit both regularly shaped nanopillars (disks/squares) as well as irregularly shaped nanopillar geometries, the concept of a \emph{mask} $M\subset \windowPlane$ is introduced. This mask can represent arbitrary freeform shapes of nanopillars. In Algorithm~\ref{alg:training_procedure}, the mask effectively restricts the placement of objects to $\windowPlane\setminus M$, thereby allowing the generation of metasurfaces that combine regular and irregular shapes. Now we summarize the algorithm assuming that a mask $M$ is given. 
In Line 6, random disks/squares are generated by determining their random size $R_i$ by simulating a normal distribution with 
mean $r_\mathrm{m}$ and standard deviation $s r_\mathrm{m}$ that is truncated on the interval $(0,r_\mathrm{max})$ with $r_\mathrm{max}=\SI{3.2}{\micro\meter}$ to avoid unreasonable disk/square sizes. The scalar values $r_\mathrm{m},s>0$ are model parameters. To model whether the $i$-th object is a disk or a square, the $p$-value of the corresponding periodic norm is determined randomly in Line 7. More precisely, a random variable $P_i$ is set to 2 with probability $p_\mathrm{{disk}}$, or to $\infty$ with probability $1-p_\mathrm{{disk}}$, where $p_\mathrm{{disk}}$ is another model parameter. 

In Line 8, random relative permittivities $\epssymbol_i$ are assigned uniformly from the interval $[1,\epsMax]$ to each object, where $\epsMax=15$ is an upper bound for the relative permittivities. The generation of random disks/squares is stopped if their joint area (quantified by the 2D Lebesgue measure $\nu_2$) would exceed a preset area fraction $\alpha_{\mathrm{target}}$ of the available area $\nu_2(\windowPlane\setminus M)$.
Once the number of objects and their sizes have been determined, their centers are identified in Line 15, by minimizing the overlap area between pairs of disks/squares as well as the overlap  between disks/squares and the mask $M$.  Note that in Line 15, the overlap area as well as the minimization are performed numerically. In addition, we point out that it cannot be guaranteed that this approach leads to a non overlapping configuration of disks/squares/mask. Nevertheless, the numerical approach employed here yields sufficiently accurate and practically relevant results.

Finally, Algorithm~\ref{alg:training_procedure} utilizes the parameters of the generated objects to derive a 2D distribution $\epsPlane \colon \windowPlane \to [1,\epsMax]$ of relative permittivities. 
Note that due to computational limitations, in practice, we determine the values of $\epsPlane$ not for all possible points within the observation window $\windowPlane$. Instead, we consider it to be a 2D distribution $\epsPlane \colon \windowPlaneDiscrete \to [1,\epsilon_\mathrm{max}]$ on a regular grid $\windowPlaneDiscrete=\{0\cdot \rho, 1 \cdot \rho,\dots,(n_\mathrm{x}-1)\cdot \rho\} \times \{0\cdot \rho, 1 \cdot \rho,\dots,(n_\mathrm{y}-1)\cdot \rho\}\subset \windowPlane$, where $n_\mathrm{x},n_\mathrm{y}>0$ define the dimensions of the grid and $\rho>0$ denotes the pixel size. 
The side lengths $w_1,w_2$ of the (continuous) observation window $\windowPlane$ are set to $w_1=w_2=\SI{6.4}{\micro\meter}$. For discretizing $\windowPlane$, we choose $n_\mathrm{x}=n_\mathrm{y}=128$ and a pixel size of  $\rho=\SI{50}{\nano\meter}$. Consequently, the grid $\windowPlaneDiscrete$ is given by $\windowPlaneDiscrete= \{\SI{0}{nm},\SI{50}{nm}, \dots, \SI{6.35}{\micro \meter}\}^2$, i.e., it ``homogeneously'' discretizes the continuous window $\windowPlane$.

\begin{algorithm}[H]
	\caption{
		Approach for generating packings of disks and squares (\textbf{Inputs:}
		$M\subset \windowPlane$: mask of ``obstacles'' for disks/spheres; $\epssymbol_M \in [1,\epsMax]$ : relative permittivity of the mask;
		$\alpha_\mathrm{target} \in [0,1]$: target area fraction in $\windowPlane\setminus M$;
		$p_\mathrm{disk} \in [0,1]$: relative frequency of disks;
		$r_\mathrm{m}>0$: mean size of disks/squares;
		$s \geq 0$: parameter controlling the variance of sizes. \textbf{Output:} $\epsPlane \colon \windowPlane \to [1,\epsMax]$: 2D distribution of relative permittivities.)}
	\label{alg:training_procedure}
	\begin{algorithmic}[1]
		\Procedure{Generate2DPermittivityField}{$M, \epssymbol_M, \alpha_\mathrm{target},p_\mathrm{disk},r_\mathrm{m},s$}
		\State $i \gets 0$ \Comment{Object index}
		\State $A_\mathrm{total} \gets 0$
		\While{True} \  \
		\State $i \gets i +1$
		\State Sample $R_i \sim \mathcal{N}(r_m, (s\,r_m)^2)$ truncated to $(0, r_{\max})$ \Comment{Generate object sizes}
		\State Set $P_i=2$ with probability $p_\mathrm{disk}$, else set $P_i=\infty$ \Comment{Determine object type}
		\State Sample $\epssymbol_i \sim \mathcal{U}(1,\epsMax)$
		\Comment{Determine relative permittivity of object}
		\State $A_\mathrm{total} \gets A_\mathrm{total} +  \nu_2(B(o,R_i,P_i))$ \Comment{Count area of all objects}
		\If{$A_\mathrm{total}  > \alpha_{\mathrm{target}} \cdot \nu_2(\windowPlane \setminus M)$}
		\State $n \gets i-1$ \Comment{Number of objects}
		\State \textbf{break}
		\EndIf
		\EndWhile
		
		\State $(\Centroid_1,\dots, \Centroid_n) = \argmin_{\centroid_1,\dots,\centroid_n \in \windowPlane}
		\sum_{i=1}^n \nu_2(B(\centroid_i,R_i,P_i) \cap M )+
		\sum_{i,j=1; i\neq j}^n \nu_2(B(\centroid_i,R_i,P_i) \cap B(\centroid_j,R_j,P_j) )$
		\Comment{Determine object centers that minimize overlap}
		\State Set $
		\tilde{\epssymbol}_\mathrm{r}^\mathrm{2D}(\rvec) = \epssymbol_M \mathbbm{1}_M(\rvec) + 
		\sum_{i=1}^n \epssymbol_i \mathbbm{1}_{B(\Centroid_i,R_i,P_i)}(\rvec) \
		$ \Comment{Relative permittivities for points $\rvec$ located within mask or objects}
		\State Set $\epsPlane(\rvec) = \min(\max(1,\tilde{\epssymbol}^\mathrm{2D}_\mathrm{r}(\rvec)),\epsMax)$
		\Comment{Set relative permittivities for points $\rvec$ outside mask/objects to 1 and apply upper bound}
		
		\State \Return $\epsPlane$
		\EndProcedure
	\end{algorithmic}
\end{algorithm}

\subsection{Excursion set masks of Gaussian random fields}\label{appendix:generation:excursion}
As outlined above, the mask $ M \subset \windowPlane $ can be deployed to model the cross section of irregularly shaped nanopillars. In this paper, we model the masks $M$ stochastically as excursion sets of Gaussian random fields (GRFs) which are parameterized by their covariance functions \cite{adler2007random}. More precisely, let \( \{ Z(\yvec) : \yvec \in \windowPlane \} \) denote a stationary Gaussian random field on the rectangular window $\windowPlane$ with normalized marginals, i.e., $Z(\yvec)\sim\mathcal{N}(0,1)$. The mask \( M \) is an excursion set that is defined as the set of points for which the field exceeds a threshold level,
\begin{equation}
	M = \{ \yvec \in \windowPlane \colon Z(\yvec) \geq \Phi^{-1}(1-\alpha_\mathrm{mask}) \}.
\end{equation}
where $\Phi$ denotes the cumulative distribution function of the standard normal distribution and $\alpha_{\mathrm{mask}} \in [0,1]$. Note that the model parameter $\alpha_{\mathrm{mask}}$ controls the expected area fraction of $M$, i.e.,
we have $\mathbb{E}\left[ \frac{\nu_2(M)}{\nu_2(\windowPlane)} \right] = \alpha_\mathrm{mask}$. To fully describe the underlying GRF it suffices to define its underlying covariance function.
In this study, we consider a (periodic) exponential covariance function of the form
\begin{equation}\label{eq:cov}
	C(\xvec,\yvec) =  \exp\!\left(-\frac{\|\xvec-\yvec\|^2_{2,\mathrm{per}}}{h_\mathrm{GRF}^2 }\right),
\end{equation}
for $\xvec,\yvec\in \windowPlane$,
where the model parameter $h_\mathrm{GRF} > 0$ controls the correlation length  of the GRF $Z$ and consequently the typical feature size of the generated mask $M$. Note that the periodic norm deployed in Eq.~\eqref{eq:cov} ensures periodicity of the resulting field $Z$ and the generated mask $M$. This will facilitate the deployment of periodic boundary conditions in subsequent numerical simulations of electromagnetic fields that solve Maxwell's equations. The periodic norm offers further advantages. For example, discretized realizations of the random field can be generated relatively efficiently using fast Fourier transformations. More specifically, we use the approach from \cite{LangPotthoff} to generate realizations, which inherently leads to periodic realizations of random fields on a bounded grid.

\subsection{Database of metasurfaces}\label{appendix:generation:database}
The packing method in Algorithm~\ref{alg:training_procedure} and the excursion set model $M$ can be deployed to create a database of discretized 3D distributions (i.e., 3D images) of relative permittivities each representing a metasurface. The 3D distributions are defined on the grid $\windowVolumeDiscrete=\{0,\dots,(n_\mathrm{x}-1)\rho\} \times \{0,\dots,(n_\mathrm{y}-1)\rho\} \times \{0,\dots,(n_\mathrm{z}-1)\rho\}$, where we set $n_\mathrm{z}=64$.

For a given vector $(\alpha_{\mathrm{mask}}, h_\mathrm{GRF},\epssymbol_M,\alpha_\mathrm{target},p_\mathrm{disk},r_\mathrm{m},s)$, a 3D distribution of relative permittivities is simulated as follows. First, using the parameters $\alpha_{\mathrm{mask}}, h_\mathrm{GRF}$, a mask $M$ is simulated as an excursion set of a GRF as described above. Then, the mask $M$ and the remaining parameters  $\epssymbol_M,\alpha_\mathrm{target},p_\mathrm{disk},r_\mathrm{m},s,\epsMax$ serve as input for Algorithm~\ref{alg:training_procedure} to generate a 2D distribution $\epsPlane \colon \windowPlaneDiscrete \to [1, \epsMax]$ of relative permittivities that corresponds to a planar section through a metasurface. Then, a 3D distribution 
$\epsVol \colon \windowVolumeDiscrete \to [1, \epsMax]$ is constructed by
\begin{equation}\label{eq:epsVolumeAppendix}
	\epsVol(x,y,z) =
	\left\{
	\begin{array}{ll}
		1, & z< z_\mathrm{substrate},\\
		\epssymbol_\text{SiO$_2$}, & z_\mathrm{substrate}\leq z < z_\mathrm{substrate}+h_\mathrm{substrate},\\
		\epsPlane(x,y), & z_\mathrm{substrate}+h_\mathrm{substrate}\leq z < z_\mathrm{substrate}+h_\mathrm{substrate}+h_\mathrm{metasurface},\\
		1, & \text{else,}
	\end{array}
	\right.
\end{equation}
for each voxel $(x,y,z)\in \windowVolumeDiscrete$. The piecewise definition of $\epsVol$ in Eq.\eqref{eq:epsVolumeAppendix}, depending on the $z$-coordinate, is motivated as follows. At height $z_\mathrm{substrate}>0$ a SiO$_2$ film with height $h_\mathrm{substrate}$ is placed that has a constant relative permittivity of $\epssymbol_\text{SiO$_2$}=1.4585$. Directly, above the SiO$_2$ film, the metasurface is placed by assigning the planar section represented by $\epsPlane$ with a thickness of $h_\mathrm{metasurface}$. At all remaining positions the values of the 3D distribution $\epsVol$ are set to 1, which corresponds to vacuum being present at these positions. In the present paper, we set the heights to $h_\mathrm{substrate}=h_\mathrm{metasurface}=\SI{350}{\nano\meter}$ and the $z$-coordinate at which the substrate is placed to $z_\mathrm{substrate}=\SI{1050}{\nano\meter}$. 

By modifying the parameter vector $(\alpha_{\mathrm{mask}}, h_\mathrm{GRF},\epssymbol_M,\alpha_\mathrm{target},p_\mathrm{disk},r_\mathrm{m},s)$ of the model, differently structured 3D distributions of relative permittivities can be generated, which will facilitate training surrogate models. Table~\ref{tab:model_parameters_transposed} lists the model parameters considered in this study for generating metasurfaces $\epsVol$. More precisely, the parameters are selected to generate metasurfaces that we categorize into five classes, denoted by 
\ScenarioMask, \ScenarioDisk, \ScenarioSquare, \ScenarioDiskAndSquare, \ScenarioAll. For example, in the scenario \ScenarioMask, the area fraction $\alpha_\mathrm{target}$ of disks/squares is set to 0, making the parameters $p_\mathrm{disk},r_\mathrm{m},s$ that placed disks/squares redundant. The remaining parameters $\alpha_{\mathrm{mask}},h_\mathrm{GRF},\epssymbol_M$, that influence the mask, are chosen from sets with cardinality of 10, respectively, such that a total $10^3=1000$ distinct parameter vectors are considered in the scenario \ScenarioMask. Then, for each parameter vector one corresponding metasurface is generated, such that we obtain 1000 metasurfaces for the scenario \ScenarioMask.
By taking all scenarios into account, a total of 5296 metasurfaces $\varepsilon$ have been generated.

For example, in the scenario \ScenarioMask, the area fraction $\alpha_\mathrm{target}$ of disks/squares is set to 0, making the parameters $p_\mathrm{disk}$, $r_\mathrm{m}$ and $s$, which define the placement of disks/squares, redundant. The remaining parameters, $\alpha_{\mathrm{mask}}$, $h_\mathrm{GRF}$ and $\varepsilon_M$, which influence the mask, are each chosen from sets with a cardinality of 10. Consequently, a total of $10^3 = 1000$ distinct parameter vectors are considered in the scenario \ScenarioMask. For each parameter vector, one corresponding metasurface is generated, resulting in 1000 metasurfaces for the scenario \ScenarioMask.

Taking all scenarios into account, a total of 5296 metasurfaces $\epsVol$ were generated.

\begin{table}[htbp]
	\centering
	\caption{Overview of model parameters across five structural scenarios.}
	\label{tab:model_parameters_transposed}
	\small
	\setlength{\tabcolsep}{4.5pt}
	\renewcommand{\arraystretch}{1.12}
	
	\begin{tabularx}{\linewidth}{L*{5}{P}}
		\toprule
		\textbf{Parameter} &
		\ScenarioMask &
		\ScenarioDisk &
		\ScenarioSquare &
		\ScenarioDiskAndSquare &
		\ScenarioAll \\
		\midrule
		$\alpha_{\mathrm{mask}}$ &
		$\LS{0}{1}{10}$ & 0 & 0 & 0 & $\LS{0.1}{0.9}{10}$ \\
		$h_{\mathrm{GRF}}$ in $\nm$ &
		$\LS{100}{600}{10}$ & \NA & \NA & \NA & $\LS{100}{600}{10}$ \\
		$\varepsilon_M$ &
		$\LS{6.25}{15}{10}$ & \NA & \NA & \NA & $\mathcal{U}(1,\varepsilon_{\max})$ \\
		$\alpha_{\mathrm{target}}$ &
		0 & $\LS{0.1}{0.7}{10}$ & $\LS{0.1}{0.7}{10}$ & $\LS{0.1}{0.7}{6}$ & $\LS{0.1}{0.5}{10}$ \\
		$p_{\mathrm{disk}}$ &
		\NA & 1 & 0 & $\LS{0.1}{0.9}{6}$ & 0.5 \\
		$r_{\mathrm{m}}$ in $\nm$ &
		\NA & $\LS{350}{600}{10}$ & $\LS{350}{600}{10}$ & $\LS{350}{600}{6}$ & $500$ \\
		$s$ &
		\NA & $\LS{0}{0.2}{10}$ & $\LS{0}{0.2}{10}$ & $\LS{0}{0.2}{10}$ & 0.2 \\
		\makecell[l]{Number of\\scenarios} &
		1000 & 1000 & 1000 & 1296 & 1000 \\
		\bottomrule
	\end{tabularx}
	
	\vspace{2pt}
	\raggedright\footnotesize
	Notes: $\LS{a}{b}{n}$ denotes a set of $n$ equally spaced values in $[a,b]$, which is given by 
	$\LS{a}{b}{n}=\left\{a+\frac{i}{n-1}(b-a) \colon i=0,\dots,n-1\right\}$. In the scenario \ScenarioAll{} the relative permittivity $\epssymbol_M$ of the mask is generated at random in the interval $[1,\epsMax]$.
\end{table}

\section{Adaptive weight adaption}\label{appendix:weights}

The weight factor $\lambda_\mathrm{Maxwell}$ is initialized with  $\lambda_\mathrm{Maxwell}^{(0)}=0$.
Starting with epoch 21, after each epoch, we determine the ratio $\gamma$ between the data and the Maxwell loss given by
\begin{equation}
	\gamma=
	\frac{
		\frac{1}{n_\mathcal{X}} \sum_{(\epsVol,\HFieldPhasorDisreteRealAlt) \in \mathcal{X}}
		L_\mathrm{data}(f(\epsVol),\HFieldPhasorDisreteRealAlt)
	}{
		\frac{1}{n_\mathcal{X}} \sum_{(\epsVol,\HFieldPhasorDisreteRealAlt) \in \mathcal{X}}
		L_\mathrm{Maxwell}(\epsVol,f(\epsVol),\HFieldPhasorDisreteRealAlt)
	}.
\end{equation}
Then, the updated weight is given by
\begin{equation}
	\lambda_\mathrm{Maxwell}^{(t+1)}
	= 0.9\,\lambda_\mathrm{Maxwell}^{(t)}
	+ 0.1\, s\, \gamma,
	\label{eq:AdaptiveLambda}
\end{equation}
where $s=0.25$ is a scaling factor  that prevents the physics-based term from dominating the total loss during training.

\section{Adaptation of WaveY-Net to the 3D scenario}\label{appendix:wavey}
The original WaveY-Net architecture described in \cite{Chen2022Stanford} is a two-dimensional convolutional neural network whose structure is based on the 2D U-Net~\cite{ronneberger2015}. WaveY-Net operates on planar sections of permittivity distributions and therefore receives 2D images as input. It predicts the corresponding two-dimensional electromagnetic field on this slice. In its implementation, WaveY-Net outputs only  $y$-components of the complex magnetic field amplitudes and its final layer therefore contains two channels representing the real and imaginary parts of these components.

To benchmark our proposed 3D surrogate model against the WaveY-Net approach, we implemented a volumetric extension, which we refer to as \emph{3D-WaveY-Net}. To transition from 2D to 3D, all convolutional, downsampling and upsampling operations in the original architecture were replaced with their 3D counterparts. Furthermore, to enable direct comparability with our model, which predicts the full complex magnetic field vector, we expand the output layer from two channels to six, corresponding to the real and imaginary parts of all three vector components. As in the original WaveY-Net, the output layer applies a hyperbolic tangent nonlinearity. In our 3D version, we use a scaled activation function as given in Eq.~\eqref{eq:activation}, which maps the outputs of the network to a predefined range of reasonable values. Overall, these modifications yield a 3D-WaveY-Net architecture that processes discretized 3D distributions of permittivities and outputs the corresponding real and imaginary parts of the full three-dimensional magnetic field.

In addition to the changes described above, further architectural modifications were necessary to ensure that the 3D-WaveY-Net remains computationally feasible in 3D. The original WaveY-Net architecture comprising blocks which consist of six convolutional layers with skip connections among them (cf.  Fig.~S1 in  \cite{Chen2022Stanford}). In comparison, the original 2D U-Net architecture uses shallower blocks containing only two convolutional layers with no internal skip connections. While such ``deep 2D blocks'' deployed in WaveY-Net are tractable for 2D input images, replicating these deeper blocks in a 3D network would immensely increase memory usage and computational costs.  
For this reason, our 3D-WaveY-Net employs shallower blocks of convolutional layers at each resolution level. More precisely, they consist of two 3D convolutional layers each of which are followed by an \texttt{InstanceNorm3d} layer and a \texttt{GELU} activation function \cite{torch}.

\section{Quantitative performance analysis with respect to E-fields}\label{appendix:efields}
\begin{figure}
	\centering
	\includegraphics[width=0.35\textwidth]{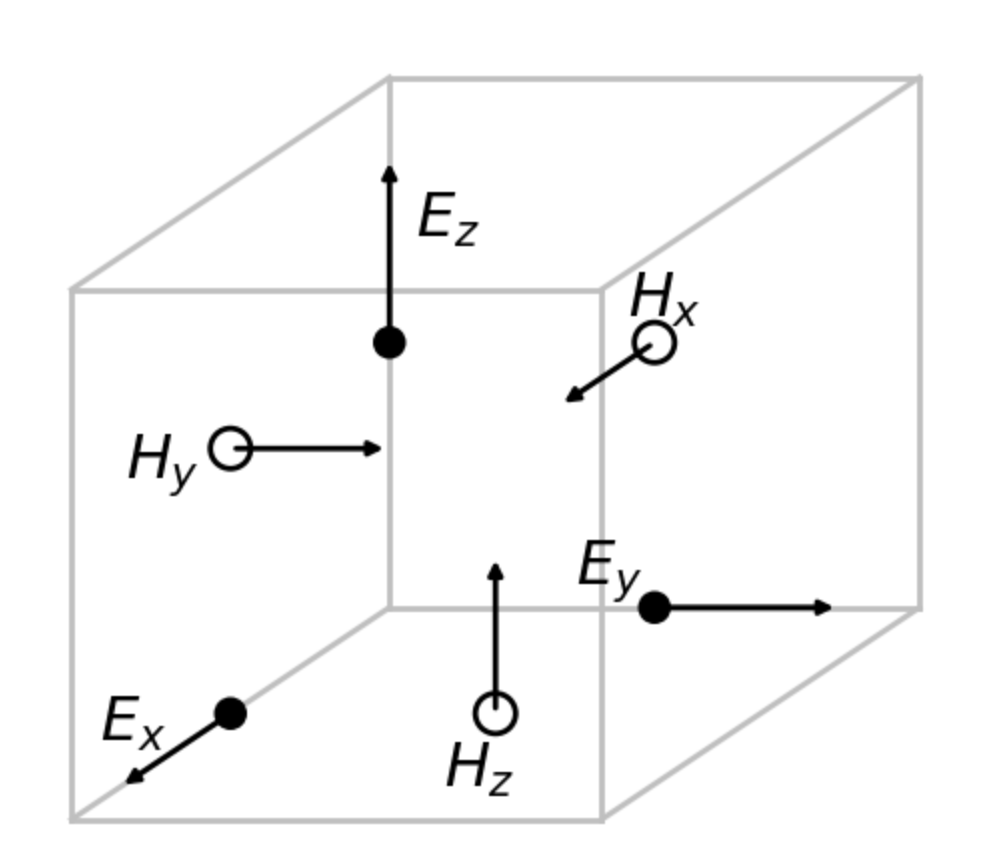}
	\caption{\textbf{Yee grid.} 
		Schematic of a single Yee cell illustrating the staggered grid of EM field components. 
		The grids of the components $E_\mathrm{x},E_\mathrm{y}, E_\mathrm{z}$ are shifted by 
		$(0.5,0,0)\rho$, $(0,0.5,0)\rho$ and $(0,0,0.5)\rho$, respectively, where $\rho$ is the grid spacing. For the grid of the components $H_\mathrm{x},H_\mathrm{y}, H_\mathrm{z}$ are shifted by $(0,0.5,0.5)\rho,(0.5,0,0.5)\rho$ and $(0.5,0.5,0)\rho$, respectively.}
	\label{fig:Yee:grid}
	
\end{figure}

A discretized field $\HFieldPhasorDisreteRealAlt=(H_\mathrm{x},H_\mathrm{y},H_\mathrm{z})$ associated with a 3D distribution $\epsVol$ is given on the staggered Yee grid, see Fig.~\ref{fig:Yee:grid}. By means of  Eq.~\eqref{eq:Maxwell:curlH} the magnetic field $\HFieldPhasorDisreteRealAlt$ can be used to determine the corresponding electric field $\EFieldPhasorDisreteRealAlt=(E_\mathrm{x},E_\mathrm{y},E_\mathrm{z})$ by
\begin{equation}\label{eq:ef}
	\EFieldPhasorDisreteRealAlt = \frac{1}{\mathrm{i}\omega \epssymbol_0 \epsVol} \nabla \times \HFieldPhasorDisreteRealAlt.
\end{equation}
Note that the partial derivatives involved in the curl operation in Eq.~\eqref{eq:ef} are approximated on the staggered Yee grid by finite difference quotients. For example,
\begin{equation}
	\frac{\partial H_\mathrm{z}}{\partial y}\!\left(i\cdot \rho ,(j+\frac{1}{2})\cdot \rho,k\cdot \rho\right)
	\;\approx\;
	\frac{
		H_\mathrm{z}\!\left(i\cdot \rho,(j+1)\cdot \rho,k\cdot \rho\right)
		-
		H_\mathrm{z}\!\left(i\cdot \rho,j\cdot \rho,k\cdot \rho\right)
	}{\Delta y}\,,
\end{equation}
for $(i\cdot \rho,j\cdot \rho,k\cdot \rho)\in \windowVolumeDiscrete$, where $\rho=\SI{50}{\nano\meter}$ is the grid spacing.

\begin{figure}[H]
	
	\subfloat[ground truth]{
		\begin{tikzpicture}
			\tikzPlotAdvanced{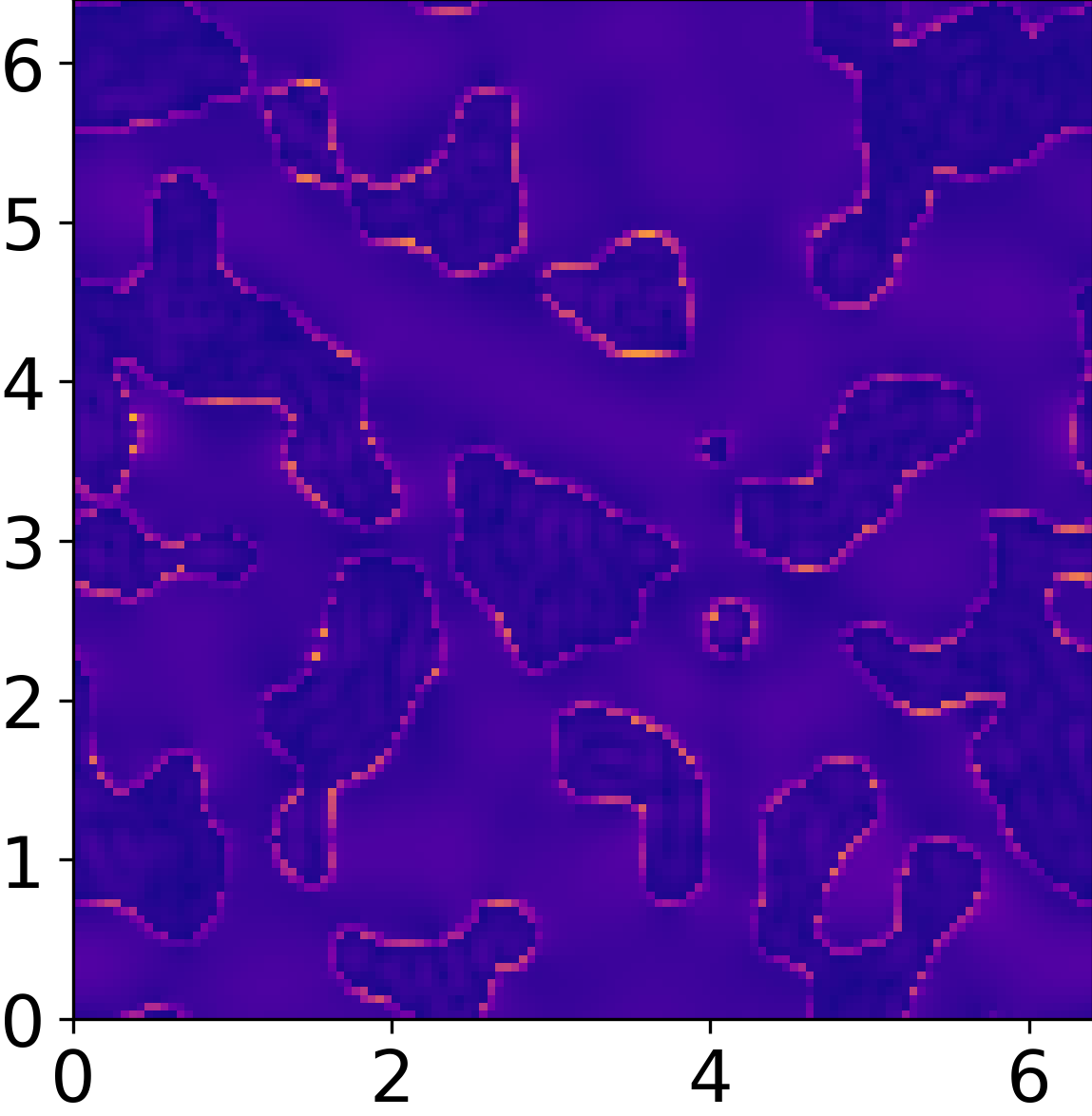}{$x$  [\si{\micro\meter}]}{$y$  [\si{\micro\meter}]}{0.25}{}{ref}{inner sep=0pt}
		\end{tikzpicture}
	}
	\subfloat[FNO]{
		\begin{tikzpicture}
			\tikzPlotAdvanced{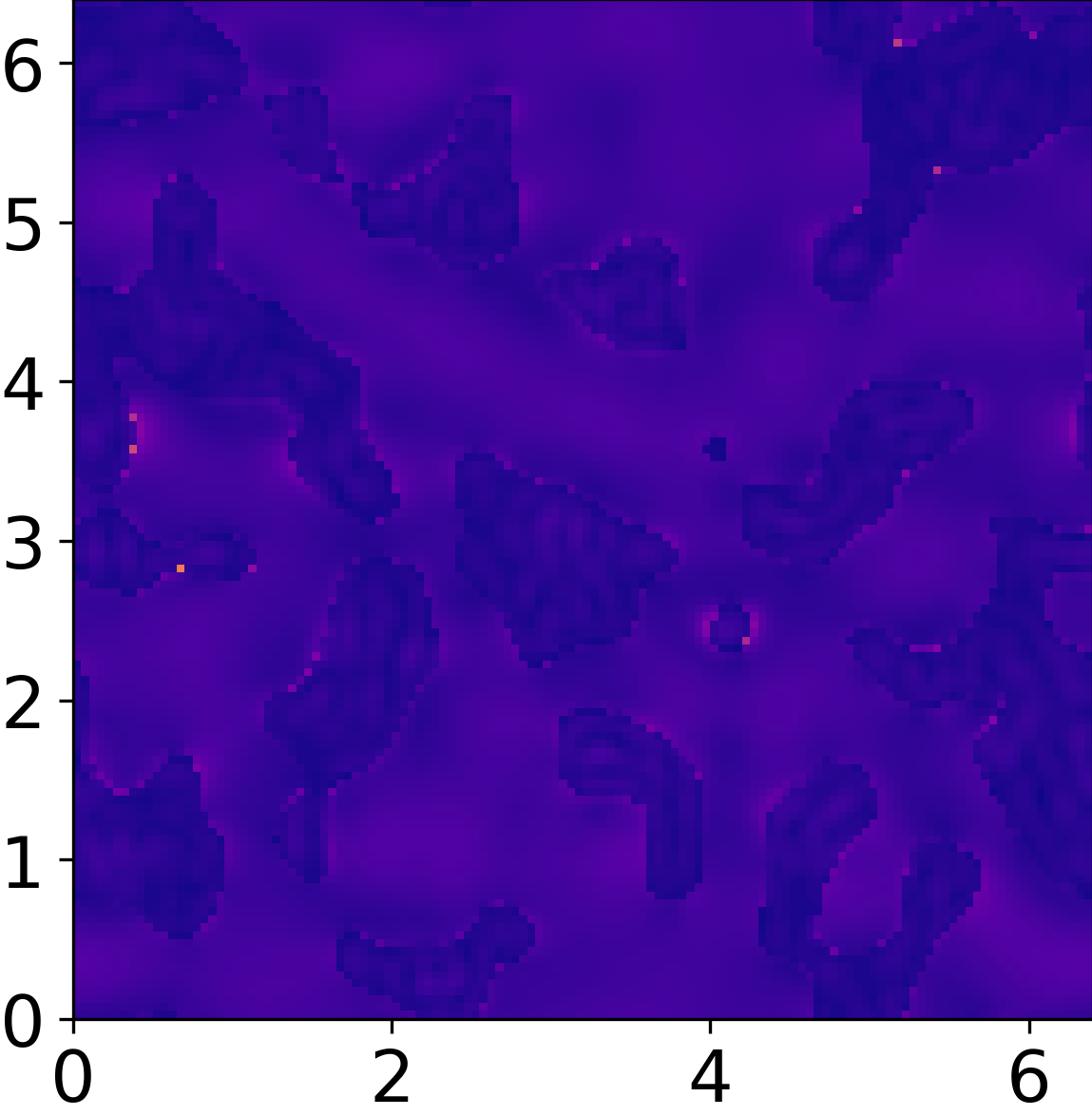}{$x$  [\si{\micro\meter}]}{$y$  [\si{\micro\meter}]}{0.25}{}{ref}{inner sep=0pt}
		\end{tikzpicture}
	}
	\subfloat[\FNOL]{
		\begin{tikzpicture}
			\tikzPlotAdvanced{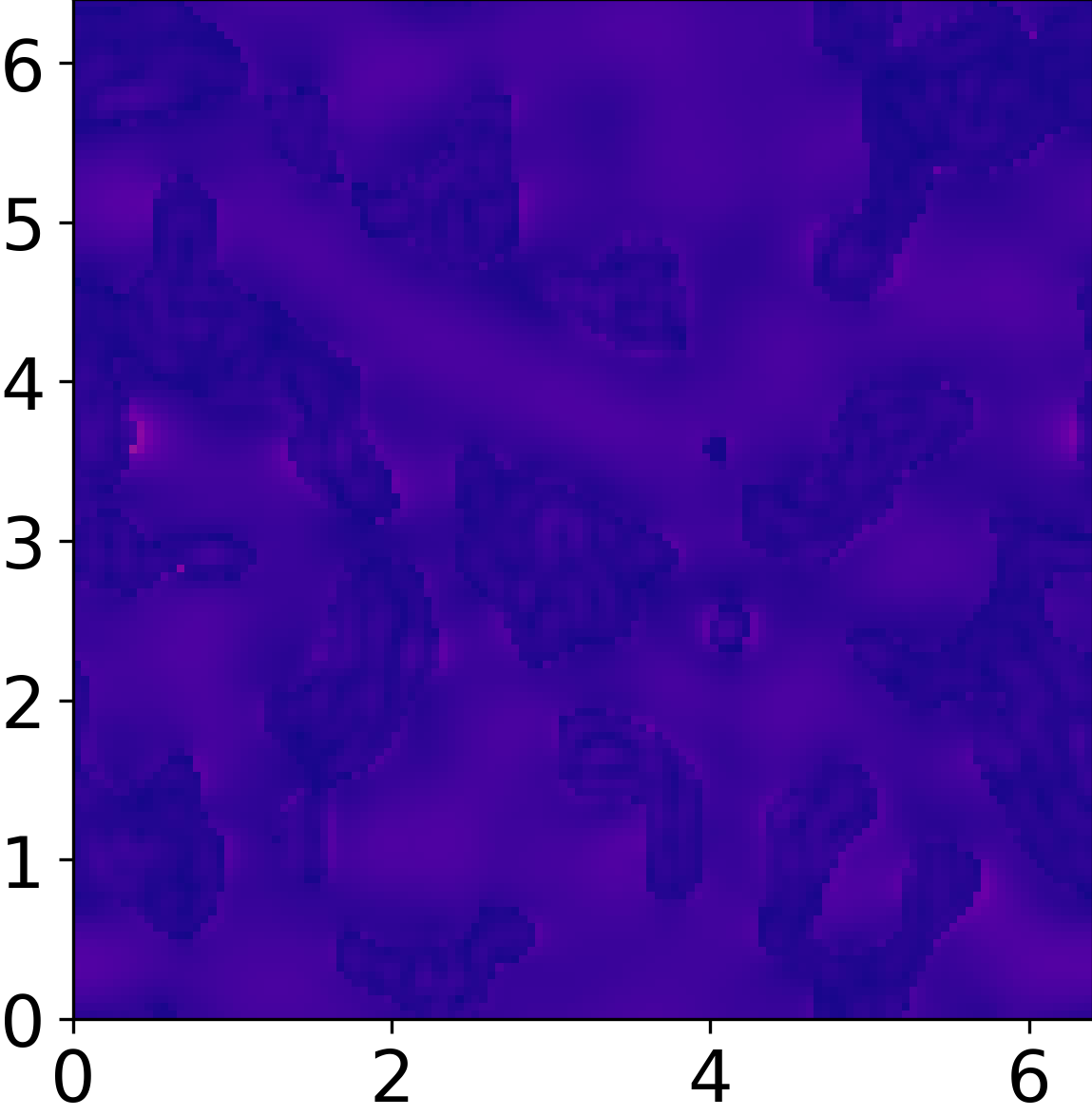}{$x$  [\si{\micro\meter}]}{$y$  [\si{\micro\meter}]}{0.25}{}{ref}{inner sep=0pt}
		\end{tikzpicture}
	}
	\vspace{-1em}
	
	\subfloat[\FNOConstr]{
		\begin{tikzpicture}
			\tikzPlotAdvanced{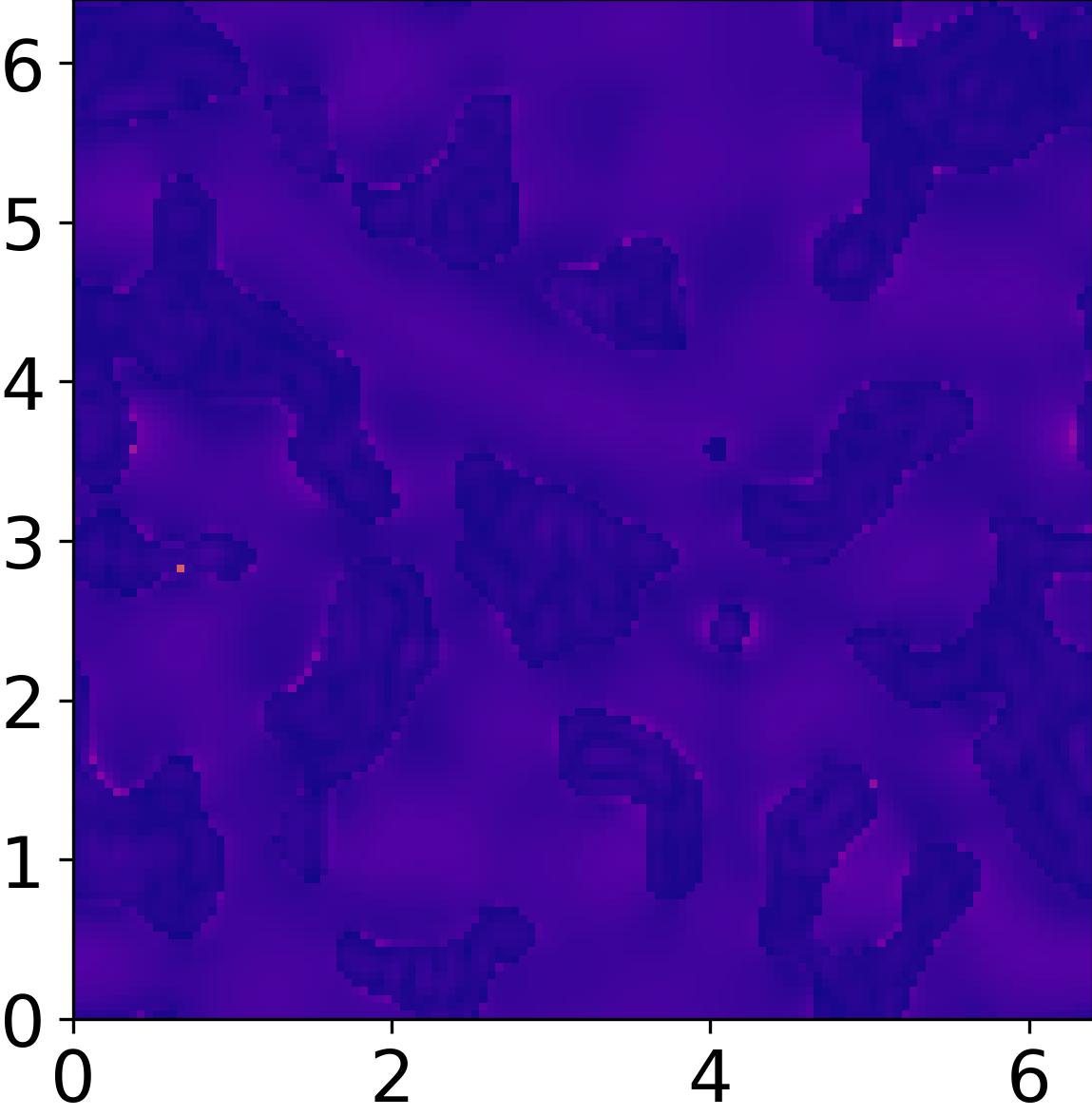}{$x$  [\si{\micro\meter}]}{$y$  [\si{\micro\meter}]}{0.25}{}{ref}{inner sep=0pt}
		\end{tikzpicture}
	}
	\subfloat[3D-WaveY-Net]{
		\begin{tikzpicture}
			\tikzPlotAdvanced{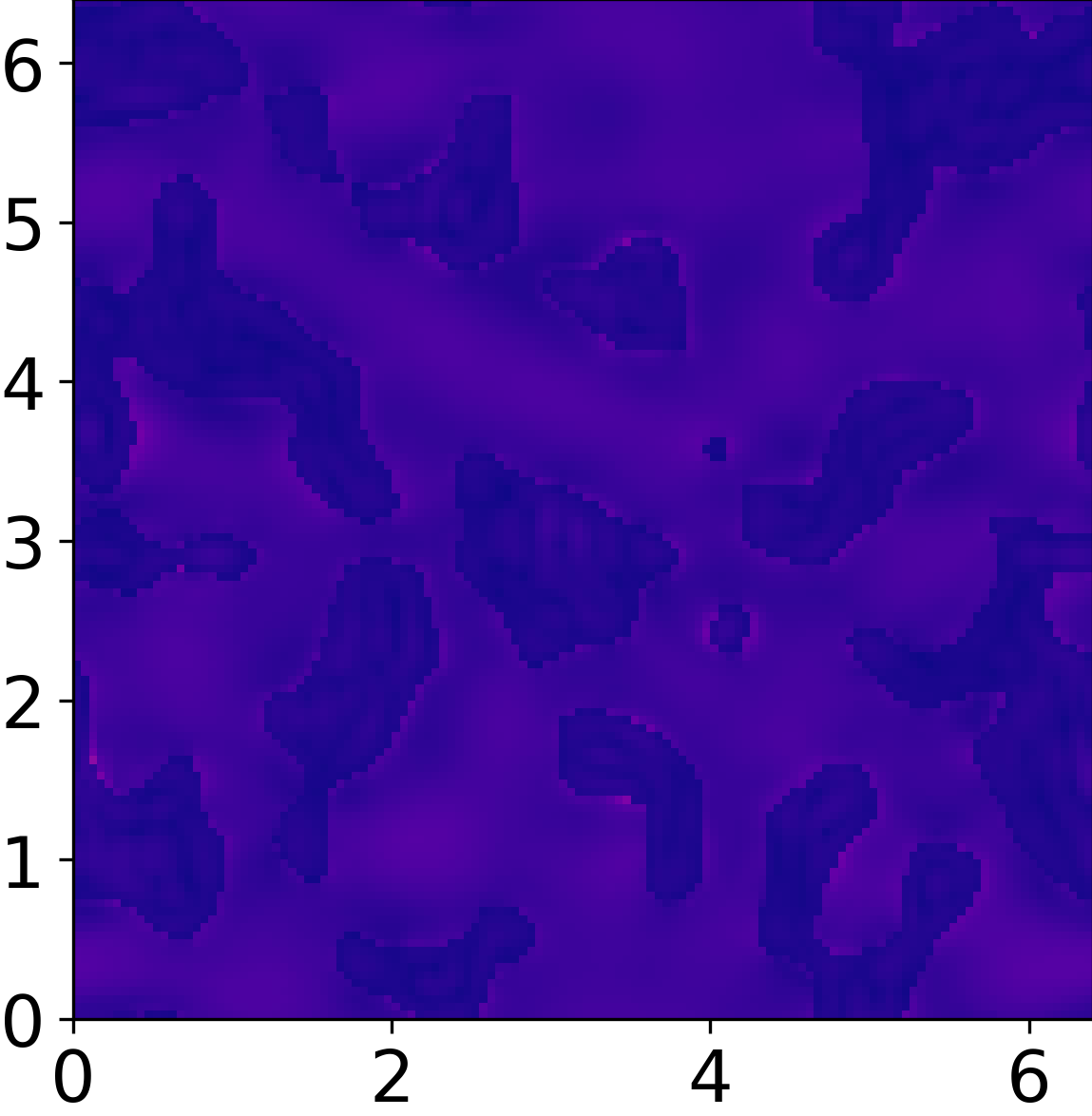}{$x$  [\si{\micro\meter}]}{$y$  [\si{\micro\meter}]}{0.25}{}{ref}{inner sep=0pt}
		\end{tikzpicture}
	}
	\subfloat[\WaveYConstr]{
		\begin{tikzpicture}
			\tikzPlotAdvanced{E_H_neural_oplimitted_data_xy.png}{$x$  [\si{\micro\meter}]}{$y$  [\si{\micro\meter}]}{0.25}{}{ref}{inner sep=0pt}
		\end{tikzpicture}
	}
	\raisebox{0.25cm}{
		\begin{tikzpicture}
			\node[inner sep=0pt ] (colorbar) {
				\includegraphics[width=1cm]{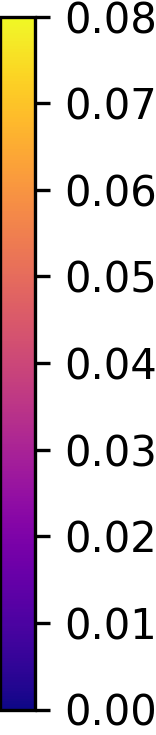}
			};
			\node[right=0.2cm of colorbar.east, anchor=west , inner sep=0pt,rotate=90 ] {
				$|\EFieldPhasorDisreteRealAlt|_2$
			};
		\end{tikzpicture}
	}

	\caption{\textbf{Visual validation.}  Ground truth of the absolute values of $\EFieldPhasorDisreteRealAlt$ in a planar section (a), and the corresponding absolute values of $\EFieldPhasorDisreteRealAlt^\mathrm{pred}$predicted by FNO (b), \FNOL{} (c), {\FNOConstr} (d), 3D-WaveY-Net (e) and \WaveYConstr{} (f).  }
	\label{fig:visual:validationE}	
\end{figure}	

Analogously, electric fields $\EFieldPhasorDisreteRealAlt^\mathrm{pred}=(E_\mathrm{x}^\mathrm{pred},E_\mathrm{y}^\mathrm{pred},E_\mathrm{z}^\mathrm{pred})$ associated with predicted magnetic fields $\HFieldPhasorDisreteRealAlt^\mathrm{pred}$ can be computed. 
Fig.~\ref{fig:visual:validationE}	visualizes a ground truth electric field $\EFieldPhasorDisreteRealAlt$ and predictions $\EFieldPhasorDisreteRealAlt^\mathrm{pred}$ determined by the five surrogate models.

\begin{figure}[h]
	\centering
	\subfloat[FNO]{
		\begin{tikzpicture}
			\tikzPlotAdvanced{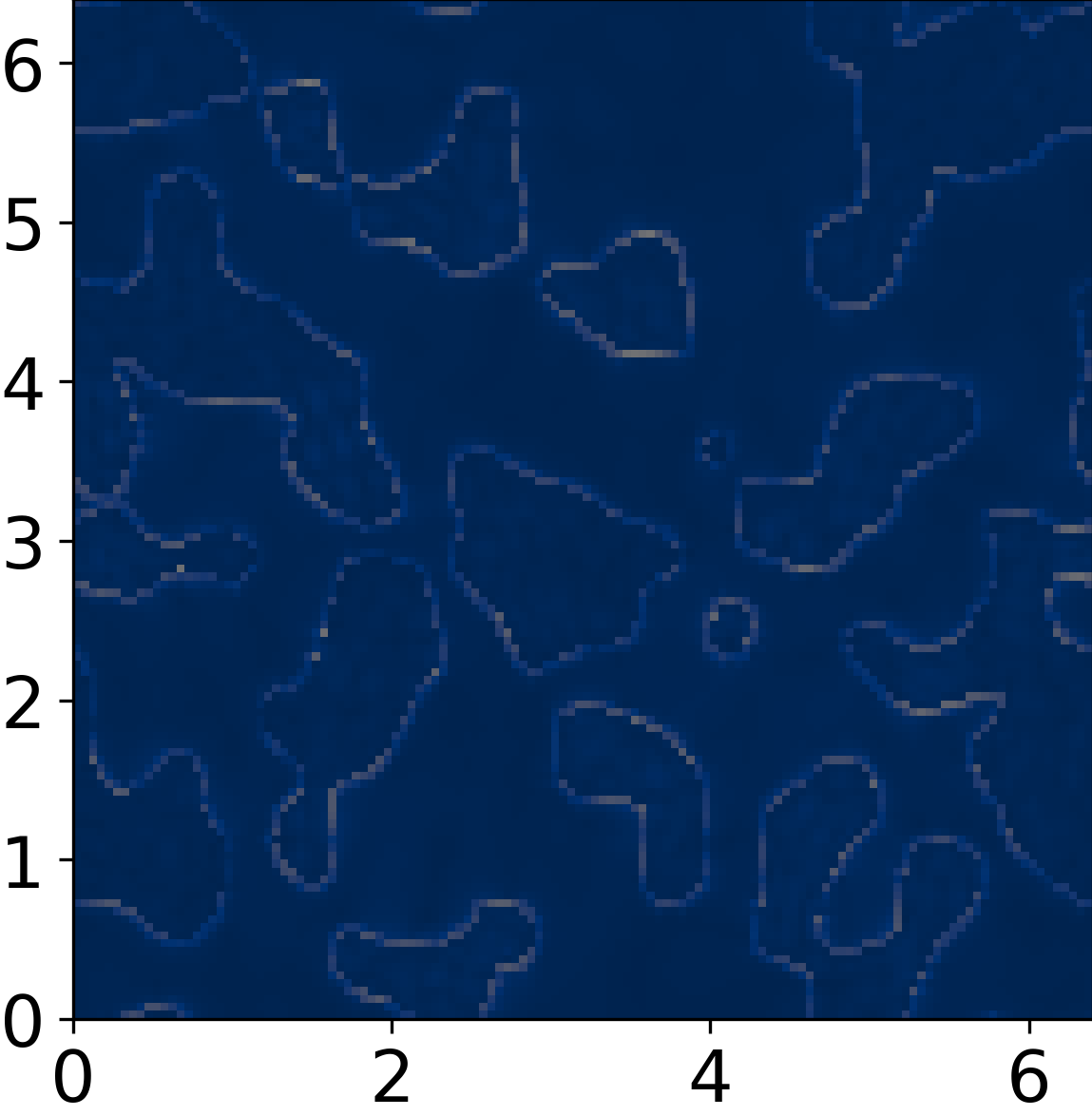}{$x$  [\si{\micro\meter}]}{$y$  [\si{\micro\meter}]}{0.25}{}{ref}{inner sep=0pt}
		\end{tikzpicture}
	}
	\subfloat[\FNOL]{
		\begin{tikzpicture}
			\tikzPlotAdvanced{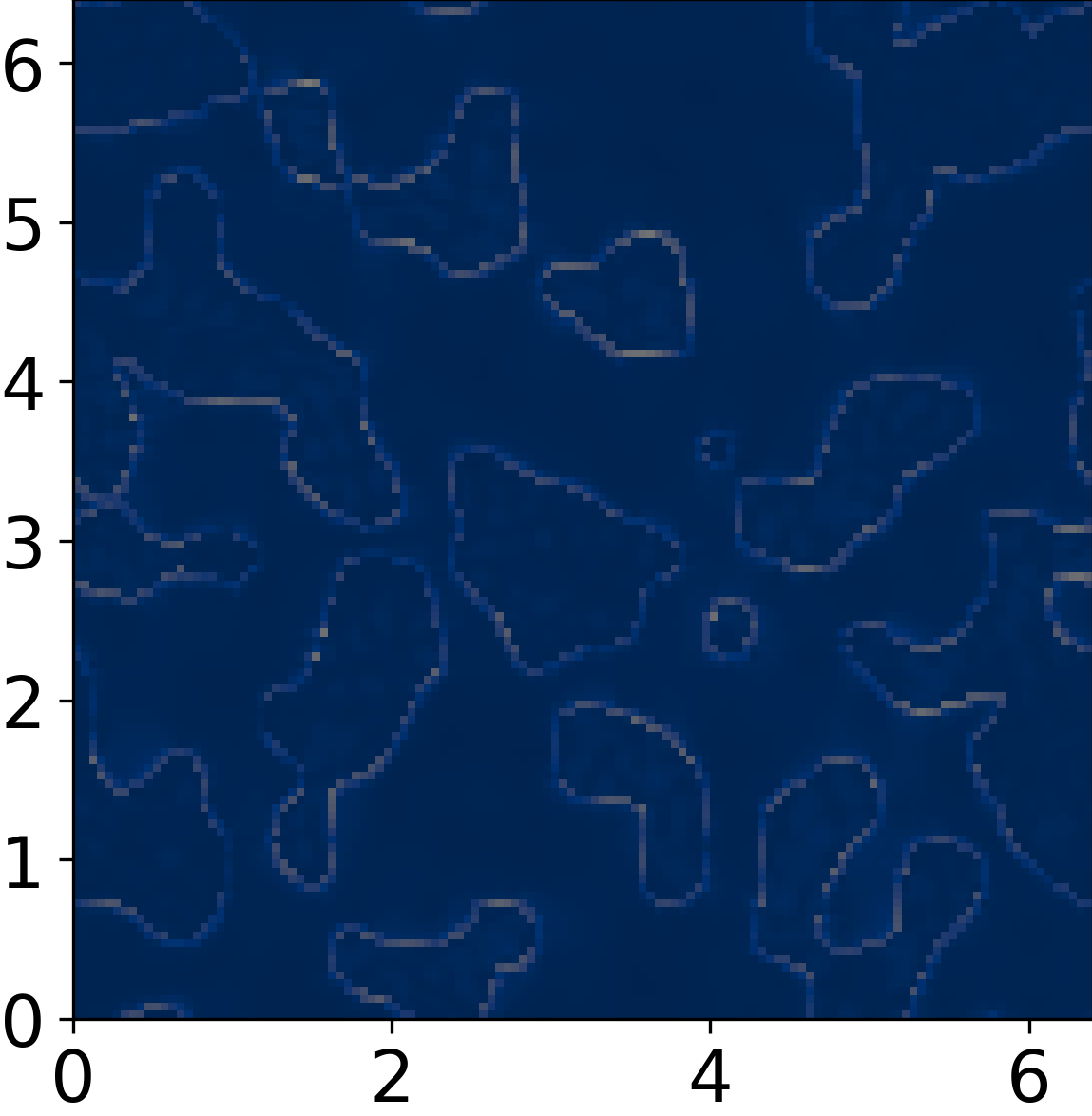}{$x$  [\si{\micro\meter}]}{$y$  [\si{\micro\meter}]}{0.25}{}{ref}{inner sep=0pt}
		\end{tikzpicture}
	}
	\subfloat[\FNOConstr]{
		\begin{tikzpicture}
			\tikzPlotAdvanced{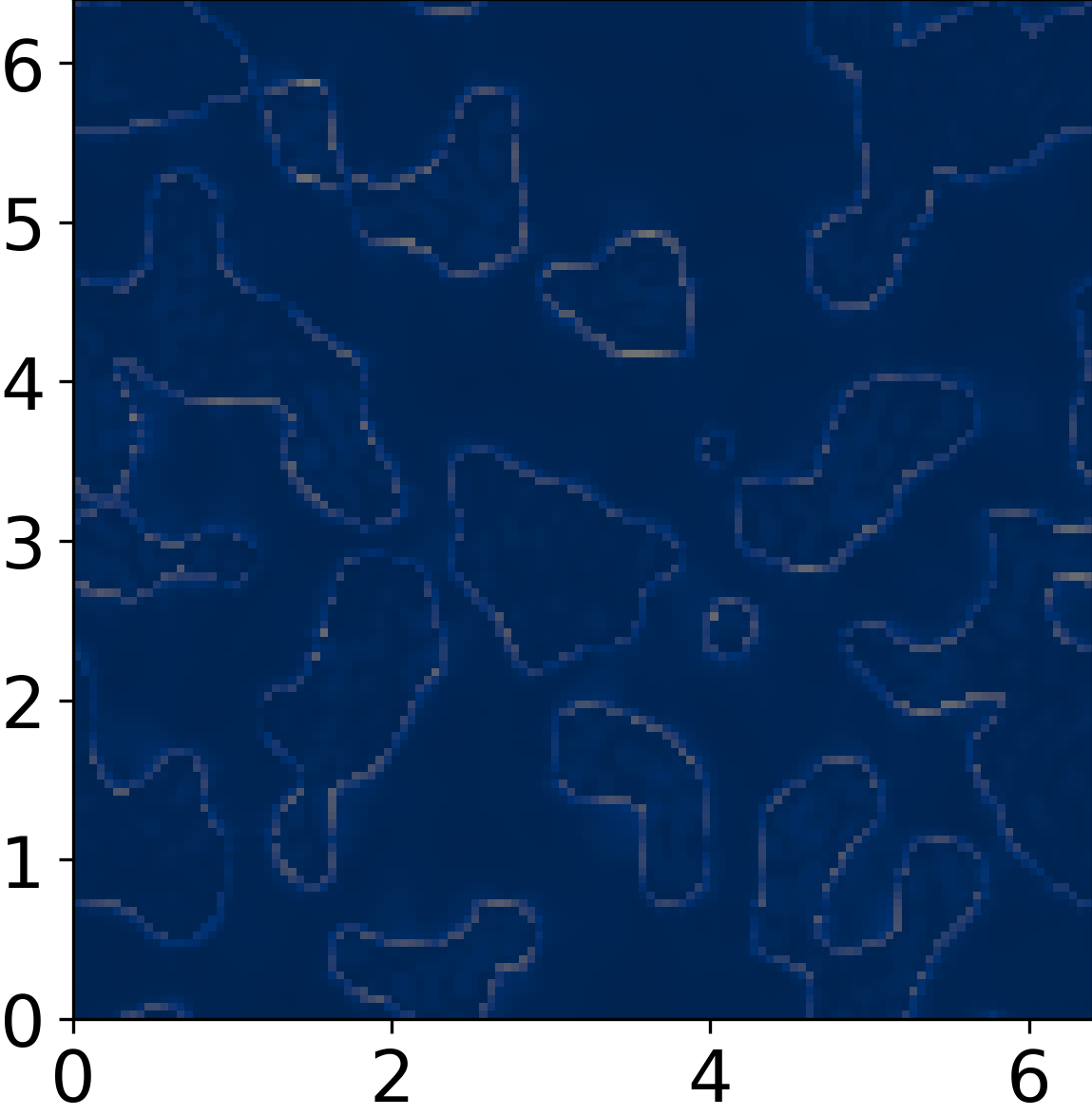}{$x$  [\si{\micro\meter}]}{$y$  [\si{\micro\meter}]}{0.25}{}{ref}{inner sep=0pt}
		\end{tikzpicture}
	}
	\vspace{-1em}
	
	\hspace{5em}
	\subfloat[3D-WaveY-Net]{
		\begin{tikzpicture}
			\tikzPlotAdvanced{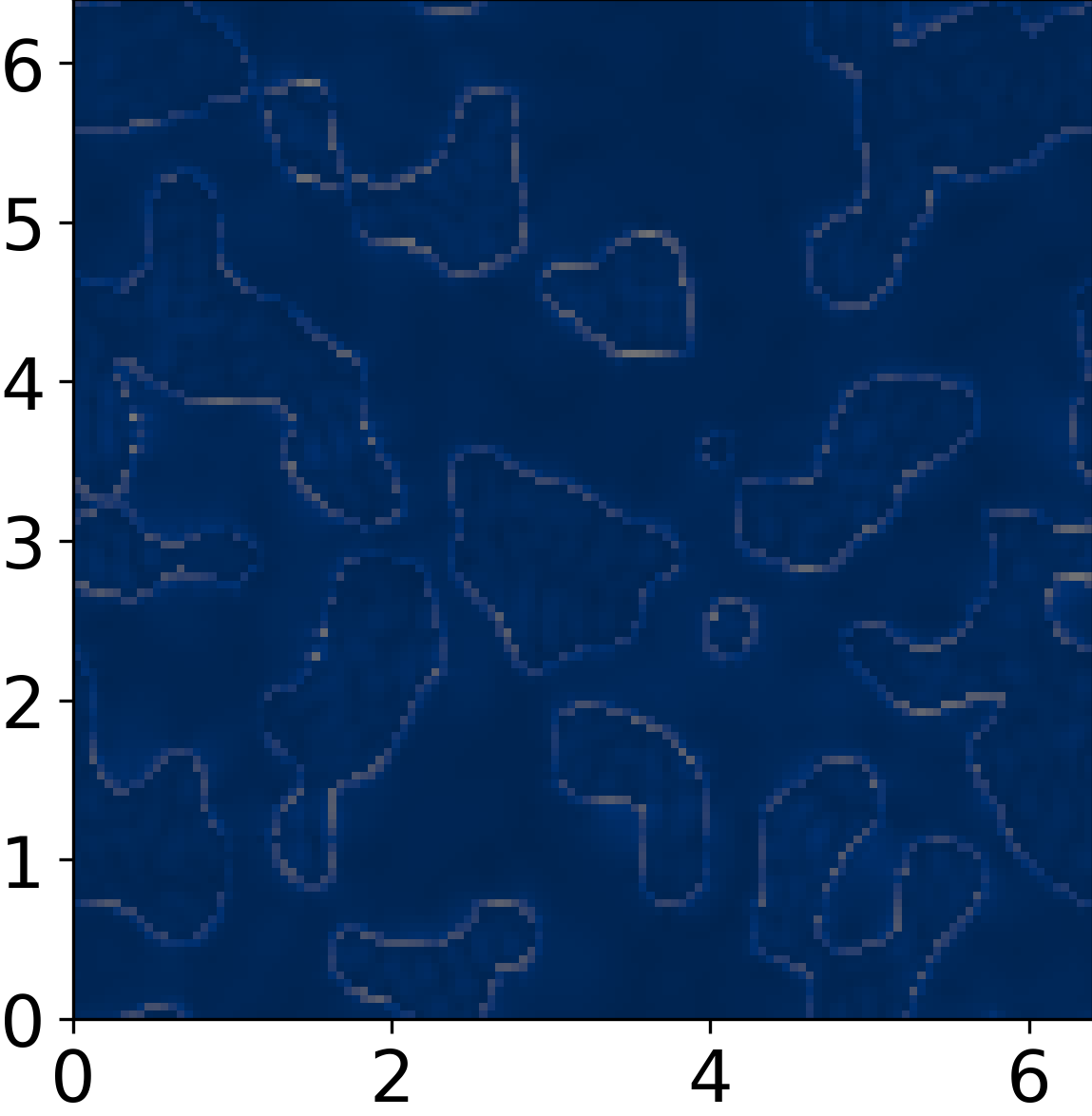}{$x$  [\si{\micro\meter}]}{$y$  [\si{\micro\meter}]}{0.25}{}{ref}{inner sep=0pt}
		\end{tikzpicture}
	}
	\subfloat[\WaveYConstr]{
		\begin{tikzpicture}
			\tikzPlotAdvanced{E_error_H_neural_oplimitted_data_xy.png}{$x$  [\si{\micro\meter}]}{$y$  [\si{\micro\meter}]}{0.25}{}{ref}{inner sep=0pt}
		\end{tikzpicture}
	}
	\raisebox{0.25cm}{
		\begin{tikzpicture}
			\node[inner sep=0pt ] (colorbar) {
				\includegraphics[width=1cm]{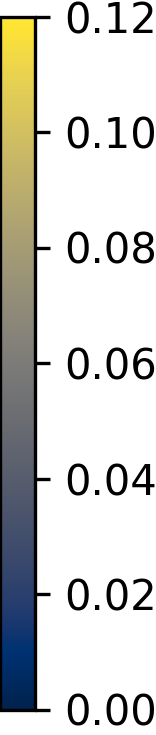}
			};
			\node[right=0.2cm of colorbar.east, anchor=west , inner sep=0pt,rotate=90,xshift=-1cm ] {
				$|\EFieldPhasorDisreteRealAlt-\EFieldPhasorDisreteRealAlt^{\mathrm{pred}}|_2$
			};
		\end{tikzpicture}
	}
	\caption{\textbf{Point-wise errors.} Absolute values of residuals $\EFieldPhasorDisreteRealAlt-\EFieldPhasorDisreteRealAlt^\mathrm{pred}$   in a planar section for FNO (a), \FNOL{} (b), {\FNOConstr} (c), 3D-WaveY-Net (d) and \WaveYConstr{} (e).  }
	\label{fig:visual:pointwiseE}	
\end{figure}

Corresponding point-wise errors of the magnetic fields shown in Fig.~\ref{fig:visual:validationE}	are visualized in Fig.~\ref{fig:visual:pointwiseE}.
To quantify the discrepancy between the predicted electric field 
$\EFieldPhasorDisreteRealAlt^{\mathrm{pred}}$ and the ground truth field 
$\EFieldPhasorDisreteRealAlt$, we use $\mathrm{nMAE}$ and $\mathrm{nRMSE}$ to quantify relative 
$L^p$ errors. More precisely, following the approach outlined in \cite{Chen2022Stanford} we consider
\begin{equation}
	e_{L^1}^{\mathbf{E}}
	(\EFieldPhasorDisreteRealAlt^\mathrm{pred},\EFieldPhasorDisreteRealAlt)
	= \frac{1}{2} \left(
	\mathrm{nMAE}(\mathrm{Re}(E_\mathrm{x}^\mathrm{pred}), \mathrm{Re}(E_\mathrm{x}))
	+
	\mathrm{nMAE}(\mathrm{Im}(E_\mathrm{x}^\mathrm{pred}), \mathrm{Im}(E_\mathrm{x}))
	\right),
\end{equation}
which we refer to as relative $L^1$ error. The relative $L^2$ error is given by
\begin{equation}
	e_{L^2}^{\mathbf{E}}
	(\EFieldPhasorDisreteRealAlt^\mathrm{pred},\EFieldPhasorDisreteRealAlt)
	= \frac{1}{2} \left(
	\mathrm{nRMSE}(\mathrm{Re}(E_\mathrm{x}^\mathrm{pred}), \mathrm{Re}(E_\mathrm{x}))
	+
	\mathrm{nRMSE}(\mathrm{Im}(E_\mathrm{x}^\mathrm{pred}), \mathrm{Im}(E_\mathrm{x}))
	\right).
\end{equation}
These metrics quantify the global deviation between 
prediction and ground truth with respect to the $L^p$ norm and normalize it by the 
corresponding $L^p$ magnitude of the ground truth field, thereby providing a 
scale-invariant measure of overall prediction accuracy.
Note that, the metrics are computed for the dominant component of the electric field, namely the $x$-component.

To evaluate how accurately the surrogate models reproduce the magnitudes of 
electric fields, we consider the relative error $e_{\mathrm{amp}}^\mathbf{E}$ of field amplitudes which is given by
\begin{equation}
	e_{\mathrm{amp}}^\mathbf{E}
	=
	\frac{1}{|W_\mathrm{d,const.}^\mathrm{3D}|}
	\sum_{\rvec \in W_{\mathrm{d,const.}}^{\mathrm{3D}}}
	\left|
	\frac{
		|\EFieldPhasorDisreteRealAlt^{\mathrm{pred}}(\rvec)|_2 - 	|\EFieldPhasorDisreteRealAlt(\rvec)|_2
	}{
		|\EFieldPhasorDisreteRealAlt(\rvec)|_2 + \tau
	}
	\right|,
\end{equation}
where 
$\tau = 10^{-10}$  is a constant that is introduced in the denominator to avoid numerical instability.

\begin{figure}[h]
	\centering
	\subfloat[]{
		\begin{tikzpicture}
			\tikzPlotAdvanced{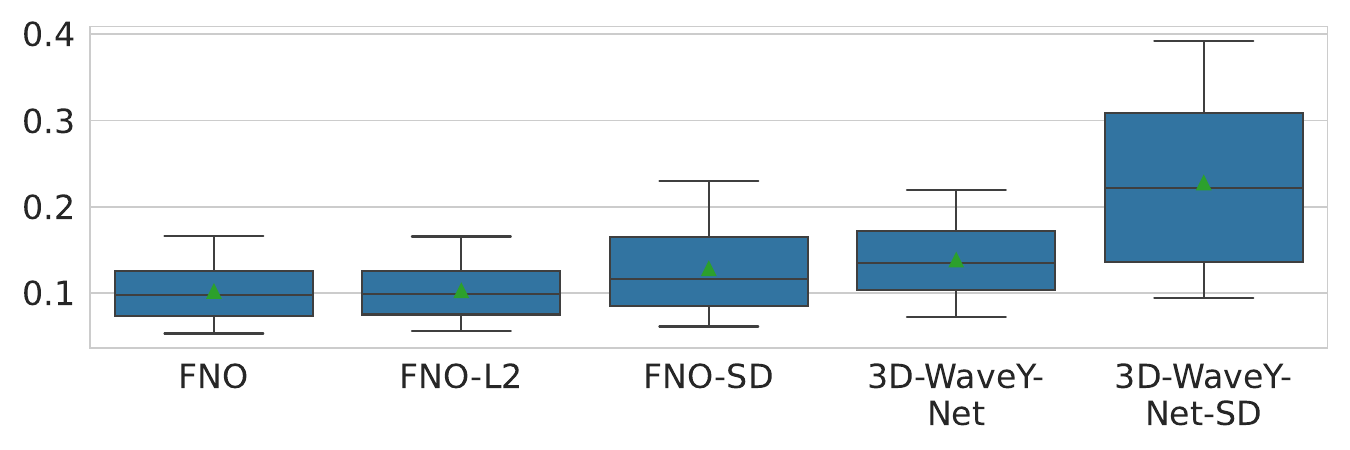}{}{$e_{L^1}^{\mathbf{E}}$}{0.45}{}{ref}{inner sep=0pt}
		\end{tikzpicture}
	}
	\subfloat[]{
		\begin{tikzpicture}
			\tikzPlotAdvanced{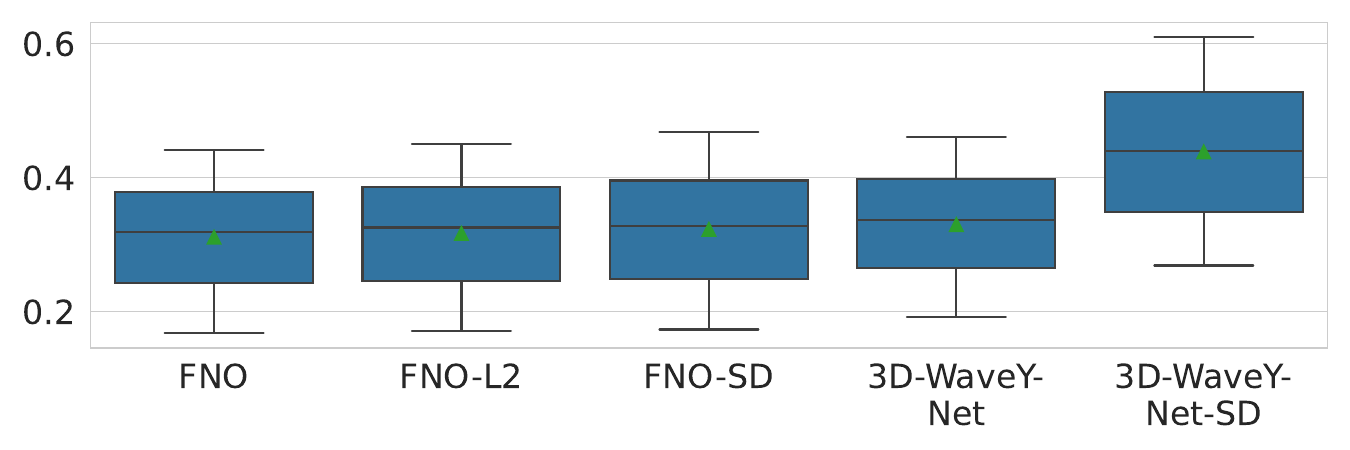}{}{$e_{L^2}^\mathbf{E}$}{0.45}{}{ref}{inner sep=0pt}
		\end{tikzpicture}
	}
	\vspace{-1em}
	
	\subfloat[]{
		\begin{tikzpicture}
			\tikzPlotAdvanced{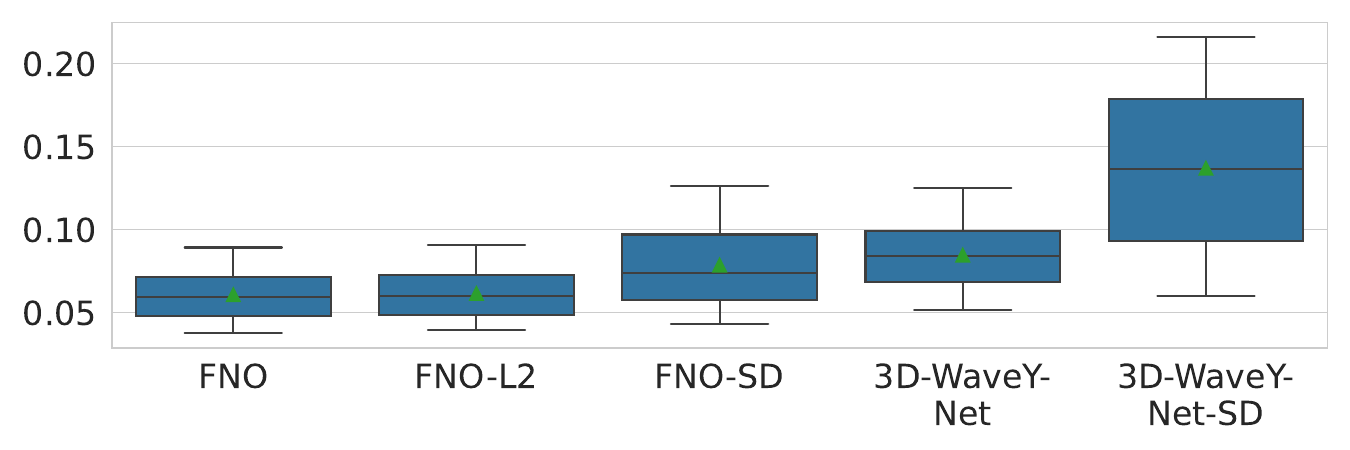}{}{$e_{\mathrm{amp}}^\mathbf{E}$}{0.6}{}{ref}{inner sep=0pt}
		\end{tikzpicture}
	}

	\caption{\textbf{Quantitative evaluation of surrogate model performance.}
		Box plots of metrics  
		(a)  relative $L^1$ error $e_{L^1}^\mathbf{E}$, 
		(b)  relative $L^2$ error $e_{L^2}^\mathbf{E}$ and 
		(c) amplitude error $e_{\mathrm{amp}}^\mathbf{E}$  computed over the electric fields derived from the test data.
		Box heights represent the interquartile range; horizontal lines within boxes mark medians; the whiskers extend to the 5th and 95th percentiles; the green triangles indicate the mean.
	}
	\label{fig:quantitativeE}
\end{figure}

\begin{figure}[H]
	\centering
	\subfloat[FNO]{
		\begin{tikzpicture}
			\tikzPlotAdvanced{L1_vs_area_fraction_smoothed.pdf}{$\varphi_\mathrm{area}$}{$e_{L^1}^\mathbf{E}$}{0.4}{}{ref}{inner sep=0pt}
		\end{tikzpicture}
	}
	\subfloat[\FNOL]{
		\begin{tikzpicture}
			\tikzPlotAdvanced{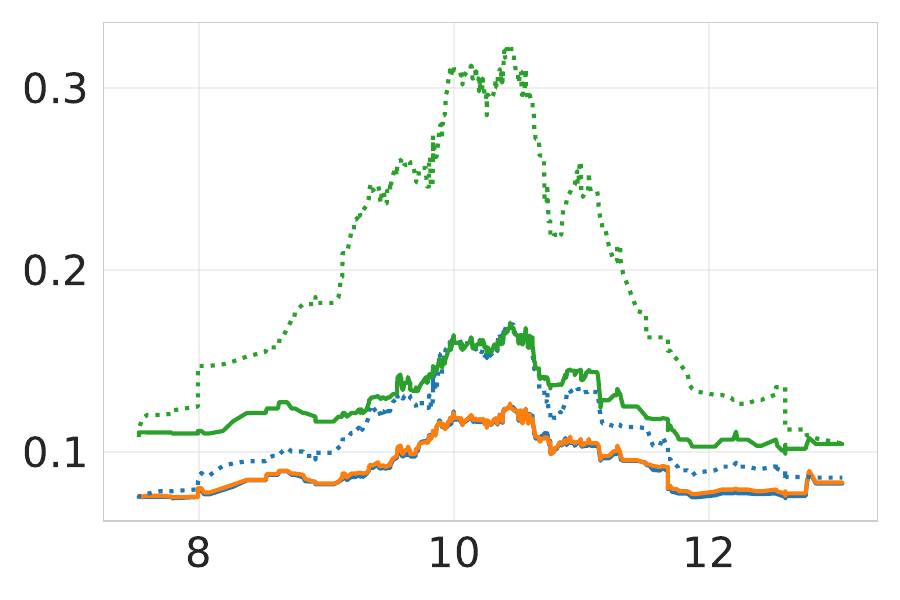}{$\varphi_\mathrm{mean}$}{$e_{L^1}^\mathbf{E}$}{0.4}{}{ref}{inner sep=0pt}
		\end{tikzpicture}
	}
	
	\begin{tikzpicture}
		\node[inner sep=0pt]{\includegraphics[width=0.5\textwidth]{legend_models.pdf}};
	\end{tikzpicture}
	\vspace{-1em}
	
	\subfloat[\FNOL]{
		\begin{tikzpicture}
			\tikzPlotAdvanced{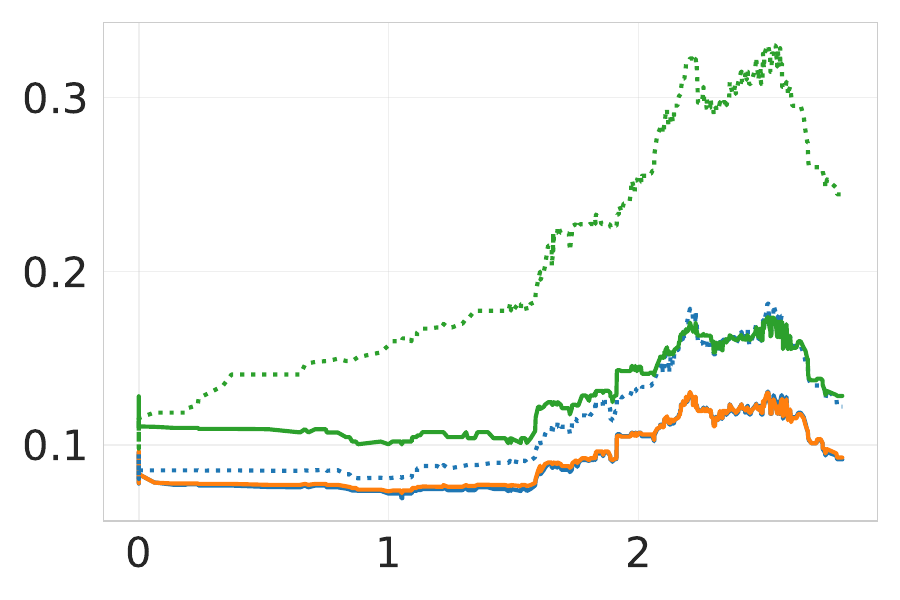}{$\varphi_\mathrm{std}$}{$e_{L^1}^\mathbf{E}$}{0.4}{}{ref}{inner sep=0pt}
		\end{tikzpicture}
	}	
	\subfloat[\FNOConstr]{
		\begin{tikzpicture}
			\tikzPlotAdvanced{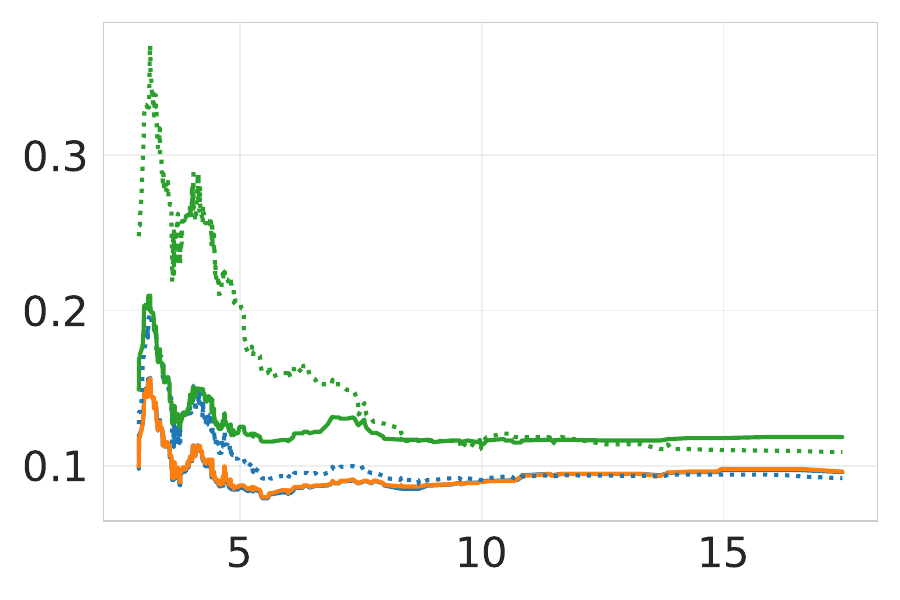}{$\varphi_\mathrm{radius}$}{$e_{L^1}^\mathbf{E}$}{0.4}{}{ref}{inner sep=0pt}
		\end{tikzpicture}
	}

	\caption{
		\textbf{Dependence of prediction error on metasurface descriptors.} Dependence of prediction error on metasurface descriptors. Moving-median curves showing the relative $L^1$ error $e_{L^1}^\mathbf{E}$ as a function of (a) area fraction $\varphi_{\mathrm{area}}$, (a) mean relative permittivity $\varphi_{\mathrm{mean}}$, (c) standard deviation $\varphi_{\mathrm{std}}$, and (d) mean radius $\varphi_{\mathrm{radius}}$ for the surrogate models. For each subfigure, the considered descriptor and the $e_{L^1}$  have been determined on the test dataset.  A moving median with a window size of 50 samples was then applied to the sorted error values to obtain smooth trends.
	}
	\label{fig:correlation:dependenceE}
\end{figure}

The metrics described above are computed for each individual sample in the test dataset and for each of the trained surrogate models. 
Statistics of the metrics computed in this manner are visualized in Fig.~\ref{fig:quantitativeE}, which summarizes the resulting distributions over test sets. Mean values and standard deviations of these metrics are listed in Table~\ref{tab:Efield}. 
The influence of metasurface descriptors on the performance of surrogate models is shown in Fig.~\ref{fig:correlation:dependenceE}.

Overall, the quantitative analysis performed on the electric fields in this section exhibits trends similar to those observed for the magnetic fields, see Section~\ref{sec:results}. Overall, the results indicate that the FNO-based surrogate models provide the most robust and consistent predictive performance across all considered metrics.

\begin{table}[h]
	\caption{Mean and standard deviation of performance metrics of surrogate models computed over the test set with respect to electric fields (instead of the magnetic fields). The lowest mean error for each metric is highlighted in bold.}\label{tab:Efield}
	\centering
	\begin{tabular}{lccc}
		\toprule
		& $e_{L^1}^{\mathbf{E}}$ & $e_{L^2}^{\mathbf{E}}$ & $e_{\mathrm{amp}}^{\mathbf{E}}$ \\
		\midrule
		FNO & \textbf{0.102 $\pm$ 0.035} & \textbf{0.311 $\pm$ 0.089} & \textbf{0.061 $\pm$ 0.018} \\
		FNO-L2 & 0.103 $\pm$ 0.034 & 0.316 $\pm$ 0.091 & \textbf{0.061 $\pm$ 0.017} \\
		FNO-SD & 0.128 $\pm$ 0.053 & 0.322 $\pm$ 0.096 & 0.078 $\pm$ 0.026 \\
		3D-WaveY-Net & 0.138 $\pm$ 0.046 & 0.329 $\pm$ 0.089 & 0.084 $\pm$ 0.023 \\
		3D-WaveY-Net-SD & 0.228 $\pm$ 0.098 & 0.438 $\pm$ 0.106 & 0.137 $\pm$ 0.050 \\
		\bottomrule
	\end{tabular}
\end{table}

\end{document}